\documentclass[12pt]{article}

\usepackage{graphicx}
\usepackage{amsmath}
\usepackage{amssymb}
\usepackage{authblk}
\usepackage{braket}
\usepackage{hyperref}
\usepackage{frontespizio}
\usepackage{color}

\usepackage{caption}

\usepackage{cite}

\usepackage{subfigure}
\newcommand{\bea}{\begin{eqnarray}}
\newcommand{\eea}{\end{eqnarray}}
\newcommand{\be}{\begin{equation}}
\newcommand{\ee}{\end{equation}}
\newcommand{\ba}{\begin{align}}
\newcommand{\ea}{\end{align}}
\newcommand{\V}{\mathcal{V}}
\newcommand{\Kahler}{\ensuremath{\text{K}\ddot{\text{a}}\text{hler}\,}}

\newcommand{\ns}{n_{\rm{s}}}

\begin{document}

\title{{\bf \LARGE{Oscillons from String Moduli}}\vspace{0.6cm}
  }

\author[1,2]{\bf\small Stefan Antusch\hspace{1pt}}
\author[1]{\bf \small Francesco Cefal\`a\hspace{1pt}}
\author[3]{\bf \small Sven Krippendorf{\hspace{2pt}}}
\author[3]{\bf\small Francesco Muia\hspace{1pt}}
\author[1]{\bf\small Stefano Orani\hspace{1pt}}
\author[4,5]{\bf\small Fernando Quevedo\hspace{1pt}\vspace{0.4cm}}
\affil[1]{\small \it Department of Physics, University of Basel, Klingelbergstr. 82, CH-4056 Basel, Switzerland}
\affil[2]{\small \it Max-Planck-Institut f\"ur Physik (Werner-Heisenberg-Institut), F\"ohringer Ring 6, D-80805 M\"unchen, Germany}
\affil[3]{\small \it Rudolf Peierls Centre for Theoretical Physics, 1 Keble Road, Oxford, OX1 3NP, UK}
\affil[4]{\small \it ICTP, Strada Costiera 11, Trieste 34014, Italy}
\affil[5]{\small \it DAMTP, Centre for Mathematical Sciences, Wilberforce Road, Cambridge, CB3 0WA, UK}
\date{\small\today}
\maketitle

\begin{abstract}
A generic feature of string compactifications is the presence of many scalar fields, called moduli. Moduli are usually displaced from their post-inflationary minimum during inflation. Their relaxation to the minimum could lead to the production of oscillons: localised, long-lived, non-linear excitations of the scalar fields. Here we discuss under which conditions oscillons can be produced in string cosmology and illustrate their production and potential phenomenology with two explicit examples: the case of an initially displaced volume modulus in the KKLT scenario and the case of a displaced blow-up K\"ahler modulus in the Large Volume Scenario (LVS). One, in principle, observable consequence of oscillon dynamics is the production of gravitational waves which, contrary to those produced from preheating after high scale inflation, could have lower frequencies, closer to the currently observable range. We also show that, for the considered parameter ranges, oscillating fibre and volume moduli do not develop any significant non-perturbative dynamics. Furthermore, we find that the vacua in the LVS and the KKLT scenario are stable against local overshootings of the field into the decompatification region, which provides an additional check on the longevity of these metastable configurations.
\end{abstract}
\newpage

\tableofcontents

\section{Introduction}
Moduli fields, scalar fields describing the size and shape of extra dimensions, are generic in string theory. It is well-known that moduli are displaced from their low-energy minimum during and after inflation~\cite{Coughlan:1983ci,Banks:1993en,deCarlos:1993wie}. As it is expected that there are $\mathcal{O}(100)$ moduli fields, it is a natural question to ask for phenomenological consequences of moduli displacement in the early Universe. For example, it has been argued that these moduli lead to a deviation from the standard thermal history of a radiation dominated Universe shortly after inflation. An additional period of matter (i.e.~moduli) domination takes place. This can lead to additional contributions of dark radiation~\cite{Cicoli:2012aq, Higaki:2012ar}, baryogenesis~\cite{Allahverdi:2016yws} and a non-thermal production of dark matter~\cite{Allahverdi:2013noa} (see~\cite{Kane:2015jia} for a review).

Moduli fields $\phi$ become dynamical when the relevant scales in the potential $V(\phi)$ are of order the Hubble scale $H$, in particular when the friction term $3H\dot{\phi}$ is of order $V'(\phi_{\rm displaced}).$ 
Phenomenologically, the absence of a cosmological moduli problem~\cite{Coughlan:1983ci,Banks:1993en,deCarlos:1993wie} sets a lower scale of roughly 30 TeV to the masses of moduli. Advances in our understanding of moduli potentials allow us to address the relevance of moduli in the early Universe. By now there are well worked out scenarios of moduli stabilisation in various corners of string theory and much attention has been devoted to the type IIB scenarios of KKLT~\cite{Kachru:2003aw} and the Large Volume Scenario (LVS)~\cite{Balasubramanian:2005zx,Conlon:2005ki}. For concreteness we concentrate in this paper on these scenarios of type IIB string theory. Here, the moduli masses are hierarchically below the string scale and are described within a four dimensional effective ${\cal N}=1$ supergravity field theory.

This property is generic in string compactifications with low-energy supersymmetry for which the presence of such light moduli can be understood from general field theory arguments in supergravity~\cite{deCarlos:1993wie, 0804.3730, 0804.1073,0706.2785,hep-th/0602246,hep-th/0606273}. The lightest non-axionic modulus, is associated to the scalar partner of the Goldstino. Its mass is lower or equal to the gravitino mass of a given setup. In string theory, the consistency of the four dimensional ${\cal N}=1$ effective field theory requires that the gravitino mass and hence at least one modulus has a mass below the compactification and the string scale.

In this article we consider  a possible phenomenological consequence of moduli displacement: the production of oscillons. Oscillons are spatially localised oscillating scalar field configurations which are long-lived with respect to the characteristic time scale of oscillations. In these regions the modulus oscillates with large amplitude. They have been studied in the context of field theory models (of inflation)~\cite{Gleiser:1993pt,Copeland:1995fq,Copeland:2002ku,Broadhead:2005hn,Farhi:2005rz,Fodor:2006zs,Graham:2006vy,Gleiser:2007te,Amin:2011hj,Achilleos:2013zpa,Gleiser:2014ipa,Antusch:2015nla,Antusch:2015ziz,Bond:2015zfa,Liu:2017hua,Antusch:2016con} including potentials that appear in axion monodromy inflation in string theory~\cite{1304.6094} (although neglecting moduli stabilisation). The production of oscillons depends on whether the shape of the potential supports such localised scalar field configurations as well as the sufficient growth of perturbations\footnote{For a review on the non-perturbative dynamics during preheating after inflation, see e.g.~\cite{Amin:2014eta}.} and  the corresponding cosmological history. 

Oscillons have been found to be present in many field theories for single real scalar fields  for which the corresponding scalar potential flattens out close to its minimum. This causes an effective attractive interaction among the scalar field particles that triggers these time-dependent energy lumps to form. When present, oscillons can substantially affect the cosmological evolution of the Universe. They can dominate the energy density  before decaying and delay thermalisation. They could also catalyse the implementation of second and relatively weak first order phase transitions, be a source of baryon asymmetry, gravitational waves, etc. It is then important to see if they have a role in string cosmology during and after inflation.

 In this article we start exploring the role of oscillons in the rich structure of scalar potentials for string theory moduli. Given the magnitude of the energy landscape with hundreds  of scalar fields,  quasi-stable field configurations could play an important role in the cosmological dynamics of the theory.
 With respect to moduli displacements in the early Universe, we present two examples where we find the production of oscillons numerically. The first example is that of a displaced volume modulus in KKLT. The second example is based on the LVS where we find oscillons from displacing so-called blow-up moduli.\footnote{For this example, our analysis extends the work in~\cite{Barnaby:2009wr} in which reheating after K\"ahler moduli inflation was studied in detail although the presence of oscillons was not discussed.}
 We also find that in the context of string moduli the production of oscillons is model dependent. Some moduli potentials do not give rise to oscillons such as the overall volume  or K\"ahler moduli associated to fibrations in the LVS. In the current analysis, which should serve as a proof of concept, we treat the initial displacement as a phenomenological parameter. Connecting with a fully-fledged UV setup including an early inflationary period (or alternatives) the initial displacement would be explicitly determined. Here we will not assume a  particular model of string inflation but consider only the dynamics of moduli fields when relaxing towards their zero-temperature vacuum. 

Depending on the type of  configuration, oscillons can produce gravitational radiation, which has been studied in the context of axion monodromy inflation~\cite{1304.6094} and more recently during preheating after hilltop inflation~\cite{Antusch:2016con}, and leads to a distinct phenomenological signal of oscillons. In general, the frequency of gravitational waves (GWs) is related to the energy scales involved in the scalar (moduli) potential. This scale can be distinct from the energy scales responsible for GWs from preheating after inflation. This hence can lead to GWs produced at lower frequencies, potentially in the range experimentally tested by current experiments (e.g.~LIGO). Both of our examples lead to a production of stochastic gravitational waves which can feature a peak at the characteristic oscillon frequency. More generally, each moduli potential has its characteristic GW spectrum from various GW production mechanisms and carries a rich information about the field dynamics in this period. We also comment on how a subsequent period of moduli domination (until the lightest modulus decays before BBN) affects the stochastic gravitational wave background produced from earlier phases of ``moduli preheating''.

In the context of hilltop inflation, it has been found in \cite{Antusch:2015nla} that the strong amplification of the perturbations can lead to localised regions which overshoot over the barrier that separates the considered minimum from other minima in the potential. For the hilltop model it turned out that such an overshooting is unproblematic since the overshooting bubbles quickly collapse. A similar situation can in principle appear for moduli potentials, e.g.\ the KKLT scenario and the LVS are examples where such overshooting over the barrier could be present.  For these two examples, we find that such overshootings do not occur and hence provide an additional check on the longevity of these metastable configurations. 

In Section~\ref{sec:general} we summarise the basic requirements for the production of oscillons, then comment on how moduli potentials provide a very promising arena to the production of oscillons, and finally we introduce the common numerical methods used for studying the evolution of the perturbations. In Section~\ref{sec:example} we present explicit examples for the production of oscillons, before concluding in Section~\ref{sec:conclusions}. In Appendix~\ref{sec:discussion and other models} we discuss further moduli potentials which only lead to a homogeneous field evolution.

\section{Introduction to Oscillons}
\label{sec:general}

Before studying the appearance of oscillons in string theory, let us first recall some conditions on the production of oscillons. We then comment on when we expect to produce oscillons from string moduli.
\subsection{Oscillons in effective field theories}
\label{sec:oscillonproduction}

We concentrate on the case of a single scalar field $\phi$ with a potential $V(\phi)$ with a Langrangian of the form:
\be
\mathcal{L}= \int d^4x\sqrt{-g}\left(\frac{M_{\rm Pl}^2}{2}\, R -\frac{1}{2}\,\partial^\mu\phi\,\partial_\mu \phi - V(\phi)\right).
\ee
Here $M_{\rm Pl}$ is the reduced Planck  mass and $R$ the Ricci scalar. We will study the cosmological evolution of the field $\phi$ on a FLRW background:
\be
ds^2=-dt^2+a^2(t)\, d{\bf x}^2\,,
\ee
with $a(t)$ the scale factor.

In order to study the production of oscillons, it is convenient to express the scalar field $\phi$ involved in the process as the sum of a (spatially averaged) homogeneous background and a perturbation: $\phi(t,\vec{x})=\phi(t) + \delta\phi(t,\vec{x})$. The homogeneous component obeys the equations of motion
\begin{equation}
\label{eq:homogeneousEOM}
\ddot{\phi}(t) + 3 H \dot{\phi}(t) + V'(\phi(t)) = 0\,,\quad\textrm{with}\quad H^2(t)=\frac{1}{3\,M^2_{\rm Pl}}\left(\frac{\dot{\phi}^2(t)}{2}+V(\phi(t))\right).
\end{equation}
The perturbation $\delta\phi$ can, in turn, be decomposed into Fourier modes $\delta \phi_k$: \begin{equation}
\delta\phi(t,\vec{x}) = \int \frac{d^{\rm D}k}{(2\pi)^{\rm D}}~ \delta \phi_k(t)~e^{-i\vec{k}\cdot\vec{x}}\,,
\label{eq:fourier}
\end{equation}
where D is the number of spatial dimensions. In $D = 3$ the equation of motion for the Fourier modes takes the familiar form
\begin{equation}
\label{eq:perturbationEOM}
\delta \ddot{\phi}_k + 3 H \delta \dot{\phi}_k + \left(\frac{k^2}{a^2(t)} + V''(\phi(t))\right) \delta \phi_k = 0 \,,
\end{equation}
with initial conditions set by vacuum fluctuations
\be
\delta\phi_k \simeq \frac{H}{\sqrt{2k^3}}\left(i+\frac{k}{a\,H}\right)e^{i\,k/(aH)}\,.
\label{eq:IC_fluctuations}
\ee
Three necessary conditions for the formation of oscillons are (see for instance~\cite{Gleiser:1993pt,Amin:2010dc}):
\begin{enumerate}
\item The field perturbations $\delta\phi$ grow as the field oscillates about the minimum of the potential.
\item The growth of the field perturbations is sufficiently strong for non-linear interactions to become important.
\item The potential \textit{opens up} away from the minimum or in other words it is \textit{shallower than quadratic} in some field space region relevant for the field dynamics\footnote{A precise condition for small amplitude oscillons including also non-canonical kinetic terms has been discussed in \cite{Amin:2013ika}}.
\end{enumerate}
The third requirement is necessary for the potential to support localised configurations of the scalar field: if this condition is not satisfied the fragmentation of the scalar field obtained as a consequence of the growth of the perturbations does not lead to the formation of oscillons (as e.g.\ in \cite{Felder:2006cc}). Intuitively, contrary to a quadratic potential, a potential shallower than quadratic (e.g. $V=m^2\phi^2-\lambda\phi^4+\cdots$ for small $\phi$) implies that there is an attractive interaction ($\lambda>0$) among the scalar particles implying that boson condensation leading to oscillon production is energetically favoured. 

In order to study the formation of oscillons it is always crucial to check the production explicitly, e.g.~through lattice simulation (see Section~\ref{sec:numerical methods}). The growth of perturbations though can be obtained in various ways. 
Discussions in the literature typically consider preheating after inflation. The same mechanisms can also occur in moduli preheating. 
The mechanisms can be classified in terms of the region of the potential probed by the background field during the growth of perturbations:
\begin{itemize}
\item \textit{Tachyonic preheating.} Tachyonic preheating occurs if the scalar field is initially displaced beyond the inflection point near the minimum and takes place when the field rolls in the region of the scalar potential where $V''(\phi(t)) < 0$ toward the minimum of the potential.  
As the field rolls through the tachyonic region, all the infrared modes $\delta \phi_k$ such that $k^2/a^2 + \partial^2 V/\partial\phi^2<0$ grow exponentially.  
The importance of tachyonic preheating depends, of course, on the initial displacement of the modulus field. It is in principle possible that non-linear interactions between different modes $\delta \phi_k$ become important already at this stage, however, for the models considered in this paper it turns out that this mechanism is subdominant. Especially for the models in which oscillon production takes place, tachyonic preheating is always followed by another, more efficient mechanism for the growth of fluctuations.
\item \textit{Tachyonic oscillations.} Tachyonic oscillations happen when the field crosses periodically the inflection point $\partial^2V/\partial\phi^2=0$ as it oscillates about the minimum, probing both the region where $V''(\phi(t)) > 0$ and the region where $V''(\phi(t)) < 0$. During each oscillation, the field perturbations grow in the tachyonic region as the field accelerates toward the minimum (downhill) and decrease as the field decelerates toward the plateau (uphill). This, combined with the expansion of the Universe and the periodic crossing of the inflection point, leads to a net growth of the field perturbations peaked at a characteristic wavenumber
\begin{equation}
\label{eq:Kpeak}
k_p \lesssim \left.\sqrt{\frac{\partial^2 V}{\partial \phi^2}}\right|_{\rm min} \equiv ~ m \,,
\end{equation}
closely related to the frequency of the oscillations of the homogeneous mode at the time of production~\cite{Brax:2010ai,Antusch:2015nla}. When the expansion rate is small compared to the frequency of the background oscillations (weak damping of oscillations) and it hence allows for a sufficient number of tachyonic oscillations, non-linear interactions between fluctuations with different wavenumbers become important, and the system enters a non-perturbative phase. Heuristically, the larger the ratio $m/H$ the smaller the damping becomes. In case the damping is large, the field quickly relaxes to the minimum and this mechanism for the growth of perturbations is not efficient (see Appendix~\ref{sec:discussion and other models}).

\item \textit{Parametric resonance.} Parametric resonance can occur when the background field $\phi(t)$ oscillates around the minimum of its potential for fluctuations $\delta\phi_k$ which have a time-dependent frequency
\be
\omega^2_k(t)  = \frac{k^2}{a^2} + \frac{\partial^2 V}{\partial\phi^2}\,.
\ee
Especially, if $\omega_k(t)$ varies non-adiabatically, i.e.\ when the adiabatic condition
\be
\left|\frac{\dot{\omega}_k(t)}{\omega^2_k(t)}\right|\ll1\,,
\ee
is violated, a window of modes can experience exponential growth and the scalar field eventually fragments. This mechanism of amplification of the perturbations is extensively explored in the literature~\cite{Kofman:1994rk,Kofman:1997yn} and the growth of perturbations in the case of axion monodromy is an example of parametric resonance at work~\cite{Zhou:2013tsa}.
\end{itemize}
Let us briefly comment on the production of gravitational waves from preheating with oscillons: During an early phase of tachyonic preheating, gravitational waves with comparatively low $k$ (limited by $H$ at this time) can be produced. When, on the other hand, the zero mode survives for several (or many more) oscillations, then tachyonic oscillations can generate a rather broad peak in the spectrum of gravitational waves around $k \sim k_p$, which then gets redshifted as the Universe expands. E.g.\ via tachyonic oscillations, or via parametric resonance, oscillons can be produced which then further contribute to the gravitational wave spectrum. In addition to gravitational waves generated during oscillon production \cite{Zhou:2013tsa}, oscillons can provide a continuous source for gravitational waves as long as they are asymmetric \cite{Antusch:2016con}. In this phase, oscillons can generate a pronounced peak in the gravitational wave spectrum at the oscillation frequency of the oscillons ($k \sim m$, with $m$ being the mass at the minimum of the potential) and also further subdominant peaks from higher harmonics at multiples of this frequency. Finally, when oscillons decay, additional gravitational waves are produced. Details will be discussed later in this Section and in Section \ref{sec:example}.

\subsection{Potential for oscillon production in string theory}
By now there is a whole plethora of moduli potentials known from string theory, in parts motivated by the construction of models of de Sitter moduli stabilisation and inflation in string theory~\cite{Baumann:2014nda}. This includes polynomial type potentials for complex structure moduli, periodic potentials for axions, and exponential potentials for K\"ahler moduli, just to name a few. In general it is expected that there are many $\mathcal{O}(100)$ moduli fields and that their dynamics are encoded in their multi-field scalar potential. The aim of this paper is only to provide a first survey of this rich class of potentials.

\begin{figure}
\begin{center}
\includegraphics[width=0.6\textwidth]{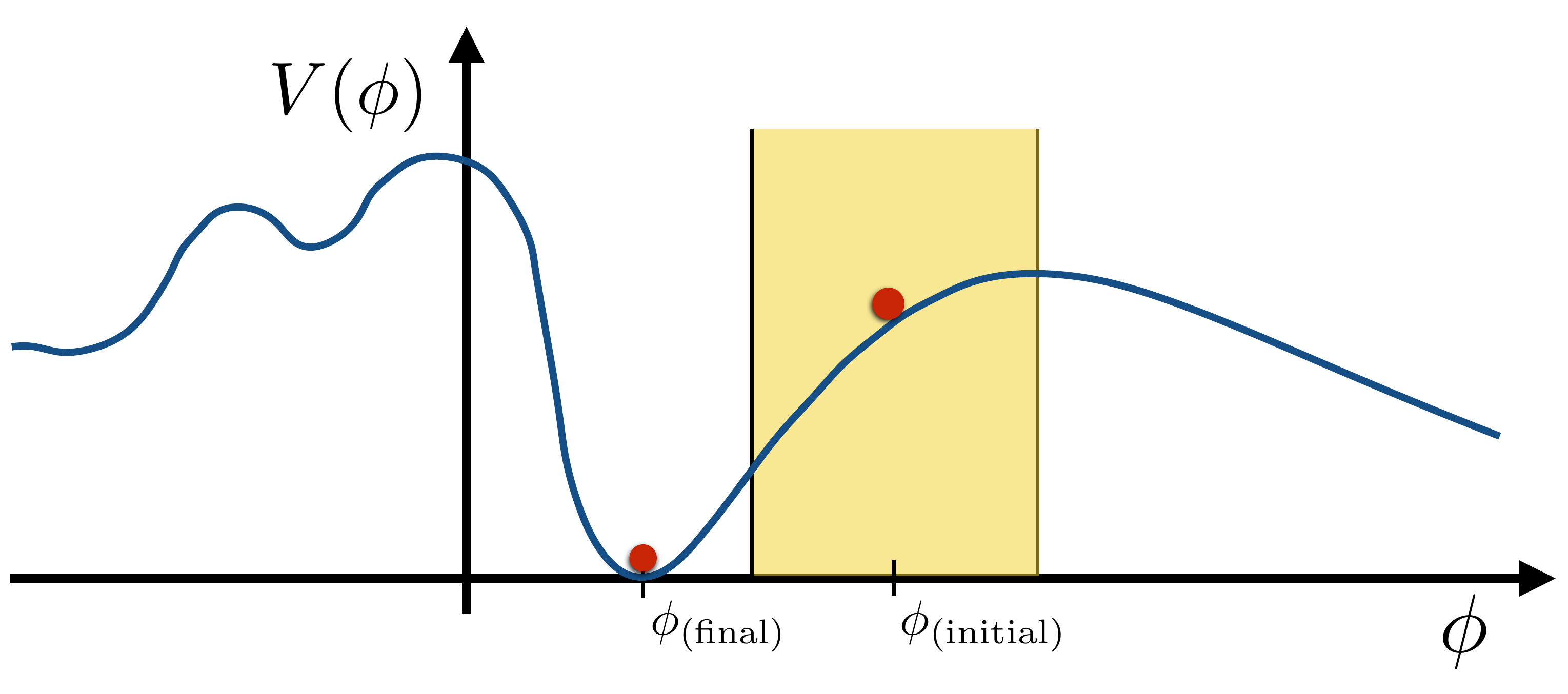}
\end{center}
\caption{Sketch of a moduli potential near its post-inflationary minimum $\phi_{\rm final}.$ If the initial displacement $\phi_{\rm initial}$ is in the yellow region the modulus is displaced in the tachyonic region of the potential and tachyonic preheating (oscillations) can take place.}
\label{fig:potentialsketch}
\end{figure}

The standard picture is that (one of) the moduli fields will drive inflation, while the other fields are trapped at their minimum of the potential. However, the minimum during inflation $\langle\phi_{\rm inf}\rangle$ for these moduli fields depends on the position of the inflaton field and generically is different from the post-inflationary minimum. After inflation ends, these moduli fields are displaced from their minimum and locked at these displaced values until Hubble friction $3H\dot{\phi}$ becomes comparable to $V'(\phi).$
The displacement is set by the inflationary potential and can be calculated in a UV model (see for instance~\cite{Cicoli:2016olq} for an example). To be independent from inflationary model building, we keep the initial displacement as a free parameter. For simplicity we also restrict ourselves to single field examples. The cartoon picture is shown in Figure~\ref{fig:potentialsketch}.

For concreteness let us concentrate on IIB string compactifications on Calabi-Yau orientifolds that lead to $\mathcal{N}=1$ supersymmetry in 4-dimensions. The set of (closed string) moduli fields are as follows:
\begin{itemize}
\item{} {\it Dilaton and complex structure moduli.} The model independent dilaton field $S$ and the many (usually of order $10^2-10^3$) complex structure moduli $U_\alpha$ measuring the size of Calabi-Yau 3-cycles. Their scalar potential is determined by fluxes of the two 3-form fields of IIB strings: $F_{MNP}, H_{MNP}$. Fluxes of these fields are quantised from Dirac quantisation condition with integers bound by general tadpole conditions but otherwise arbitrary. Varying the corresponding integers is the source of the huge landscape of string vacua. The potential for these fields is determined by the Gukov-Vafa-Witten superpotential $W=\int G_3\wedge \Omega$ where $G_{MNP}= F_{MNP}+SH_{MNP}$ and $\Omega $ is the unique $(3,0)$ form of Calabi-Yau manifolds. Their K\"ahler potential is $K_{\rm cs}+K_S$ with $K_{cs}=-\log{\left(\int\Omega\wedge \Omega^*\right)}$ and $K_S=-\log(S+S^*)$. They give rise to a scalar potential of order string scale (or below) and are minimised by the condition $D_UW=D_SW=0$. We will assume these fields are sitting in their minimum but in a more general discussion they may be considered as potential sources of oscillons during their cosmological evolution.

\item{} {\it K\"ahler moduli.} These are $T_i=\tau_i+{\mathrm i} a_i$ where $\tau_i$ measure the size of the 4-cycles of the Calabi-Yau manifold and $a_i$ are axion-like fields. For concreteness we will only concentrate on the potential for the fields $\tau_i$ where $i=1,\cdots, h_{11}$ where the Hodge number $h_{11}$ measures the number of non-trivial 4-cycles which ranges from one up to several hundreds for known manifolds. Among these fields there are three classes that are usually distinguished: the overall volume modulus, blow-up modes (measuring the size of cycles that can be collapsed to zero size) and fibre moduli (measuring the size of fibrations which are generic in Calabi-Yau manifolds). We will discuss potentials for each of these classes generated by non-perturbative superpotentials and K\"ahler potentials that will be written explicitly below. Again their axionic partners could also be sources of oscillons that we will not consider at this stage.

\end{itemize}
To be specific, let us focus on the dynamics of K\"ahler moduli in type IIB string theory. For such moduli, the potential takes the schematic form
\begin{equation}
V(\phi)=\sum_{j=0}^n C_j e^{-a_j  \phi}+\ldots~,
\end{equation}
where the number of relevant terms $n,$ the coefficients $C_j$ and $a_j$ are model dependent,  the dots include terms with multiple exponentials $e^{-e^\phi}.$ The coefficients $C_j$ and $a$ can have a mild (i.e.~polynomial) dependence on the moduli. The characteristic exponential structure for the potential arises due to two distinct reasons:
\begin{enumerate}
\item The shift symmetry of the K\"ahler moduli forbids perturbative terms in the superpotential depending on the K\"ahler moduli. Only non-perturbative contributions, hence exponentials, can arise in the superpotential. An example of such a generation of terms is the case of blow-up moduli in the LVS which we discuss in Section~\ref{sec:KMI}.
\item The K\"ahler potential of the overall volume modulus $K/M^2_{\rm Pl}=-2\log{{\mathcal V}}$ leads to a canonically normalised volume of the form $\phi/M_{\rm Pl}\simeq \log{{\mathcal V}}.$ Although ${\cal V}$ appears polynomially in the potential, the potential for the canonically normalised field is exponential. Examples of such terms arise in our discussion of the overall volume modulus in KKLT (cf. Section~\ref{sec:KKLT}) and in the LVS (cf. Section~\ref{sec:LVS}).
\end{enumerate}
The minimum in these potentials arises from a balancing of various effects which scale differently with respect to moduli (e.g.~$\alpha'-$corrections and non perturbative corrections in the LVS). This distinct scaling on both sides of the potential generically leads to the appearance of asymmetric potentials. For example in the context of blow-up moduli in the LVS (cf.~Section~\ref{sec:KMI}), the potential schematically takes the following form:
\begin{equation}
V=V_0(1-\kappa e^{-\alpha\phi})^2~.
\label{eq:pkm}
\end{equation}
This potential is asymmetric about the minimum, features an inflection point at $\phi_{\rm inf}\simeq\ln(2\kappa)/\alpha$ and is clearly shallower than quadratic.\footnote{Note that a similar opening up is expected in the case of flux potentials for complex structure moduli where the back-reaction of the fluxes on the geometry leads to an effective flattening of the potential~\cite{Landete:2017amp}.} The precise form of the prefactor, e.g.~a polynomial moduli dependence of $\kappa$, and the coefficient in the exponential $\alpha$ are model dependent. In fact the potential shape is qualitatively similar to the case of hilltop-like potentials of the form:
\begin{equation}
V(\phi)=V_0\left(1-\left(\frac{\phi}{v}\right)^p\right)^2\,.
\label{eq:phi}
\end{equation} 
Such potentials have been studied in the context of inflation~\cite{Linde:1981mu,Izawa:1996dv,Izawa:1997df,Senoguz:2004ky,Boubekeur:2005zm} and are known to support oscillons in large regions of their parameter space~\cite{Antusch:2015nla,Antusch:2015vna,Antusch:2015ziz,Antusch:2016con}. Given the similar qualitative features of potentials (\ref{eq:pkm}) and (\ref{eq:phi}), we expect (and will show) similar conclusions to be valid for K\"ahler moduli. 

In potentials of the form of Equations~(\ref{eq:pkm}) and (\ref{eq:phi}) the necessary conditions that we have listed above are satisfied. For a sufficient initial displacement, the growth of field perturbations happens via tachyonic preheating and tachyonic oscillations: the amplitude of the Fourier modes $\delta \phi_k$ grows when the homogeneous field $\phi(t)$ accelerates through a tachyonic region of the potential, i.e.~for regions of the potential such that $k^2/a^2(2) + V''(\phi(t)) < 0$. In particular, for the potential in Equation~\eqref{eq:pkm} the growth of perturbations takes place for field values $\phi > \log(2 \kappa)/\alpha$.

Generically such regions where perturbations grow can be found in potentials for K\"ahler moduli near the minimum. Whether the perturbations grow sufficiently large depends on the coefficients in the potential $C_j,~a_j$ and is hence model dependent. Similarly whether the conditions for parametric resonance are satisfied depends on the parameter in the moduli potential.

In the case that perturbations grow sufficiently, gravitational radiation is produced. After its production, the spectrum of gravitational radiation is redshifted according to the subsequent evolution of the Universe. The presence of many moduli in string models could lead to a modification of the standard big bang picture, that features a single period of radiation domination lasting from reheating after inflation to radiation/matter equality. In fact, a modulus $\Phi$ oscillating around its minimum can lead to a phase of matter domination which lasts until the modulus decays when the Hubble parameter becomes comparable to the decay rate of the modulus itself $\Gamma_\Phi\simeq m_{\Phi}^3/M_{\rm Pl}^2$. To avoid the cosmological moduli problem this lightest modulus has to decay before BBN, setting an upper bound on the duration of this phase of matter domination. Given the existence of many moduli in string models, in general there can be many periods of matter domination. Concerning oscillon and GW production, we can have two different situations:
\begin{itemize}
\item Oscillons and consequently GWs are produced by the dynamics of the lightest mod- ulus and are followed by a radiation dominated phase.
\item After oscillon and GW production, there is at least one additional displaced lighter modulus which can dominate the energy density of the Universe, leading to at least one additional epoch of matter domination.
\end{itemize}
Since during matter domination the Universe expands more rapidly than during radiation domination, the abundance of GWs produced by oscillons is more diluted when matter domination lasts longer. In order to compute the actual dilution, it is necessary to evaluate the evolution of the scale factor. For concreteness, let $\Phi$ be the lightest modulus. Oscillons and GWs are produced either by $\Phi$ itself or by a heavier modulus. Let us also assume that matter domination lasts from GW production to the decay of $\Phi$\footnote{In the most generic picture there could be several alternate matter and radiation domination eras.}. We denote by $t_{\rm e}$ the time of production of GWs, by $t_{*}$ the time of decay of $\Phi$ determined by $\Gamma_\Phi \simeq m_\Phi^3/M_{\rm Pl}^2 \sim H(t_{*})$ and by $t_0$ the current epoch. We assume that after the decay of $\Phi$ the thermal bath of relativistic particles obeys entropy conservation, so that $g(T) a^3 T^3 \sim \rm const.$, where $g(T)$ is the number of relativistic degreees of freedom at temperature $T$. Then, the evolution of the scale factor can be written as
\begin{equation}
\label{eq:scalefactorevolution}
\mathcal{R} \equiv \frac{a\left(t_{\rm e}\right)}{a\left(t_{0}\right)} = \frac{a\left(t_{\rm e}\right)}{a\left(t_{*}\right)} \frac{a\left(t_{*}\right)}{a\left(t_{0}\right)} = \frac{\rho_{*}^{1/3}}{\rho_{\rm e}^{1/3}} \left(\frac{g_{ 0}}{g_{*}}\right)^{1/3} \frac{T_{0}}{T_{*}} = \left(\frac{g_{0}}{g_{*}}\right)^{1/12} \frac{\rho_{*}^{1/12}}{\rho_{\rm e}^{1/3}}~\rho_{0, \rm rad}^{1/4}\,,
\end{equation}
where we have used that during matter domination $\rho \sim a^{-3}$, while during radiation domination $\rho \sim g(T) T^4$. In general $\rho_{\rm e}$ can be inferred by the numerical simulations, while $\rho_{*}$ is determined by the reheating temperature $\rho_{*} = \frac{\pi^2}{30} g(T) T_{*}^4$, where $T_{\rm dec} = \sqrt{\Gamma_\Phi M_{\rm Pl}}$. Finally $\rho_{0, \rm rad}$ is the energy density in radiation in the current epoch, and can be written in terms of the critical density $\rho_{\rm crit} \sim 10 \,\rm GeV/m^3$: $\rho_{0, \rm rad} = 4.3 \times 10^{-5} \, \rho_{\rm crit}$. From Equation~\eqref{eq:scalefactorevolution} it is evident that the longer the matter domination epoch (i.e. the smaller the energy density at the decay of the lightest modulus $\Phi$) the smaller is the ratio between the scale factors at the GW production epoch and today.

Since GWs redshift as radiation, the abundance of GWs today is suppressed by the fourth power of the ratio $\mathcal{R}$
\begin{equation}
\Omega_{\rm GW, 0} = \mathcal{R}^4 \, \Omega_{\rm GW, e}\,.
\label{eq:rescaling1}
\end{equation}
On the other hand, the frequency $f$ of GWs gets redshifted as usual
\begin{equation}
f_0 = \mathcal{R} \, f_{\rm e}\,.
\label{eq:rescaling2}
\end{equation}

\subsection{Numerical methods}
\label{sec:numerical methods}
The dynamics of a scalar field in an expanding Universe essentially depends on the exact shape of its potential. For the various models considered in this paper, the latter typically depends on multiple parameters. Although the shape of the potentials can be qualitatively similar for different choices of parameter sets, the overall dynamics can vary significantly. While for some regions of the parameter space, the field may be well described by the homogeneous evolution, others may support a strong amplification of scalar field perturbations eventually leading to non-perturbative dynamics.

To study the evolution of the scalar field perturbations we use different common methods which we describe below in this section. In general, the procedure is as follows: To determine whether and at which scales the fluctuations can potentially experience a phase of rapid growth, for a given model, we use Floquet theory. To get a rough impression on whether the dynamics are expected to be mild or rather violent we may also consider model specific quantities like, for instance, $m/H$ specifically for models where the situation is less clear (e.g.\ when the parameter space is larger). Based on the results of the Floquet analysis we either solve the fully non-linear equations of motion in an expanding Universe using lattice simulations, or we solve the linearised equations for the perturbations Equation~\eqref{eq:perturbationEOM} in cases where no non-linear effects are expected.

\subsubsection{Floquet analysis}
\label{sec:floquet analysis}
A Floquet analysis is only applicable during the linear regime, i.e.\ when the fluctuations are small compared to the motion of the background $\phi(t)$ and only if the motion of the background is periodic. The latter is certainly not the case in an expanding Universe. Nevertheless, for a first exploration of the dynamics we can use Floquet theory to compute the growth rate of the fluctuations in Minkowski space. If the growth rate is large compared to the expansion rate of the Universe this is typically a reliable indicator for the perturbations to experience a phase of rapid growth.

In Minkowski space the linearised equations for the modes $\delta\phi_k$ Equation~\eqref{eq:perturbationEOM} reduce to
\be
\delta\ddot{\phi}_k(t) + \left(k^2 + \frac{\partial^2V(\phi(t))}{\partial \phi(t)^2}\right)\delta\phi_k(t) = 0\,,
\label{eq:Minkowski_fluct}
\ee
where $\phi(t)$ is the homogeneous background field. Since $\partial^2V(\phi)/\partial \phi^2$ is now periodic, Equation~\eqref{eq:Minkowski_fluct} has the form of Hill's equation and according to the Floquet theorem the solutions can be written as
\be
\delta\phi_k(t) = P_+(t)e^{\mu_k\,t} + P_-(t)e^{-\mu_k\,t}\,,
\ee
where the $\mu_k$ are called Floquet exponents and the $P_{\pm}$ are periodic functions with the same period as $\partial^2V(\phi)/\partial \phi^2$. For fluctuations $\delta\phi_k$ for which the Floquet exponent exhibits a non-vanishing real part $\Re[\mu_k]$ the solution will be exponentially growing. Moreover, if the growth rate is much larger compared to the expansion rate $|\Re[\mu_k]|/H \gg 1$, this is a reliable indicator for the corresponding mode to get significantly amplified in an expanding Universe. The Floquet exponents are typically computed numerically via a Floquet analysis. We can rewrite Equation~\eqref{eq:Minkowski_fluct} as a first-order system of linear differential equations 
\be
\dot{x}(t)=U(t)x(t)\,,
\ee
with
\be
U(t) = \begin{pmatrix}0 & 1\\ -k^2-\frac{\partial^2V(\phi)}{\partial \phi^2} & 0 \end{pmatrix}\,,
\ee
and defining $x(t)=(\delta\phi_k , \delta\pi_k)^T$ where
\be
\delta\pi_k \equiv \delta\dot{\phi}_k\,.
\ee
The Floquet exponents can then be computed in three steps as follows (see Ref.\cite{Amin:2014eta} for more details)
\begin{enumerate}
\item We first compute the period $T$ of the system assuming that the field is at rest at $t=0$, i.e.\ $\dot{\phi}_{\rm initial}\equiv\dot{\phi}(0)=0$:
\be
T = 2\int_{\phi_{\rm minimal}}^{\phi_{\rm maximal}} \frac{d\phi}{\sqrt{2V(\phi_{\rm maximal}) - 2V(\phi)}}\,,
\ee
where $\phi_{\rm initial}\equiv\phi(0)=\phi_{\rm minimal}$ (or $\phi_{\rm maximal}$) is the initial field value and the maximal and minimal field amplitudes $\phi_{\rm maximal}$ and $\phi_{\rm minimal}$ are calculated by solving $V(\phi_{\rm minimal}) = V(\phi_{\rm maximal})$. 
\item After having computed the period $T$ we solve the equation for the homogeneous background in Minkowski space and simultaneously Equation~\eqref{eq:Minkowski_fluct} from $t=0$ to $t=T$ and for two sets of initial conditions $\{\delta\phi_{k,1}(0) = 1,\delta\dot{\phi}_{k,1}(0) = 0\}$ and $\{\delta\phi_{k,2}(0) = 0,\delta\dot{\phi}_{k,2}(0) = 1\}$.
\item Finally, the Floquet exponents are computed as
\be
\Re[\mu^{\pm}_k] = \frac{1}{T}\ln |\sigma^{\pm}_{k}|
\ee
where
\be
\sigma^{\pm}_{k} = \frac{1}{2}\left(\delta\phi_{k,1} + \delta\dot{\phi}_{k,2} \pm \sqrt{\left[\delta\phi_{k,1} - \delta\dot{\phi}_{k,2}\right]^2 + 4\,\delta\phi_{k,2}\delta\dot{\phi}_{k,1}}\right)
\ee
with all quantities being evaluated at $t=T$.
\end{enumerate}

\subsubsection{Lattice simulations}
\label{sec:lattice simulations}
To properly capture the evolution of the field in models where we expect a strong growth of perturbations we perform numerical lattice simulations using a modified version of LatticeEasy~\cite{Felder:2000hq}. The program is originally written to solve the fully non-linear equations of motion of scalar fields in an expanding Universe. For the case of a single scalar field the program solves the following set of equations on a discrete spacetime lattice
\begin{eqnarray}
\ddot{\phi}\, + \, 3H\dot{\phi}\, - \,\frac{1}{a^2}\nabla^2\phi\, + \,\frac{\partial V}{\partial\phi} \, = \, 0\,, \label{eq:EOM_fld} \\
H^2\, = \,\frac{1}{3M^2_{\rm{Pl}}}\left( V\,  + \,\frac{1}{2}\dot{\phi}^2\, + \frac{1}{2a^2}\left|\nabla\phi\right|^2 \right)\,, \label{eq:EOM_hubble}
\end{eqnarray}
with initial fluctuations given by quantum vacuum fluctuations \cite{Polarski:1995jg,Khlebnikov:1996mc}. On the lattice this is realised by initializing the fluctuations as random variables with a Rayleigh distributed  magnitude and with randomly uniformly distributed phase (see Ref.~\cite{Felder:2000hq}).

In addition to Equation~\eqref{eq:EOM_fld} and \eqref{eq:EOM_hubble} the code has been extended to compute the evolution of the transverse and traceless (TT) part of the metric perturbation in the synchronous gauge, where the line element takes the form
\be
ds^2 = -dt^2 + a^2(t)(\delta_{ij} + h_{ij})dx^idx^j\,.
\ee
The gravitational waves, $h_{ij}$, are sourced by TT-part of the anisotropic stress of the scalar field
\begin{equation}
\Pi_{ij}^{\rm TT} =\frac{1}{a^2}\left[\partial_i\phi\partial_j\phi\right]^{\rm TT}\,,
\label{eq:effective_source_term}
\end{equation}
where on the right hand side of Equation~\eqref{eq:effective_source_term} we kept only the terms that are linear in the gravitational coupling. The evolution of the GW is given by
\begin{equation}
\ddot{h}_{ij}\, + \,3H\dot{h}_{ij}\, - \,\frac{1}{a^2}\nabla^2h_{ij}\, = \,\frac{2}{M^2_{\rm Pl}}\Pi_{ij}^{\rm TT}~,
\label{eq:EOM_metric_perturbation}
\end{equation} 
where the initial conditions for the metric perturbations are $h_{ij}(0)=\dot{h}_{ij}(0)=0.$
Instead of directly solving for the $h_{ij}$, however, our code computes the evolution of the (non-transverse and non-traceless) tensor $u_{ij}$ 
\be
\ddot{u}_{ij}\, + \,3H\dot{u}_{ij}\, - \,\frac{1}{a^2}\nabla^2u_{ij}\, = \,\frac{2}{M^2_{\rm Pl}}\frac{1}{a^2}\partial_i\phi\partial_j\phi~,
\label{eq:EOM_metric_perturbation_NON-TT}
\ee
and projects out the TT part of $u_{ij}$ only when output is generated. This procedure saves a considerable amount of computation time but is mathematically equivalent to directly solving Equation~\eqref{eq:EOM_metric_perturbation} \cite{GarciaBellido:2007af}. The TT projection is performed in Fourier space according to
\be
h_{ij}(\textbf{k},t) = \left(P_{il}(\textbf{\textbf{\^{k}}})P_{jm}(\textbf{\textbf{\^{k}}}) - \frac{1}{2}\,P_{ij}
(\textbf{\textbf{\^{k}}})P_{lm}(\textbf{\^{k}})\right)u_{ij}(\textbf{k},t)~,
\ee
where $P_{ij}$ is the projection tensor defined as
\be
P_{ij}(\textbf{\^{k}}) \equiv \delta_{ij} - \hat{k}_i\hat{k}_j\,,\qquad\textrm{with}\quad\hat{k}_i \equiv k_i/|\textbf{k}|\,.
\ee
Ultimately, the spectrum of GW per logarithmic momentum interval is computed
\be
\Omega_{\rm GW}(k) \, = \,\frac{1}{\rho_{\rm c}}\,k\,\frac{d\,\rho_{\rm GW}}{dk}\,,
\ee
where $\rho_{\rm c} = 3\,M^2_{\rm Pl}H^2\,$ is the critical density of the Universe. The energy density of the GW is 
\be 
\rho_{\rm GW}(t) = \frac{M^2_{\rm Pl}}{4}\,\left\langle\dot{h}_{ij}(\textbf{x},t)\dot{h}_{ij}(\textbf{x},t)\right\rangle_{\rm{V}}\,,
\ee
with $\langle...\rangle_{\rm V}$ denoting an average over a sufficiently large volume, containing several wavelengths of the GW.

The spectrum $\Omega_{\rm GW, \rm e}(k)$ refers to the spectrum at the time when the GWs are emitted. In order to get the spectrum which would be observable today $\Omega_{\rm GW,0}(f)$, where $f$ is the frequency in Hz of the GWs, we have to take into account the expansion history of the Universe between the emission of the GWs and today. This leads to the following rescalings for the power spectrum (cf.~Equation~\eqref{eq:rescaling1})
\be
\Omega_{\rm GW,0} = \frac{4.3}{10^5}\left(\frac{a_{\rm e}}{a_*}\right)^{1-3w}\,\left(\frac{g_*}{g_0}\right)^{-1/3}\,\Omega_{\rm GW,\rm e}~,
\label{eq:OmegaGW_today}
\ee
where $a_{\rm e}$ is the scale factor at the moment of emission, $a_*$ is the scale factor at the end of reheating and $w$ is the average equation of state parameter between the moment of emission and the end of reheating. We use $g_*/g_0=100$, where $g_*$ and $g_0$ are the number of relativistic degrees of freedom at the end of reheating and today, respectively. For the frequency of the GW we have (cf.~Equation~\eqref{eq:rescaling2})
\be
f_0 = 4\times10^{10}\,\left(\frac{a_{\rm e}}{a_*}\right)^{\frac{1-3w}{4}} \frac{k}{a_{\rm e}\rho_{\rm e}^{1/4}}\,\textrm{Hz}~,
\label{eq:GW_freq_today}
\ee
where $\rho_{\rm e}$ is the energy density at the time where the GWs are emitted.

\section{Oscillons from K\"ahler  moduli}
\label{sec:example}
Let us now turn to the discussion of concrete examples where we find the formation of asymmetric oscillons which lead to a production of a stochastic gravitational wave background.

\subsection{KKLT}
\label{sec:KKLT}
The first example is that of the overall volume modulus in KKLT~\cite{Kachru:2003aw} with a single overall \Kahler modulus $T.$ After integrating out the complex structure moduli and the dilaton at their supersymmetric minimum, the four dimensional effective supergravity is described by the following K\"ahler and superpotential  
\begin{equation}
K/M^2_{\rm Pl} = -3 \log\left(T + \overline{T}\right) - K_{\rm cs}\,, \quad W/M^3_{\rm Pl} = W_0 + A e^{-a T}\,,
\end{equation}
where $K_{\rm cs}$ denotes the vacuum expectation value (VEV) of the complex structure K\"ahler potential and the dilaton, $W_0$ is the vacuum expectation value of the Gukov-Vafa-Witten flux superpotential~\cite{Gukov:1999ya}. The non-perturbative superpotential contribution $A e^{-aT}$ can be generated by gaugino condensation of D7-branes or Euclidean ED3-branes. Both the flux superpotential $W_0$ and the prefactor of the non-perturbative effect $A$ are functions of complex structure moduli and dilaton, while $a$ is a coefficient which depends on the source of the non-perturbative effect (e.g.~$a=2\pi/N$ for a gaugino condensate from $N$ D7 branes). This setup has a supersymmetric AdS minimum at $D_T W=0.$ To obtain a minimum in the region where the effective field theory is applicable ${\rm Re}(T)\gg 1,$ the flux superpotential has to have hierarchically small values. Such hierarchies can be obtained by stabilising complex structure moduli close to so-called conifold points~\cite{Giddings:2001yu}. This potential is uplifted to a dS minimum by adding some additional source (e.g.~anti D3-branes or matter fields), effectively a standard uplifting term $V_{\text{up}} \simeq \left(\frac{T + \overline{T}}{2}\right)^{-2}$ is added to the potential.\footnote{Different uplfiting mechanisms can give rise to different powers of $T$ appearing in the uplifting potential. For simplicity we just consider one example.} The total scalar potential is given by
\begin{equation}
V/M^4_{\rm Pl} = \frac{e^{K_{\rm cs}}}{6\tau^2}\left(a A^2(3+a\tau)e^{-2a\tau}-3 a A e^{-a\tau}W_0\right)~.
\label{eq:potential_KKLT_tau}
\end{equation}
This then leads to a minimum where supersymmetry is broken and the field value is shifted by a factor $\sim \log\left(M_{\rm Pl}/m_{3/2}\right)$~\cite{LoaizaBrito:2005fa} compared to the AdS supersymmetric minimum. Such a shift influences the ratio between the mass of the field at the minimum of the potential and the height of the barrier. As we see in due course, this allows for multiple oscillations in the tachyonic region of the scalar potential. However, as we will discuss below, parametric resonance is the dominant mechanism for the growth of fluctuations (and not tachyonic oscillations). In our numerical analysis of the KKLT potential we concentrate on the following standard parameter ranges
\be
10^{-12} \le W_0 \le 10^{-5}\,,\qquad 1\le A \le 10\,,\qquad 1\le a \le 2\pi~.
\ee
$W_0$ is chosen to be hierarchically smaller than unity to allow for a stabilisation with ${\rm Re}(T)\gg 1$ and the lower bound is chosen such that gravitino mass (and hence the moduli masses) are large enough to safely avoid the cosmological moduli problem. The range for $A$ is chosen that we do not assume any hierarchical suppressions from this contribution. For $a$ we start from the largest possible value and the lower limit still corresponds to a moderate number of D7 branes. The prefactor $e^{K_{\rm cs}}$ which rescales the overall potential is set to unity.

\begin{figure}
\centering
\includegraphics[width=\textwidth]{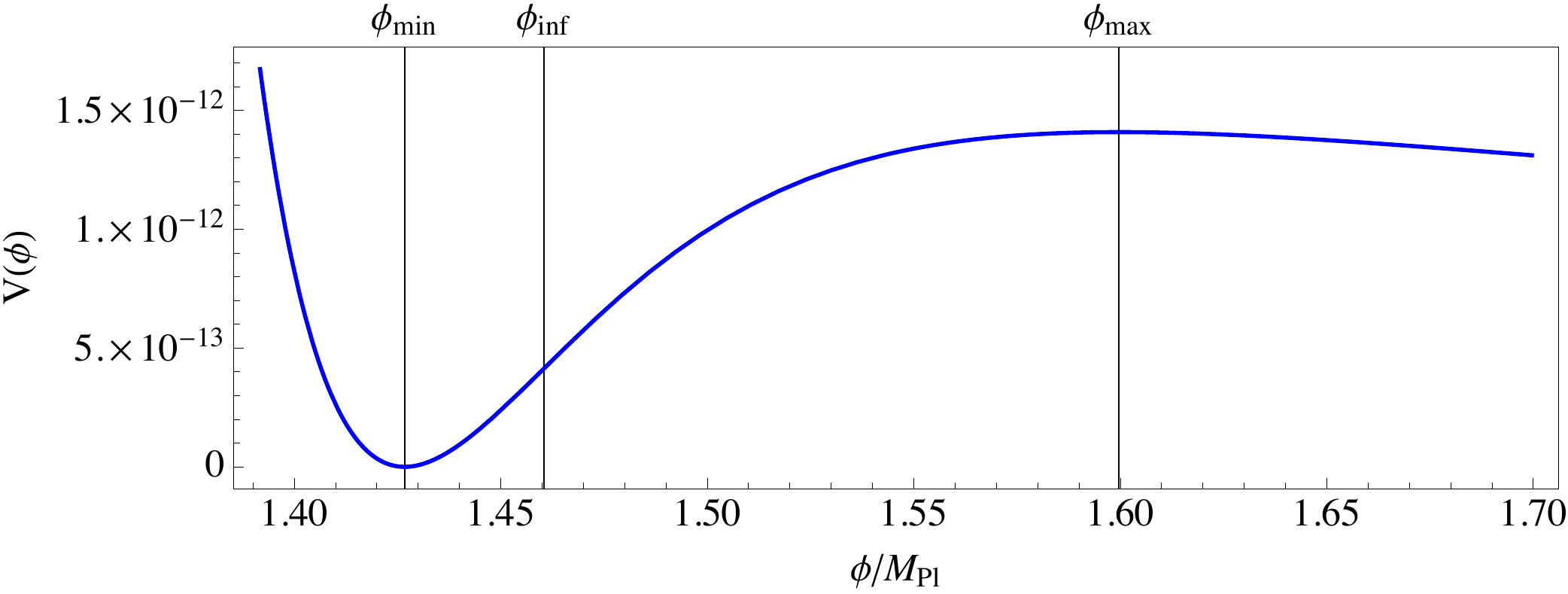}
\caption{ Example potential of the \Kahler modulus in KKLT for $W_0=10^{-5}$ $A=10$ and $a=2\pi$. The solid black lines denote the field value at the minimum of the potential $\phi_{\rm min}$, the field value at the inflection point of the potential $\phi_{\rm inf}$, and the field value at the local maximum of the potential $\phi_{\rm max}$.}
\label{fig:KKLT_potential}
\end{figure}

The canonically normalised field $\phi$ of the real part of $T$ is given by
\begin{equation}
\phi/M_{\rm Pl}=\frac{\sqrt{3}}{2}\log{\left(T+\bar{T}\right)}\,.
\end{equation}
An example of the potential can be found in Figure~\ref{fig:KKLT_potential} for  $W_0=10^{-5}$, $A=10$, and $a=2\pi$.

We would like to begin our discussion with the results of Floquet analyses of the KKLT model with $a=2\pi$, $A=10$, and for two different values of $W_0$: $W_0=10^{-12}$ and $W_0=10^{-5}$. In both cases, the Floquet exponents were calculated in Minkowski space, as a function of the initial field value $\phi(0)\equiv\phi_{\rm initial}$ , essentially corresponding to different amplitudes of oscillation. For $\phi_{\rm initial}$ we assumed values within the following range
\be
\phi_{\rm min} < \phi_{\rm initial} \le \phi_{\rm min} + \frac{\phi_{\rm max}-\phi_{\rm inf}}{2}\,,
\ee
where $\phi_{\rm min}$ is the field value at the minimum of the potential, $\phi_{\rm inf}$ is the field value at the inflection point, and $\phi_{\rm max}$ the field value at the local maximum of the potential (see Figure~\ref{fig:KKLT_potential}).

\begin{figure}
\centering
\subfigure{\includegraphics[width=7.5cm]{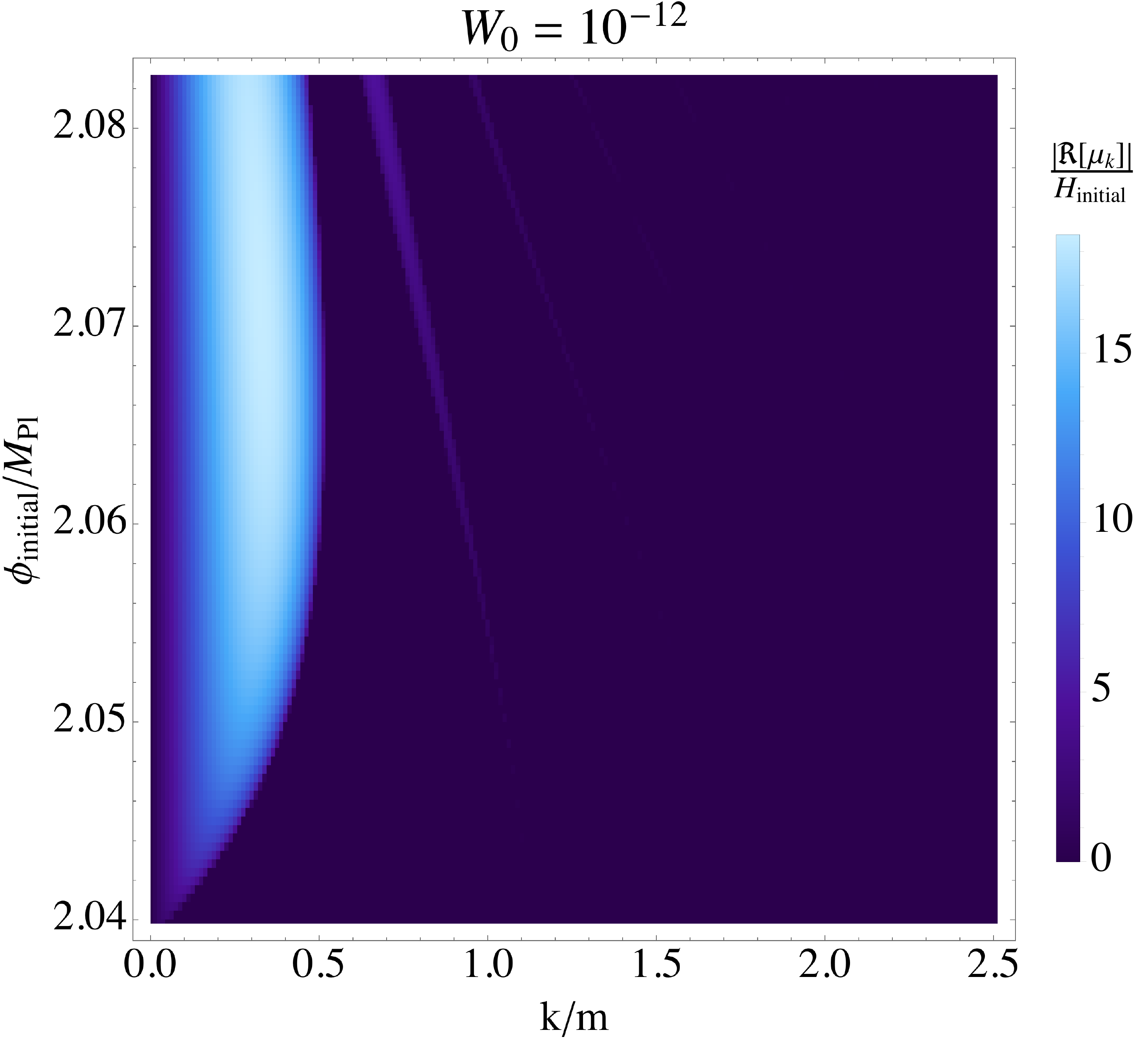}}
\hfill
\subfigure{\includegraphics[width=7.5cm]{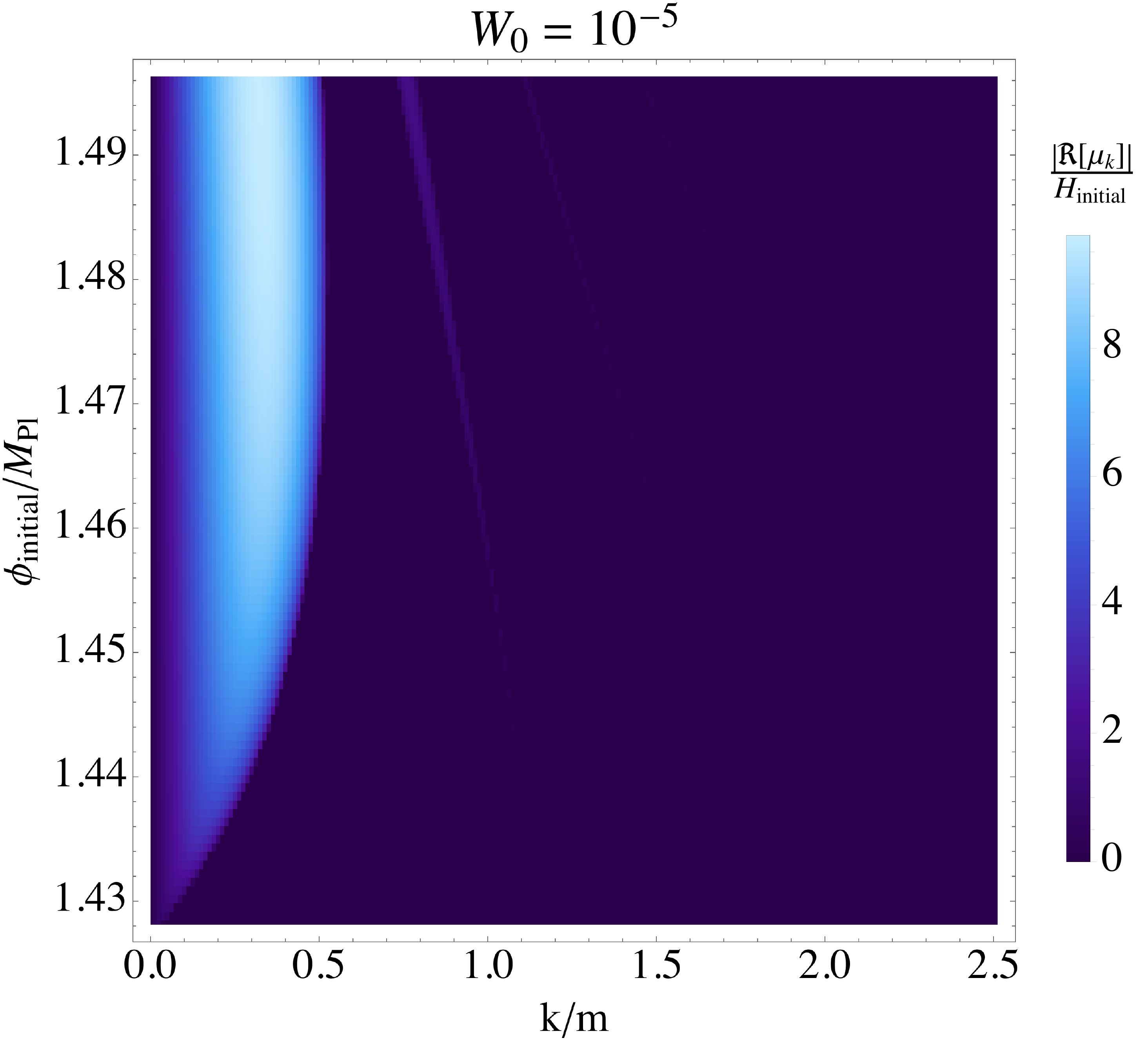}}
\caption{Instability bands for the KKLT model with $a=2\pi$, $A=10$ and $W_0=10^{-12}$ (left) and $W_0=10^{-5}$ (right). The figure shows the real part of the Floquet exponent $|\Re[\mu_k]|$ in units of the Hubble parameter $H_{\rm initial}$ as a function of the amplitude of the homogeneous field $\phi_{\rm initial}$.}
\label{fig:floquet_KKLT}
\end{figure}

The results of our analyses are presented in Figure~\ref{fig:floquet_KKLT} for $W_0=10^{-12}$ (left) and $W_0=10^{-5}$ (right). The Figure shows the real part of the Floquet exponent compared to the Hubble parameter $|\Re[\mu_k]|/H_{\rm initial}$, where
\be
H_{\rm initial}=\frac{1}{M_{\rm Pl}}\sqrt{\frac{V(\phi_{\rm initial})}{3}}\,.
\ee
In both cases, there is a broad instability band with $|\Re[\mu_k]|/H_{\rm initial}\sim\mathcal{O}(10)$ for $k\lesssim0.5m$. Two other thin and weaker  bands are also visible for $k>0.5m$. They are, however, narrower and also weaker than the first band. In view of these results, we would expect a noticeable amount of growth for modes with comoving wavenumbers $k\lesssim0.5m$ in an expanding universe. To investigate the evolution of the fluctuations in greater detail we performed lattice simulations. The results are presented in the next section.

\subsubsection{Results from lattice simulations}
\label{sec:KKLT_lattice simulations}
Lattice simulations of the evolution of \Kahler modulus in KKLT were performed for the two sets of parameters which were also used to perform the Floquet analyses, i.e. for $W_0=10^{-12}$ and $W_0=10^{-5}$, and in both cases with $a=2\pi$ and $A=10$. The initial field value was considered as a phenomenological parameter which we chose to be half way between the field value at the inflection point $\phi_{\rm inf}$ and the field value at the local maximum of the potential $\phi_{\rm max}$. The initial field velocity $\dot{\phi}_{\rm initial}$ was set to zero. In summary, we used the following initial conditions for the homogeneous background
\be
\phi_{\rm initial}=\phi_{\rm min} + \frac{\phi_{\rm max}-\phi_{\rm inf}}{2}\,,\qquad \dot{\phi}_{\rm initial}=0\,,\qquad H_{\rm initial}=\frac{1}{M_{\rm pl}}\sqrt{\frac{V(\phi_{\rm initial})}{3}}\,.
\ee
For the initial conditions of the fluctuations $\delta\phi$ of the \Kahler modulus and their derivatives we used quantum vacuum fluctuations as discussed in Section~\ref{sec:lattice simulations}.

\subsubsection*{KKLT setup with $W_0=10^{-12}$}
\label{sec:KKLT_1e-12}

\begin{figure}
\centering
\subfigure{\includegraphics[width=7.5cm]{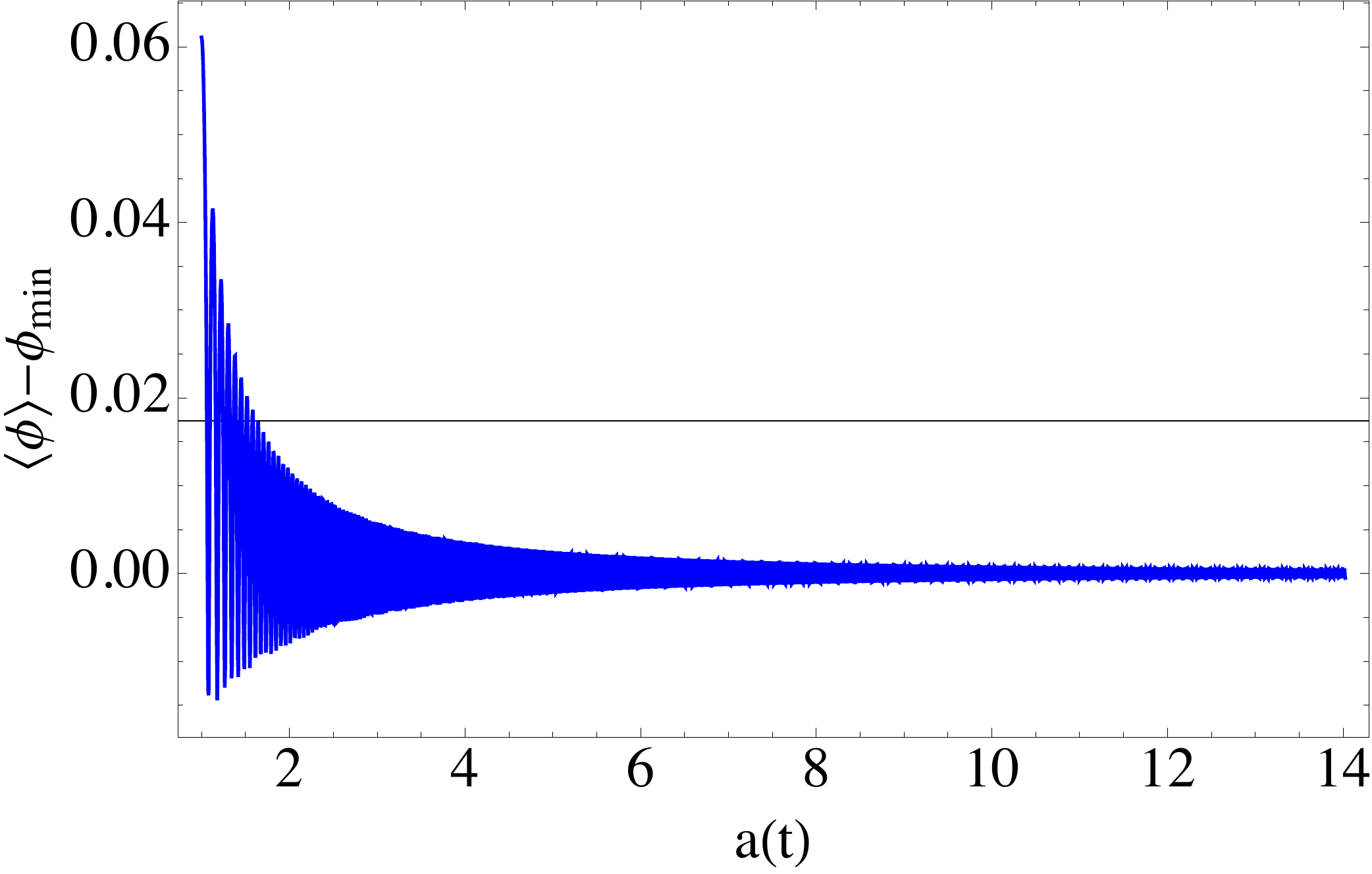}}
\hfill
\subfigure{\includegraphics[width=7.5cm]{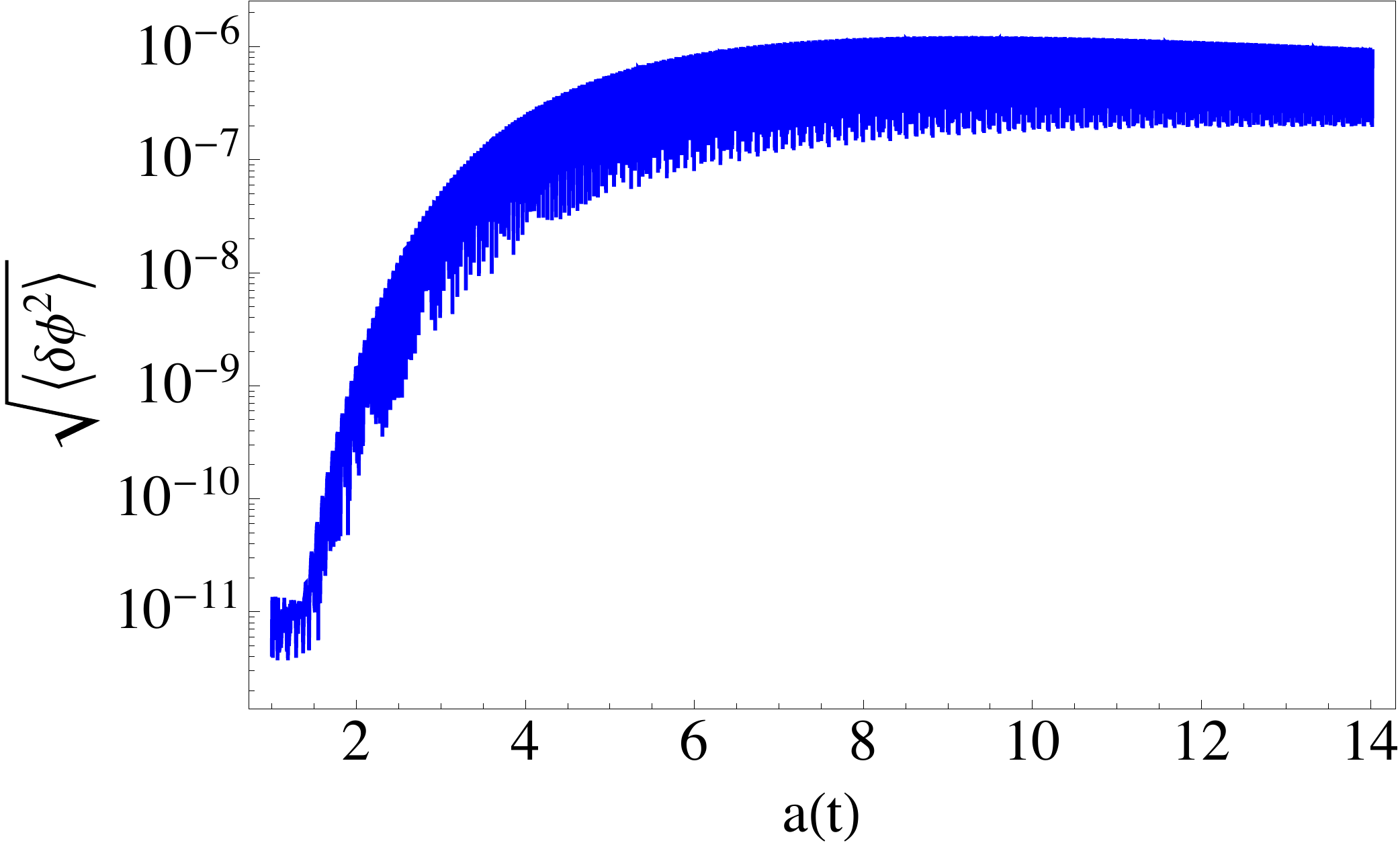}}
\caption{Evolution of the mean $\langle\phi\rangle$ (left) and the variance $\sqrt{\langle\delta\phi^2\rangle}$ (right) as a function of the scale factor $a(t)$. The solid black line in the left plot corresponds to the field value at the inflection point $\phi_{\rm inf}$. The results were obtained from a lattice simulation of the KKLT model with $W_0=10^{-12}$ with $256^3$ lattice points.}
\label{fig:mean_variance_KKLT_1e-12}
\end{figure}

\begin{figure}
\begin{center}\includegraphics[width=\textwidth]{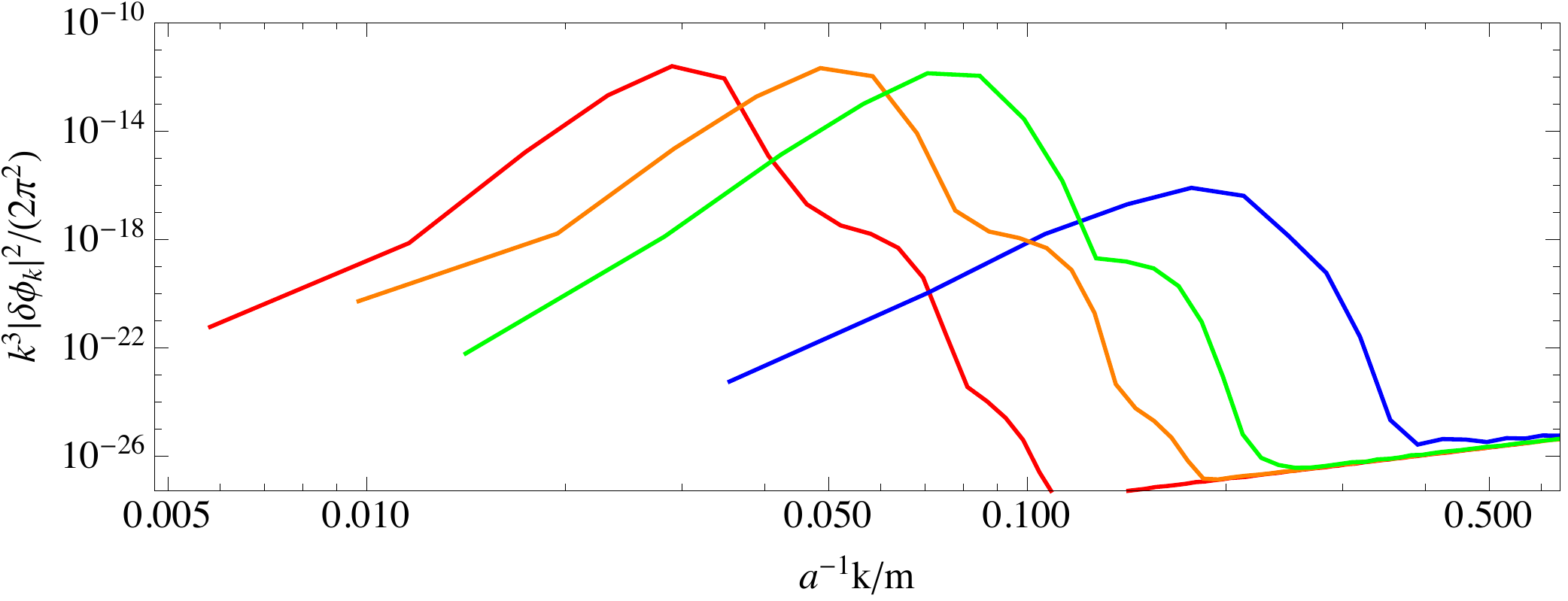}
\end{center}
\caption{The spectrum of $\phi$ fluctuations as a function of the \textit{physical} momentum $a^{-1}k/m$ obtained from a lattice simulation of the KKLT model with $W_0=10^{-12}$. The different colours correspond to different moments in time: $a=2.29$ (blue), $a=5.76$  (green), $a=8.36$ (orange) and $a=14.03$ (red).}
\label{fig:KKLT_spec_1e-12}
\end{figure}

The simulation of the KKLT model with $W_0=10^{-12}$ was performed in three spatial dimensions in a box with \textit{comoving} volume $V_{\rm lattice} = L^3 \simeq (0.7/H_{\rm initial})^3$ and $256$ lattice points per dimension. The results of the simulation are demonstrated in Figures~\ref{fig:mean_variance_KKLT_1e-12} and \ref{fig:KKLT_spec_1e-12}. 

The left part of Figure~\ref{fig:mean_variance_KKLT_1e-12} shows the time evolution of the mean $\langle\phi\rangle$ as a function of the scale factor $a(t)$. The solid black line denotes the field value  at the inflection point of the potential $\phi_{\rm inf}$. One can see that the field crosses the inflection point during the first seven oscillations. Modes for which $k^2<|\partial^2V/\partial\phi^2|$ can be amplified during these tachyonic oscillations. However, as can be seen from the evolution of the variance $\sqrt{\langle\delta\phi^2\rangle}$ on the right part of the plot, both the initial phase of tachyonic preheating, while the field rolls down its potential towards the minimum, and also the subsequent phase of tachyonic oscillations, is not very efficient within this model setup. In fact, the amplitude of the fluctuations starts growing rather when the field falls below the inflection point at $a\sim1.5$. The growth happens due to a parametric (self-)resonance which lasts for many oscillations of $\langle\phi\rangle$. As the amplitude of the background decreases the growth becomes less and less efficient (see Figure~\ref{fig:floquet_KKLT}). At some point (roughly at $a\sim8$) the fluctuations eventually stop growing and are subsequently redshifted due to the expansion. When the variance stops growing the amplitude of the homogeneous component is still about three orders of magnitude larger then the amplitude of the fluctuations. Therefore, although the fluctuations grow by five orders of magnitude the field remains dominated by the homogeneous background and the fluctuations remain linear. 

Figure~\ref{fig:KKLT_spec_1e-12} shows the spectrum $k^3|\delta\phi_k|^2/(2\pi^2)$ of the field fluctuations as a function of the \textit{physical} momentum in units of the mass of the \Kahler modulus $m$. The spectrum is shown at different moments in time represented by the scale factor $a$. The blue curve corresponds to the spectrum at $a=2.29$, the green curve at $a=5.76$, the orange curve at $a=8.36$, and the red curve shows the spectrum at the end of the simulation at $a=14.03$. The spectrum forms a peak at $k/a\sim0.2 m$ which initially continues to grow until $a\sim6$, while the position of the peak is continuously moved to lower values of $k/a$ due to the expansion. One can see that the shape of the spectrum remains nearly the same which is exactly what is expected if non-linear interactions between different $k$ modes are absent and the resonance becomes inefficient.

\subsubsection*{KKLT setup with $W_0=10^{-5}$}
\label{sec:KKLT_1e-5}

The situation is different in the scenario with $W_0=10^{-5}$. In fact, as we will shortly discuss in greater detail, the dynamics lead to non-linear interactions between fluctuations and ultimately to the formation of oscillons. The results presented in this part of the section were obtained from a three dimensional lattice simulation of the KKLT model with $W_0=10^{-5}$ and $512$ points per spatial dimension. The simulation was performed in a box with \textit{comoving} volume $V_{\rm lattice} = L^3 \simeq (0.7/H_{\rm initial})^3$.

\begin{figure}
\centering
\subfigure{\includegraphics[width=7.5cm]{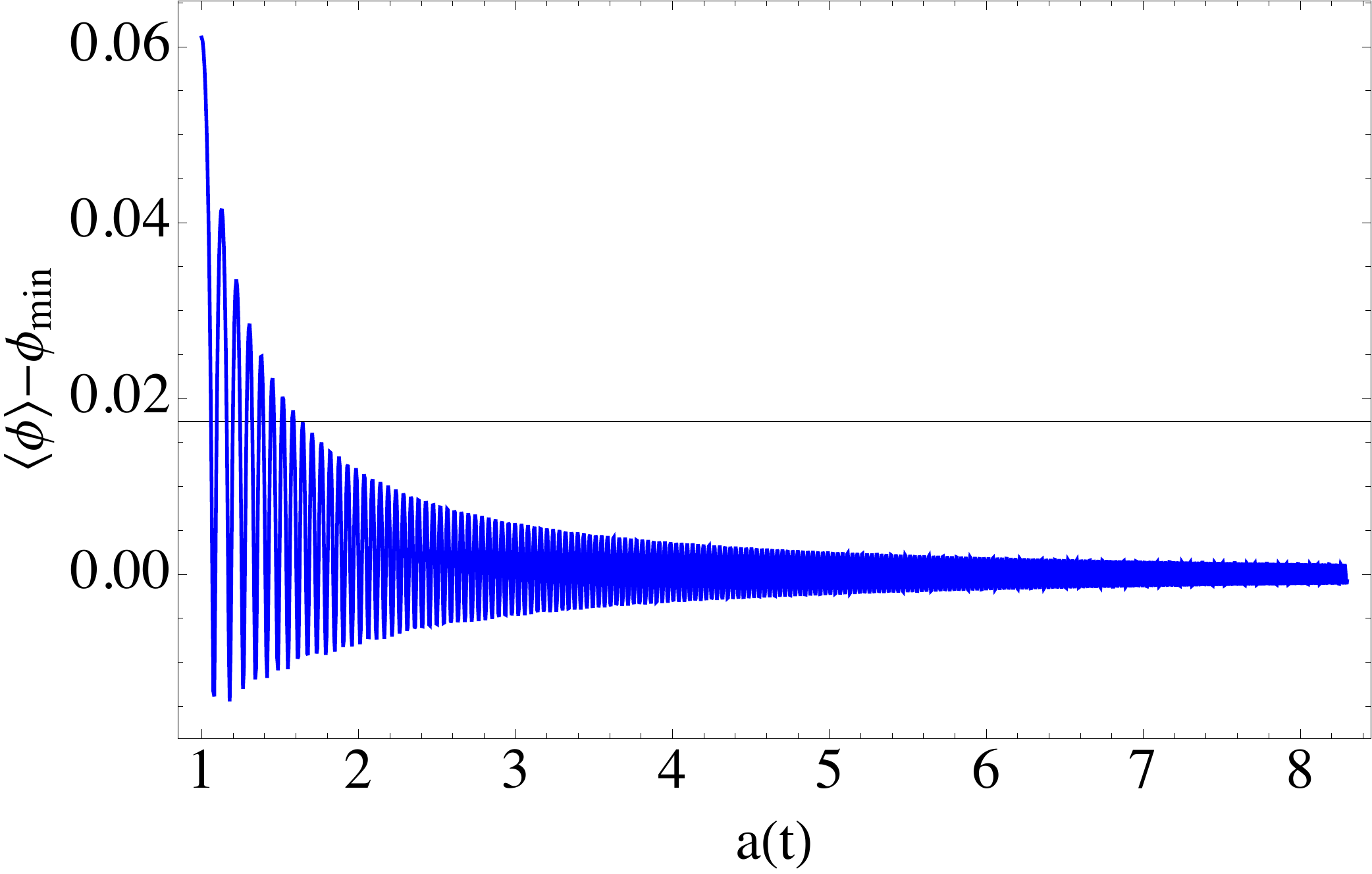}}
\hfill
\subfigure{\includegraphics[width=7.5cm]{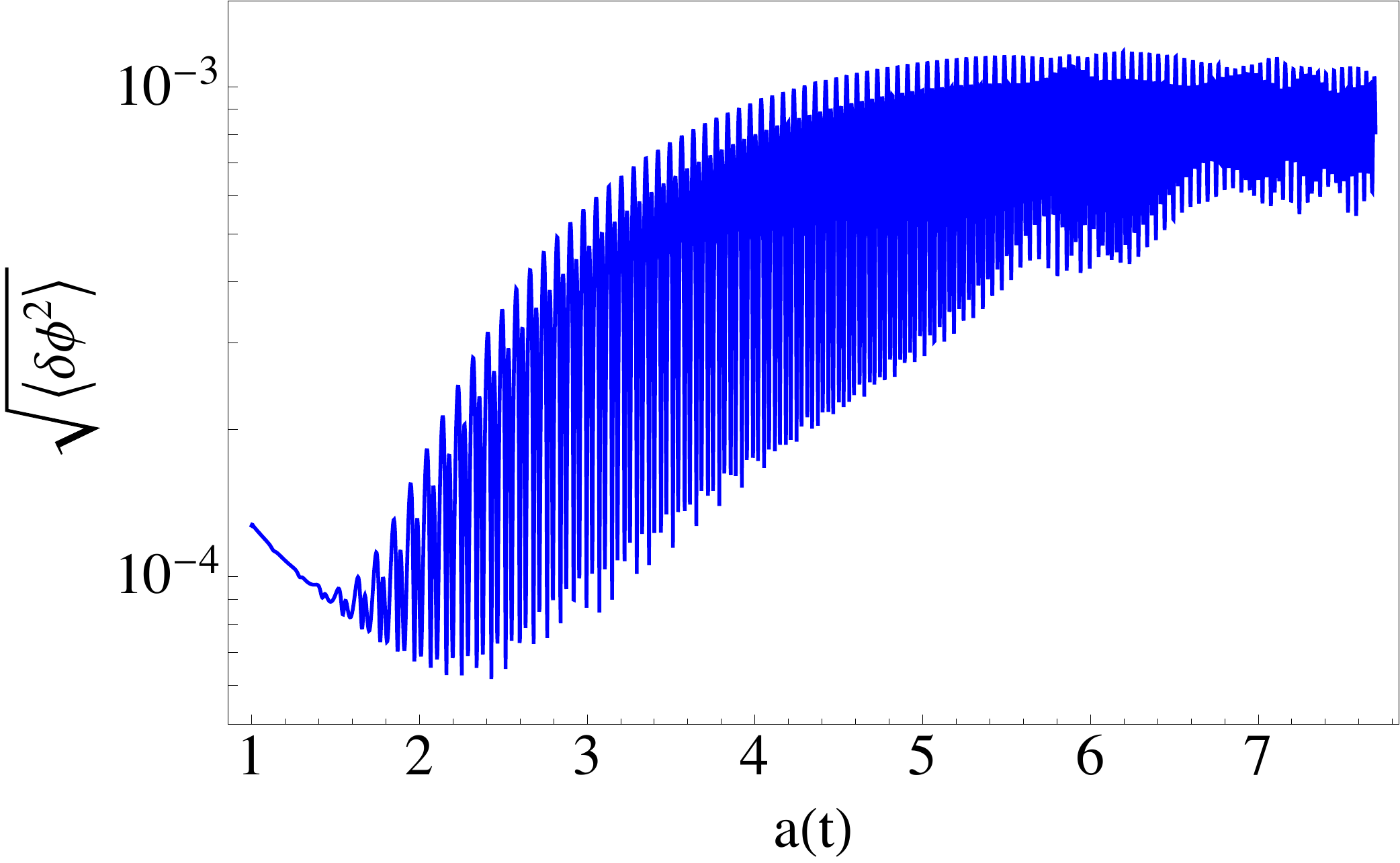}}
\caption{Results from a lattice simulation of the KKLT model with $W_0=10^{-5}$ with $512^3$ lattice points. \textit{Left:} The homogeneous mode $\phi(t)$ as a function of the scale factor $a(t)$. The solid black line marks the field value at the inflection point of the potential $\phi_{\rm inf}$. One can see that only the first three oscillations are tachyonic. \textit{Right:} The variance $\sqrt{\langle\delta\phi^2\rangle}$ as a function of the scale factor $a(t)$. The dynamics start showing non-linear effects when the amplitude of the homogeneous component $\langle\phi\rangle$ becomes comparable to the variance at $a\sim5$.}
\label{fig:mean_variance_KKLT_1e-5}
\end{figure}

Let us first discuss the evolution of the mean $\langle\phi\rangle$ and the variance $\sqrt{\langle\delta\phi^2\rangle}$ shown in Figure~\ref{fig:mean_variance_KKLT_1e-5}. The left part of the figure shows $\langle\phi\rangle$ as a function of the scale factor $a(t)$ where the solid black line corresponds to the field value $\phi_{\rm inf}$ at the inflection point of the potential. One can see that within this setup, only the first three oscillations are tachyonic. However, similar to the setup with $W_0=10^{-12}$ the initial phase of tachyonic preheating and also the subsequent tachyonic oscillations are not very efficient as can be seen from the variance on the right part of the plot in Figure~\ref{fig:mean_variance_KKLT_1e-5}. An efficient growth of fluctuations does not happen until the amplitude of the homogeneous mode falls below the inflection point. As in the setup with $W_0=10^{-12}$, the dominant mechanism which leads to the growth of fluctuations is a parametric resonance which starts to become efficient at $a\sim1.6$. The amplitude of the fluctuations grows by one order of magnitude during the parametric resonance until non-linear effects start to become important at $a\sim5$. As expected from our Floquet analysis (see Figure~\ref{fig:floquet_KKLT}), the fluctuations grow less within a certain period of expansion than in the KKLT setup with $W_0=10^{-12}$. However, since the amplitude of the initial vacuum fluctuations is now considerably larger, while the amplitudes of the averaged fields are comparable in both cases, the necessary growth for fluctuations to become affected by non-linear effects is significantly smaller. 

\begin{figure}
\begin{center}\includegraphics[width=\textwidth]{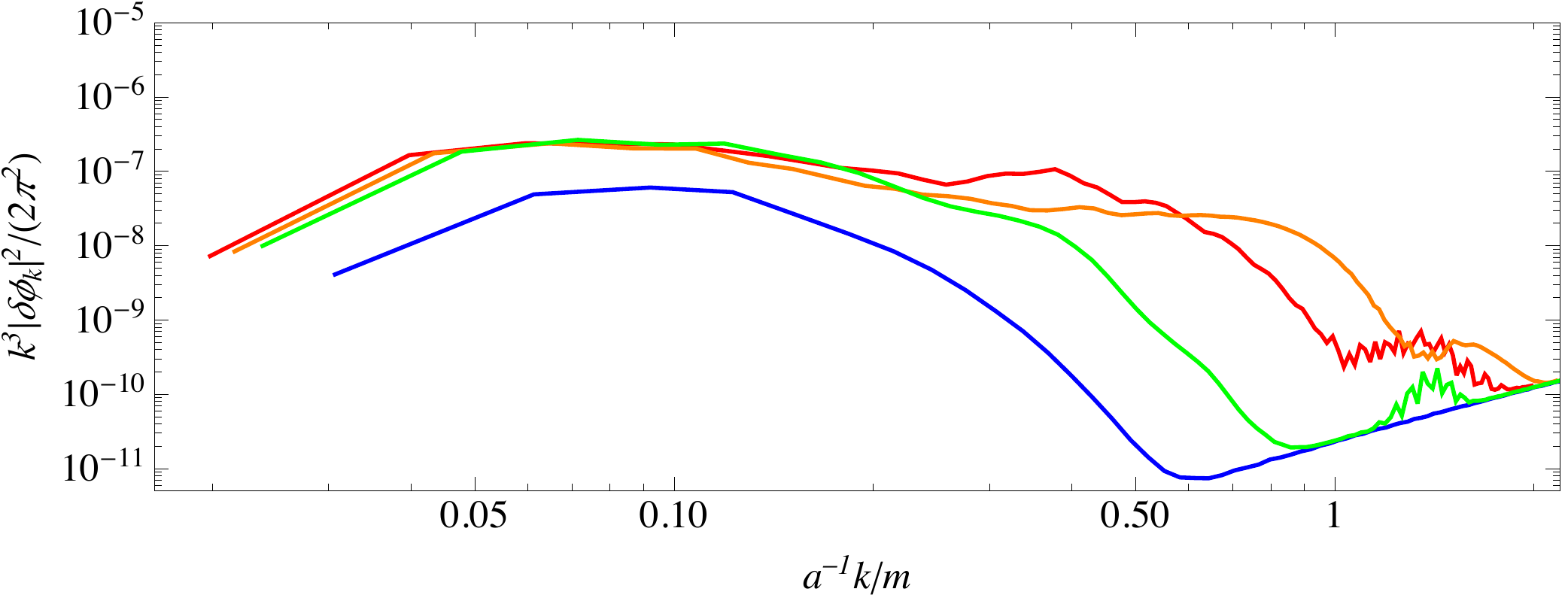}
\end{center}
\caption{The spectrum of field fluctuations $k^3|\delta\phi_k|^2/(2\pi^2)$ as a function of the \textit{physical} momentum $a^{-1}k/m$. The results originate from a lattice simulation of the KKLT model with $W_0=10^{-5}$ with $512^3$ points. The different colours correspond to the following moments in time: shortly after the end of the linear regime $a=4.98$ (blue), $a=6.41$  (green), $a=7.07$ (orange) and at the end of the simulation $a=8.3$ (red).}
\label{fig:KKLT_spec_1e-5}
\end{figure}

The non-linear evolution becomes more apparent when looking at the evolution of the field spectrum at different times. The latter is shown in Figure~\ref{fig:KKLT_spec_1e-5}. Compared to the results in Section~\ref{sec:KKLT_1e-12} (cf.\ Figure~\ref{fig:KKLT_spec_1e-12}), where the shape of the spectrum remains nearly unchanged throughout the evolution of the field, the shape of the spectrum changes significantly. The green curve in Figure~\ref{fig:KKLT_spec_1e-5}, for example, shows the spectrum at $a=6.41$. Comparing the spectrum to an earlier time at $a=4.98$ (blue curve) one can clearly see that the peak has extended to a larger range of $k$ modes and developed a smaller, second peak at $a^{-1}k \lesssim2 m$. The other two spectra correspond to times when $a=7.07$ (orange curve) and $a=8.3$ (red curve) at the end of the simulation.

\begin{figure}
\begin{center}
\includegraphics[width=0.45\textwidth]{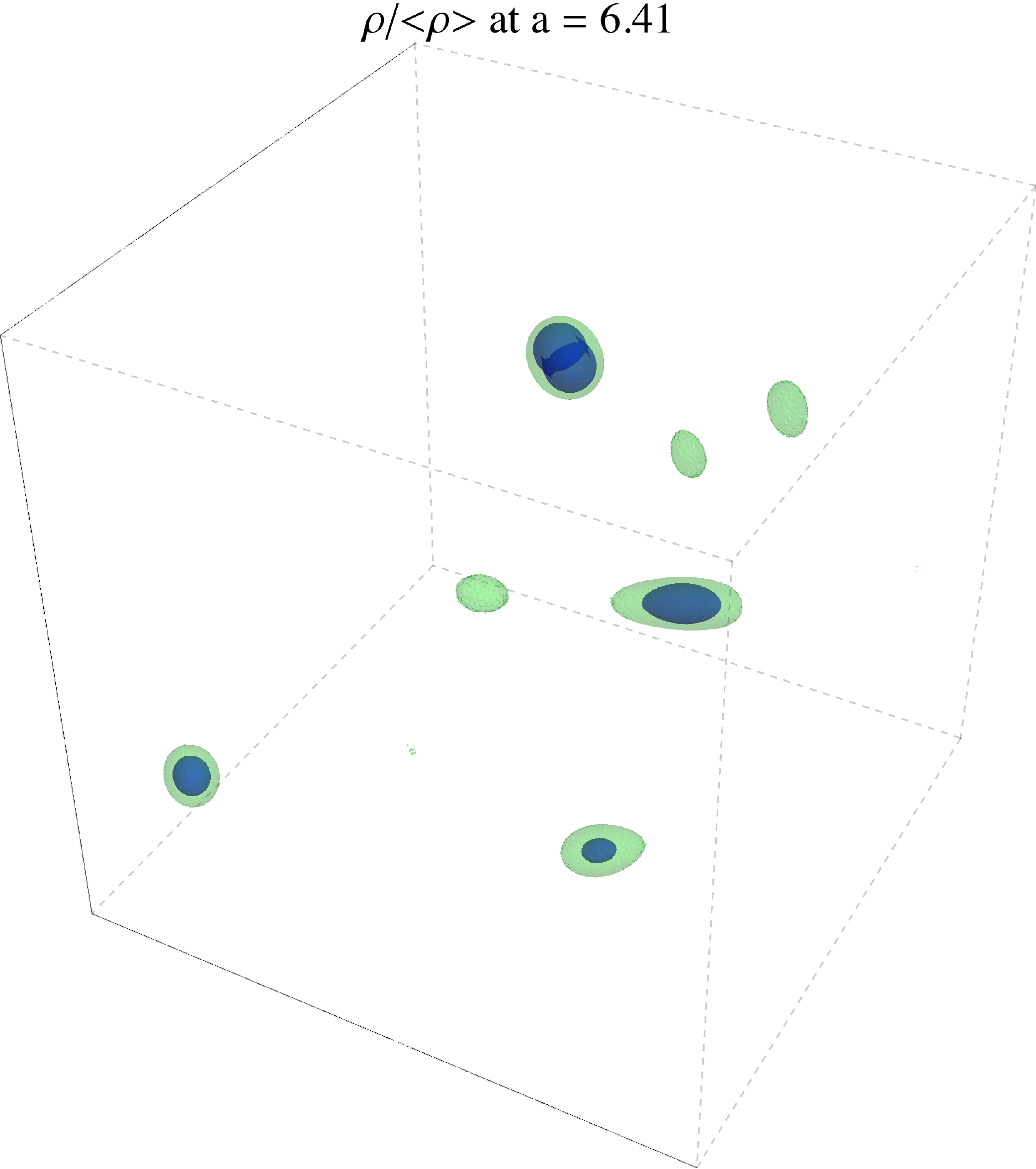}
\includegraphics[width=0.45\textwidth]{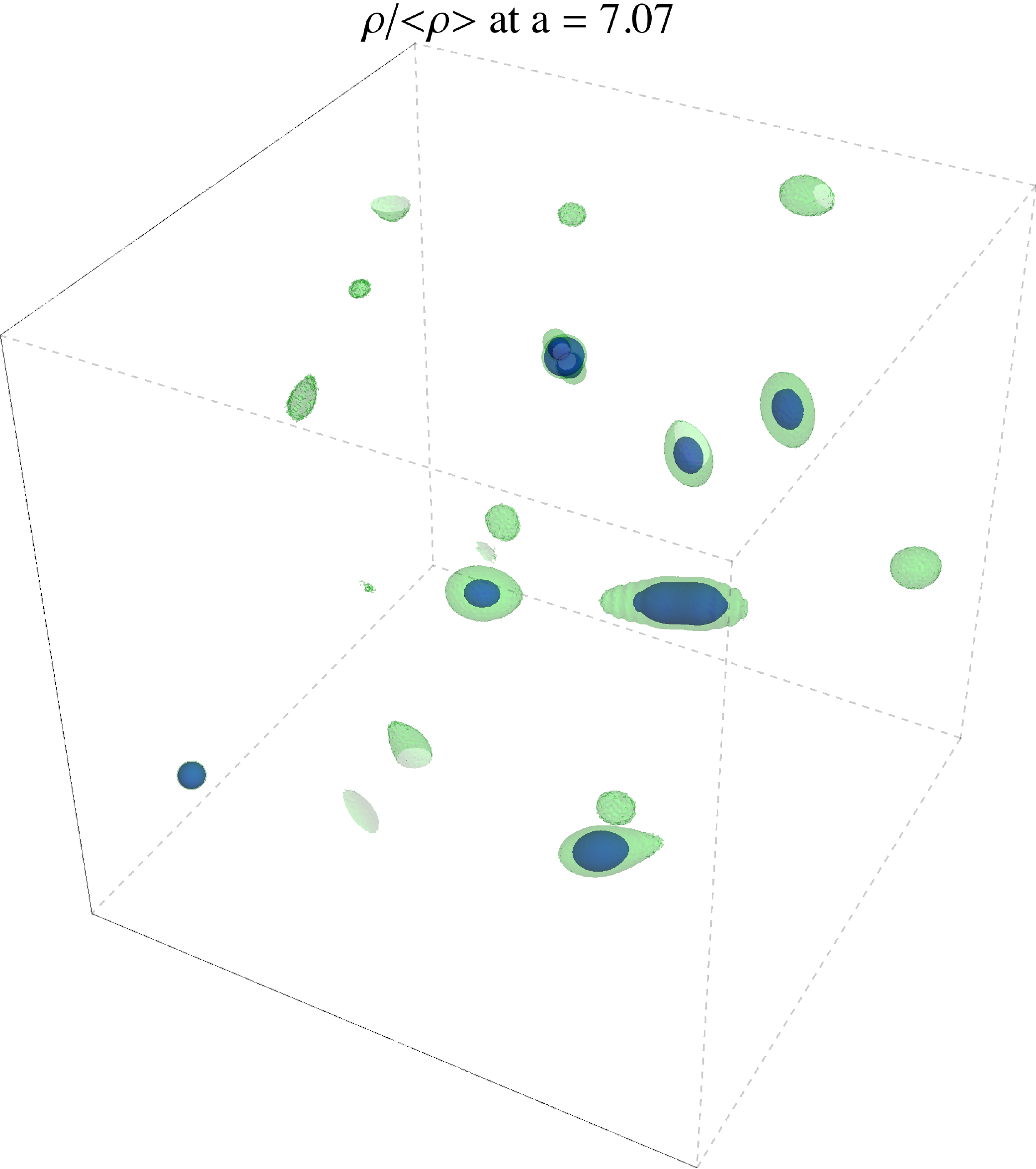}
\includegraphics[width=0.45\textwidth]{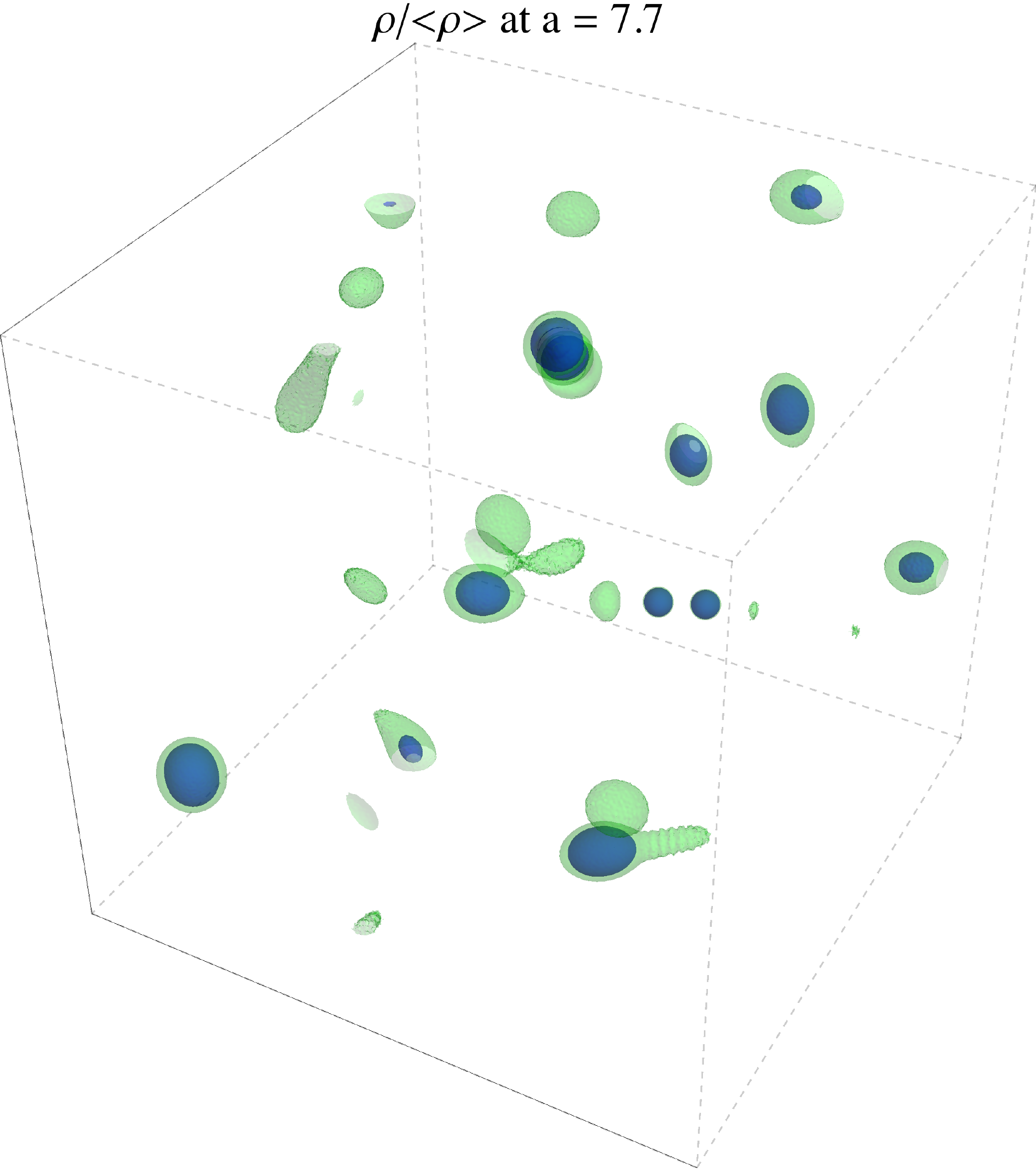}
\includegraphics[width=0.45\textwidth]{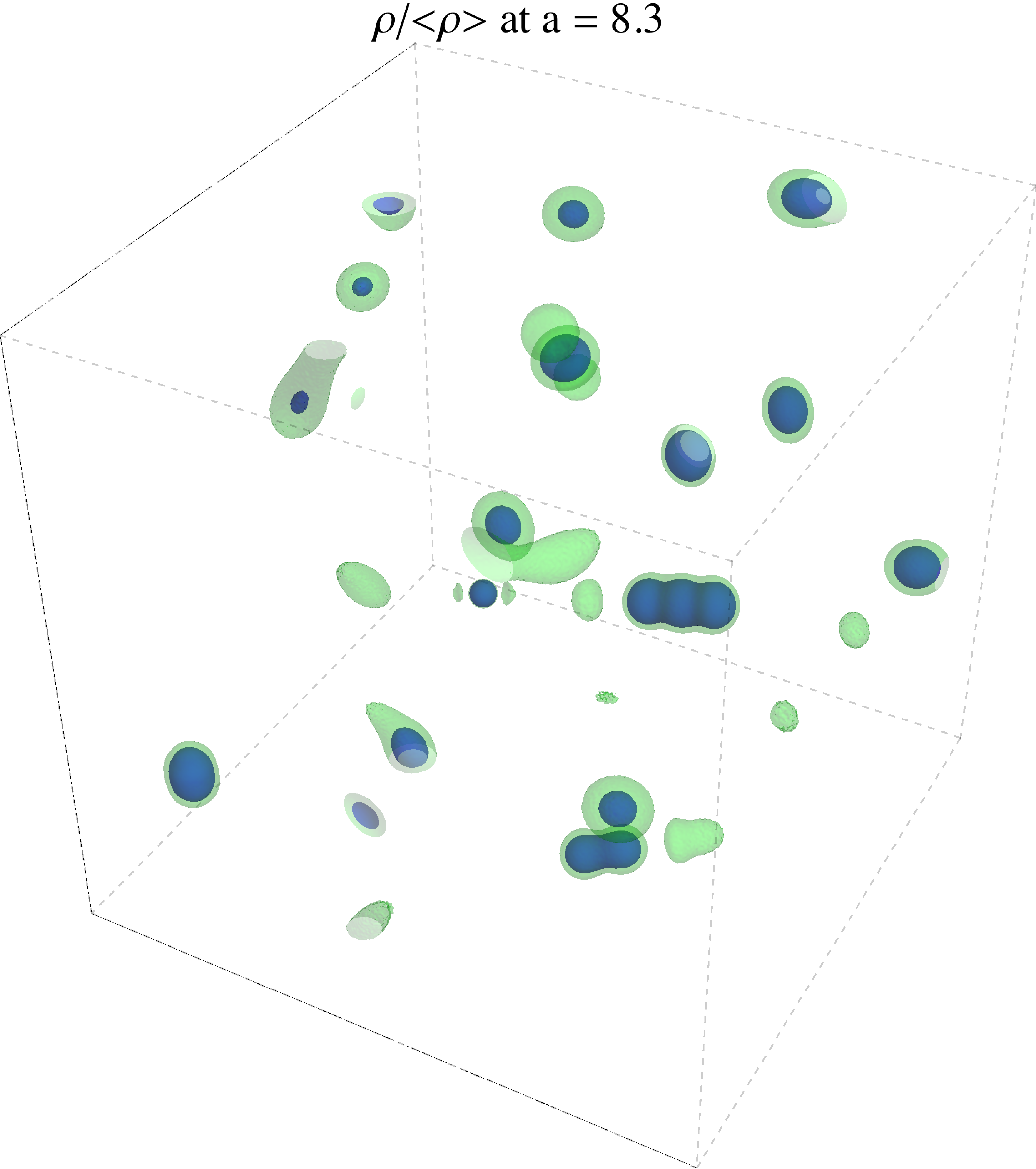}
\end{center}
\caption{Results from a lattice simulation of the KKLT model with $W_0=10^{-5}$ and $512$ points per dimension. The figure shows the three dimensional energy density distribution. The green surfaces correspond to regions with six times the average energy density $\langle\rho\rangle$ while the blue surfaces indicate $\rho/\langle\rho\rangle = 12$. The energy density distribution is shown at four different moments in time denoted by the corresponding scale factor.}
\label{fig:energydensity_KKLT}
\end{figure}

Figure~\ref{fig:energydensity_KKLT} shows the spatial energy density distribution normalised to the average energy density $\langle\rho\rangle$ at $a=6.41$ (upper left), $a=7.07$ (upper right), $a=7.7$ (lower left), and $a=8.3$ (lower right). The green surfaces correspond to regions where $\rho = 6\langle\rho\rangle$ and blue surfaces to $\rho = 12\langle\rho\rangle$. As one can see, stable overdensities which are present on each of the figures are produced. These overdensities correspond to localised, large amplitude oscillations in field space, i.e.\ oscillons. By looking carefully at Figure \ref{fig:energydensity_KKLT} one can see that sophisticated dynamics take place: individual oscillons interact among each other, non-spherical and non-ellipsoidal structures emerge and disappear. It also seems that the number of oscillons constantly increases with time. We therefore expect that the phase of oscillon production has not ended by the end of our simulation.

Although the fluctuations become sufficiently large for non-linear effects to become important and inhomogeneities develop in the form of oscillons, we observe that the homogenous mode has not yet (completely) decayed, as can be seen from the mean $\langle\phi\rangle$ in Figure~\ref{fig:mean_variance_KKLT_1e-5}. Moreover, an overshooting over the potential barrier into the decompactifying vacuum, which can happen when fluctuations are amplified strong enough (see Ref.~\cite{Antusch:2015nla}) was also not observed in the KKLT setups considered in this paper. 

\begin{figure}
\begin{center}\includegraphics[width=\textwidth]{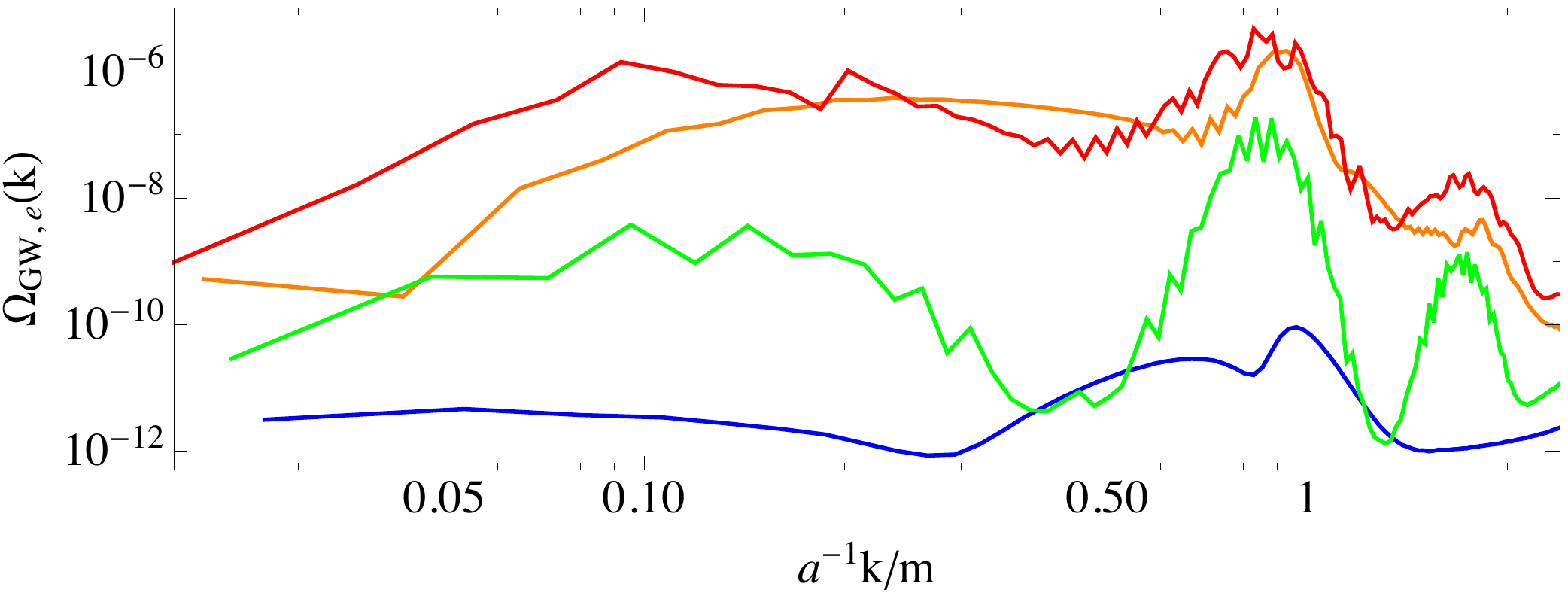}
\end{center}
\caption{Spectrum of gravitational waves $\Omega_{\rm GW,\rm e}(k)$ for KKLT model as a function of the $\textit{physical}$ momentum $a^{-1}k/m$ at $a=5.72$ (blue), $a=6.41$  (green), $a=7.07$ (orange) and at the end of the simulation $a=8.3$ (red). One can clearly see two peaks at $k/a \lesssim 1$ and $k/a \lesssim 2$. The simulation which led to the results was performed with $W_0=10^{-5}$ and $512^3$ points.}
\label{fig:KKLT_GW_1e-5}
\end{figure}

The production of oscillons and the consequent rich dynamics which is apparent in Figure~\ref{fig:energydensity_KKLT}, bring us to the last result of this simulation. In Figure~\ref{fig:KKLT_GW_1e-5} we present the spectrum of GW $\Omega_{\rm GW, \rm e}(k)$ as a function of the \textit{physical} momentum $k/a$. The different colours correspond to different times at $a=5.72$ (blue), $a=6.41$ (green), $a=7.07$ (orange), and at $a=8.3$ (red). The spectrum develops a characteristic peak structure with peaks at $k/a\lesssim m$ and $k/a\lesssim 2m$ and a broad plateau for $k/a<m$. The peaks are attributable to the oscillon dynamics and their position essentially corresponds to the oscillation frequency of the oscillons (and the next harmonic).\footnote{Note that the position of such a peak would be different in a symmetric potential due to the shift in the resonant frequencies.} Oscillons continue to be constantly formed until the end of our simulation and we therefore expect that the the results we show are not yet the final. Moreover, to have successful reheating the oscillons will have to decay. The decay of the oscillons is an additional source for GW production. However, if we assume that the Universe instantly reheats at the end of our simulation i.e.\ that the whole energy content is immediately converted into radiation, we can calculate the spectrum of the GW as it would be observable today according to Equation\eqref{eq:OmegaGW_today} and \eqref{eq:GW_freq_today} with $a_{\rm e}=a_*$. We find 
\be
\Omega_{\rm GW,0}(f_{0, \rm peak}) \sim3\times10^{-11} \,,\quad \textrm{with} \quad f_{0, \rm peak}\sim 10^9\,\textrm{Hz}\,.
\ee
Lower frequencies are in principle also possible. For example, we assumed $e^{K_{\rm cs}}=1$ for the overall rescaling of the potential from the VEV of the complex structure \Kahler potential $K_{\rm cs}$. Values smaller than unitiy would lead to smaller frequencies, since 
\be
f \propto \frac{m}{\rho^{1/4}_{\rm e}} \propto \frac{(e^{K_{\rm cs}})^{1/2}}{(e^{K_{\rm cs}})^{1/4}}\propto (e^{K_{\rm cs}})^{1/4}\,.
\ee
We note that, however, decreasing $e^{K_{\rm cs}}$ means also decreasing the magnitude of the initial vacuum fluctuations and that too small values of $e^{K_{\rm cs}}$ could lead to milder dynamics, since the fluctuations would have to grow for a longer period of time for non-perturbative effects to show up. This, in turn leads to less GW production.

\subsection{Blow-up moduli in LVS}
\label{sec:KMI}

Next we dicuss the standard potential for K\"ahler moduli in the Large Volume scenario of moduli stabilisation~\cite{Balasubramanian:2005zx,Conlon:2005ki} in type IIB string theory. It is realised in so-called Swiss-cheese type Calabi-Yau manifolds where the volume can be written as
\begin{equation}
\V = \alpha \left(\tau_{\rm b}^{3/2} - \sum_{i=2}^N \alpha_{i} \tau_{{\rm s}, i}^{3/2}\right)\,,
\end{equation}
where $\alpha,$ $\alpha_i$ are numerical constants, $\tau_i$ are the K\"ahler moduli which describe the size of four-cycles in the Calabi-Yau threefold. In the following we will focus on the \textit{small} blow-up cycles $\tau_s$ which are cycles that can shrink to zero size while the volume stays finite -- the holes of the Swiss-cheese. As in KKLT, the complex structure moduli and the dilaton are stabilised using fluxes. The potential for the K\"ahler moduli is generated by $\alpha'-$corrections and non-perturbative corrections. The K\"ahler and superpotential are given by
\begin{eqnarray}
K/M^2_{\rm Pl} &=& - 2 \log\left(\V + \frac{\xi s^{3/2}}{2}\right),\\
W/M^3_{\rm Pl} &=& W_0 + \sum_{i=2}^N A_i e^{-a_i T_{s,i}}\,,
\end{eqnarray}
where $\xi \sim\chi({\rm CY})$ is proportional to the Euler characteristic of the CY considered and parametrises the leading $\alpha'-$corrections, $s=1/g_s$ is the inverse string coupling. $W_0$ is the VEV of the flux superpotential, $A_i$ are $\mathcal{O}(1)$ coefficients, $a_i$ are constants depending on the non-perturbative effects (e.g.~for gaugino condensation caused by $N$ D7-branes,  $a_i = \frac{2 \pi}{N}$).
In the LVS, the scalar potential for the K\"ahler moduli can be organised in an inverse volume expansion and the leading contributions are given by:
\begin{equation}
\label{eq:LVSscalarpotential}
\frac{V_{\mathcal{O}\left(\V^{-3}\right)}}{M^4_{\rm Pl}} = \frac{g_s}{8 \pi} \left[\sum_{i=1}^N \left(\frac{8}{3} \left(\frac{a_i A_i}{\alpha_i}\right)^2 \sqrt{\tau_{s, i}} \,\frac{e^{-2 a_i \tau_{s, i}}}{\V} -  4 W_0  \,a_i A_i \tau_{s, i} \,\frac{e^{-a_i \tau_{s, i}}}{\V^2} \right) + \frac{3 \hat{\xi} |W_0|^2}{4 \V^3}\right] + V_{\text{dS}} \,,
\end{equation}
where $\hat{\xi}=\xi/g_s^{3/2},$ $V_{\text{dS}}=D/\V^\gamma$ ($1\leq\gamma< 3$) is an additional contribution from localised sources where $D$ is fine-tuned to uplift the potential to an approximate Minkowski minimum. In addition, there is an overall rescaling of the entire potential arising from the VEV of the complex structure \Kahler potential $e^{K_{\rm cs}}$, which, unless otherwise stated, we set to unity. The minimum with respect to the small moduli is given by
\begin{equation}
a_i A_i e^{-a_i\tau_i}=\frac{3\alpha\alpha_i}{\V}\frac{(1-a_i\tau_i)}{(1-4a_i \tau_i)}\sqrt{\tau_i}~.
\end{equation}
In the limit $\V\to \infty$ this gives
\begin{equation}
a_i \tau_i\approx \log{(\V)}~.
\end{equation}
The volume and $D$ are fixed by minimising the potential with respect to the volume and demanding the vanishing of the vacuum energy:
\begin{equation}
\frac{\partial V}{\partial \V}=0~,\qquad V=0~.
\end{equation}
This leads to an exponentially large value of the volume in the minimum (in string units)
\begin{equation}
\log{\V}\approx \frac{\hat{\xi}}{2}\left(\sum\frac{\alpha_i \alpha}{a_i^{3/2}}\right)^{-1}~.
\end{equation}
The value of $D$ can be determined numerically. The masses for the volume and small moduli in the minimum are given by
\begin{eqnarray}
m_{\tau_i}^2&\simeq & M^2_{\rm Pl}\frac{W_0^2 (\log{\V})^2}{\V^2}~,\\
m_{\V}^2&\simeq & M^2_{\rm Pl}\frac{W_0^2 }{\V^{3}\log{\V}}~.
\end{eqnarray}
The canonical normalisation of the blow-up modulus is given by $\sigma=\sqrt{\frac{4}{3{\cal V}}}\tau_2^{3/4}$.

Here we consider the case where one of the blow-up moduli is displaced from the minimum, while keeping all the other fields, in particular the volume at its minimum. 

Notice that in terms of the canonically normalised field $\sigma$ this potential is approximately of the form:

\be
V\sim V_0\left(1-\kappa(\sigma) e^{-\alpha \sigma^{4/3}}\right)^2\,,
\ee
so its behaviour is similar to the exponential potentials mentioned before. Notice also that the coefficient of the exponential $\alpha$ is hierarchically large since 
$\alpha\sim \mathcal{O}(\mathcal{V}^{2/3})$. In this case the coefficient $k$ is field-dependent: $k(\sigma) \sim \sigma^{4/3}$. This makes the scalar potential for blow-up modes to behave differently from the potentials for other moduli, such as fibre moduli for which $\alpha\sim \mathcal{O}(1)$. In particular the ratio $m/H_{\rm initial}\sim \alpha$ plays an important role in the production of oscillons. For instance in the models of ~\cite{Antusch:2016con} oscillons were found for $m/H_{\rm initial}\sim \mathcal{O}(100)$. Having the coefficient of the exponential $\alpha\sim\mathcal{O}(100)$ is natural for potentials of blow-up modes but not for other moduli.

\subsubsection{Results from lattice simulations}
\label{sec:latticesimulations}
We simulate the evolution of the canonically normalised blow-up modulus $\sigma=\sqrt{\frac{4}{3{\cal V}}}\tau_2^{3/4}$ in its potential Equation~\eqref{eq:LVSscalarpotential} using the following parameter set: $W_0=A_i=\xi=a_2=1$, $a_1=\pi$, $g_s=0.2$, $\gamma=2$ and $N=10$. The corresponding potential for the field $\sigma$, centered around the minimum of the potential at $\sigma_{\rm min}\simeq0.075\,M_{\rm Pl}$, is shown in Figure~\ref{fig:potential} in units of the plateau $V(\sigma \gg \sigma_{\rm min})\equiv V_0 \simeq 1.794\times10^{-12}\,M^4_{\rm Pl}$. The simulation was performed in $3+1$ dimensions with 512 points per spatial dimension in a box with \textit{comoving} volume $V_{\rm lattice}=L^3=(15\pi/m)^3$. The homogeneous field and its derivative were initialised as
\be
\sigma_{\rm initial} = 0.025\,M_{\rm Pl}\,,\qquad \dot{\sigma}_{\rm initial} = 0\,,\qquad H_{\rm initial} = \frac{1}{M_{\rm pl}}\sqrt{\frac{V(\sigma_{\rm initial})}{3}} = 7.477\times10^{-7} M_{\rm Pl}\,.
\ee
The initial field fluctuations were initialised as quantum vacuum fluctuations (cf.\ Section~\ref{sec:KKLT_lattice simulations}) with an ultraviolet (\textit{comoving}) momentum cutoff at $k_{\rm UV} = m$. The reason for the cutoff is simply to avoid an unphysical contribution to vacuum energy while keeping the resolution benefits of a small lattice spacing.

\begin{figure}
\centering
\subfigure{\includegraphics[width=7.5cm]{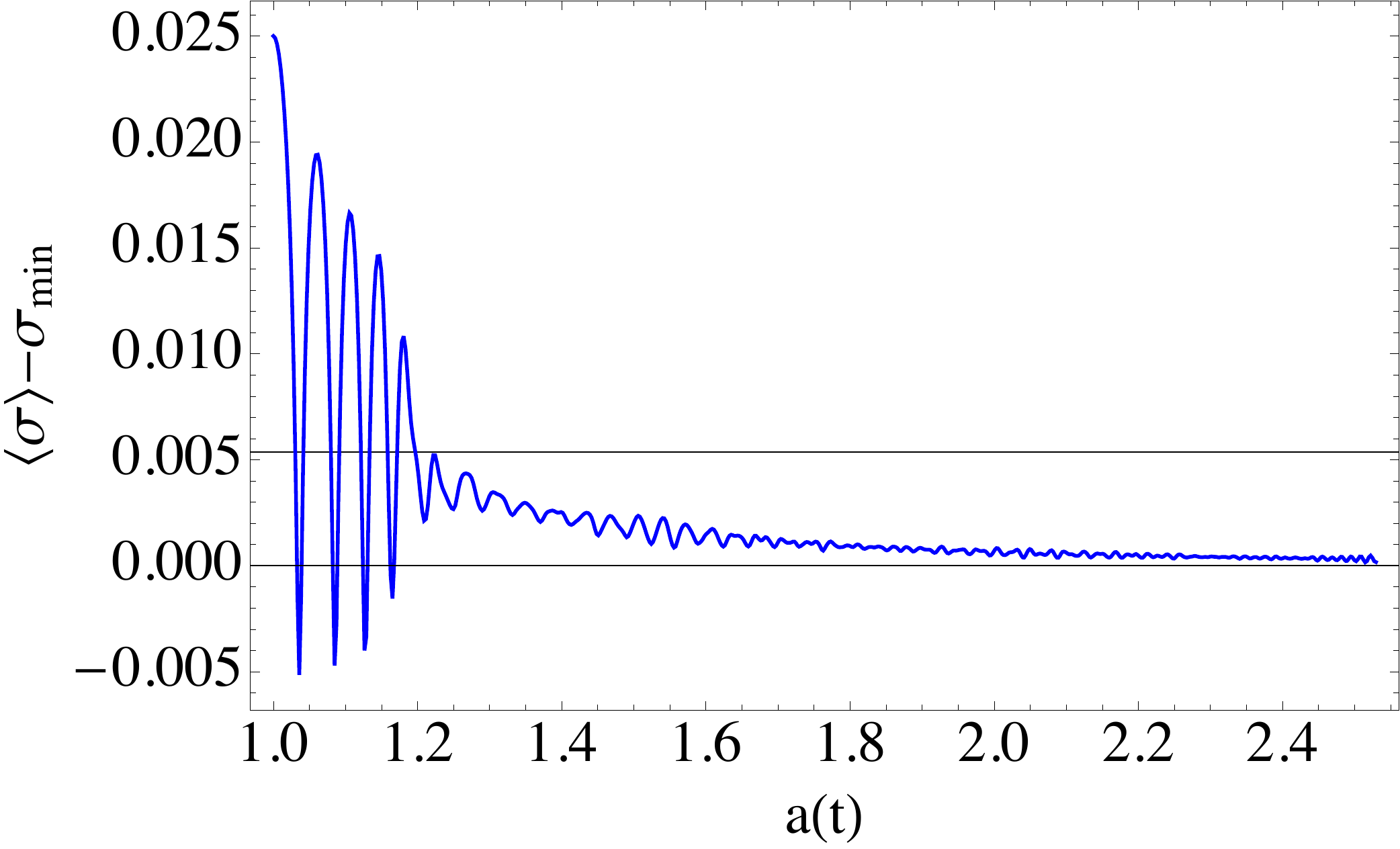}}
\hfill
\subfigure{\includegraphics[width=7.5cm]{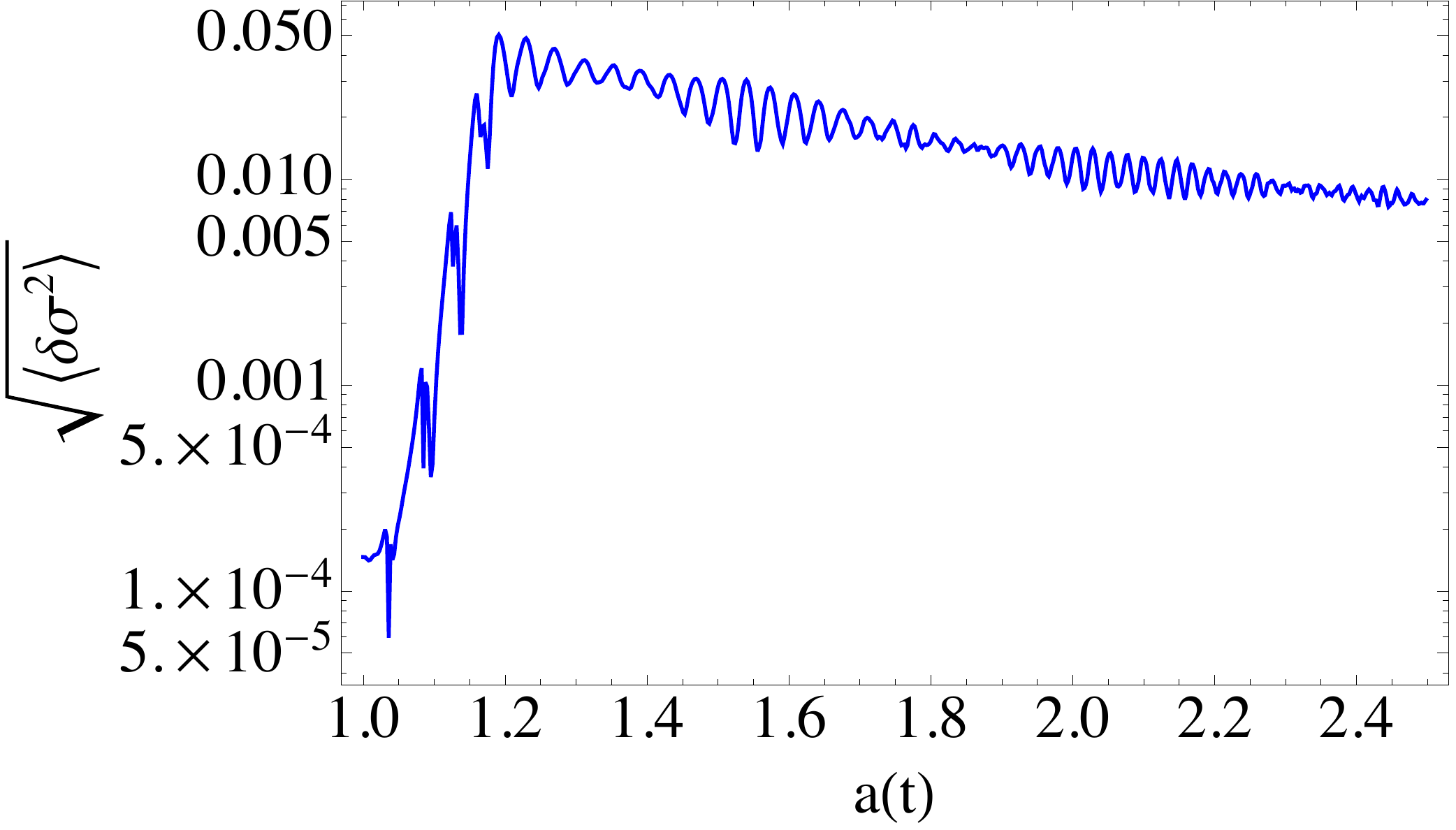}}
\caption{\textit{Left:} Evolution of the mean of the blow-up modulus $\langle\sigma\rangle$ as a function of the scale factor $a(t)$. The solid black line denotes the field value at the inflection point of the potential. One can see that the initially homogeneous field decays into inhomogeneous fluctuations within four oscillation. \textit{Right:} Evolution of the variance $\langle\delta\sigma^2\rangle^{1/2}$. The evolution becomes non-linear when $\langle\delta\sigma^2\rangle^{1/2}$ becomes comparable to the amplitude of oscillation of  homogeneous component $\langle\phi\rangle$.}
\label{fig:mean_variance}
\end{figure}

Figure~\ref{fig:mean_variance} shows the mean of the blow-up modulus $\langle\sigma\rangle$ (left) and of the variance $\langle\delta\sigma^2\rangle^{1/2}$ (right) both as a function of the scale factor $a(t)$. The solid black line on the left part of the figure denotes the field value at the inflection point of the potential. From the evolution of $\langle\sigma\rangle$ one can clearly see that the field becomes quickly inhomogeneous within the first 3-4 oscillations. The first few oscillations are tachyonic oscillations i.e.\ oscillations during which the field reaches values beyond the inflection point. As can be seen from the evolution of $\langle\delta\sigma^2\rangle^{1/2}$, on the right part of Figure~\ref{fig:mean_variance}, tachyonic oscillations are the dominant mechanism for the growth of fluctuations. The fluctuations stop growing when non-linear interactions become important.

\begin{figure}
\begin{center}\includegraphics[width=\textwidth]{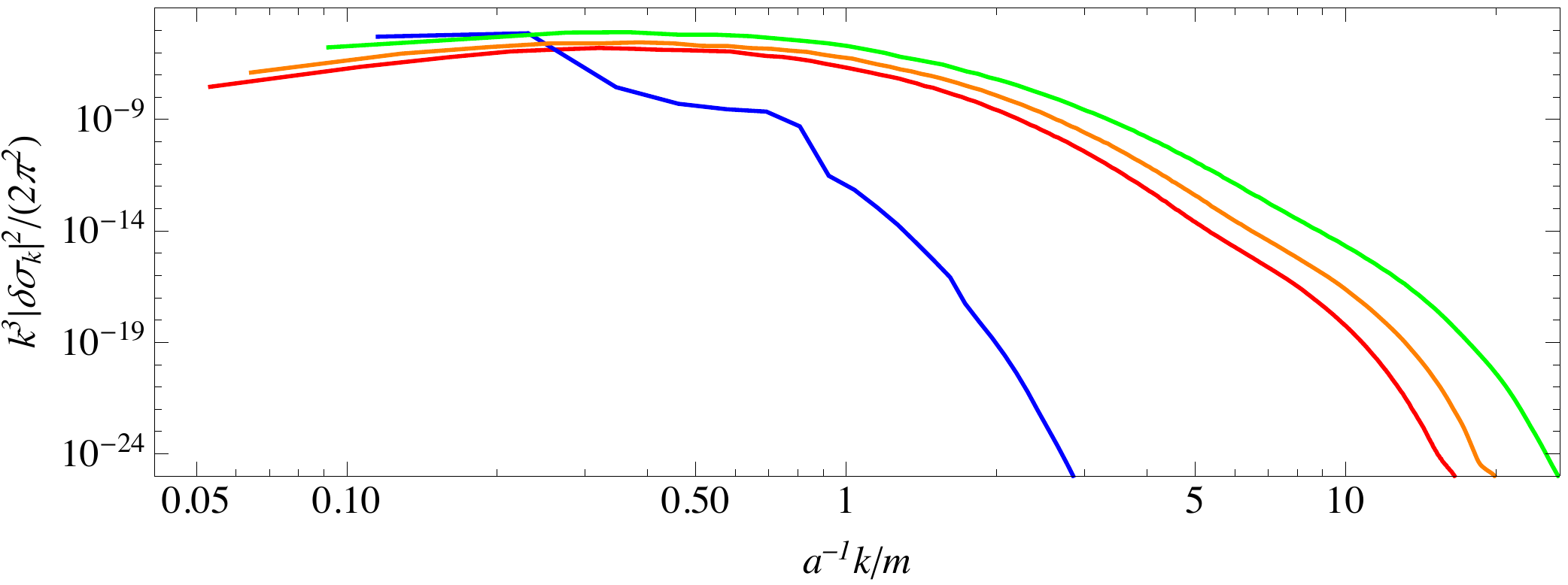}
\end{center}
\caption{Spectrum of fluctuations at different moments in time: at the end of linear preheating at $a\simeq1.16$ (blue), shortly after the beginning of the non-linear regime at $a\simeq1.45$ (green), at $a\simeq2.1$ (orange), and at the end of the simulation $a\simeq2.5$ (red).}
\label{fig:KMI_spec}
\end{figure}

The next result we want to show is the power spectrum of the fluctuations of $\sigma$. It is shown in Figure~\ref{fig:KMI_spec} as a function of the \textit{physical} wavenumber $k/a$. The different colours correspond to different times during the evolution of the field. The blue line shows the spectrum at the end of linear preheating at $a\simeq1.16$. While $a\lesssim1.2$ the fluctuations are constantly amplified due to tachyonic oscillations leading to a peak in the spectrum of $\delta\sigma_k$ for modes with $k/a\lesssim 0.5m$. This growing phase stops as different modes eventually start interacting with each other. The energy carried by the fluctuations is then distributed among different modes leading to a widening of the spectrum towards the UV. The green line shows the spectrum shortly after the beginning of the non-linear regime at $a\simeq1.45$. The other two lines show the spectrum at $a\simeq2.1$ (orange), and at the end of the simulation $a\simeq2.5$ (red).

\begin{figure}
\begin{center}\includegraphics[width=\textwidth]{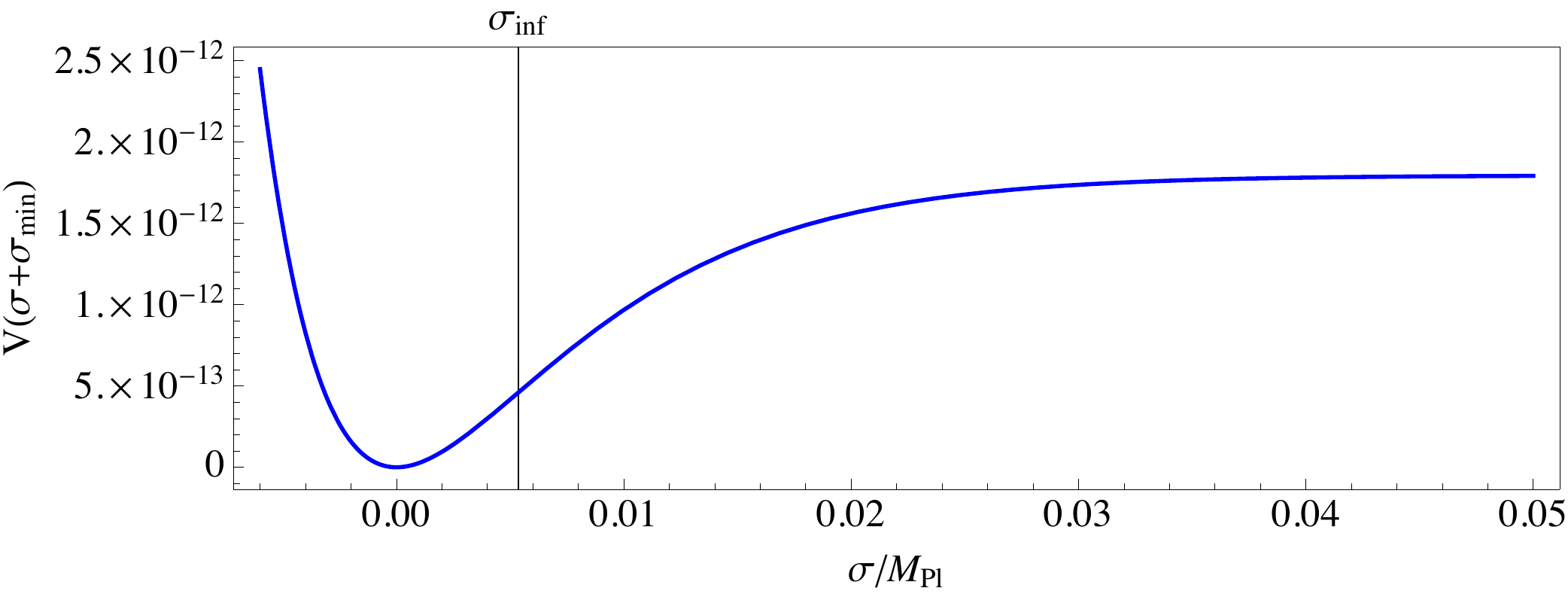}
\end{center}
\caption{The potential of the blow-up K\"ahler modulus in terms of the canonically normalised field, normalised to the height of the plateau $V_0 \simeq 1.794\times10^{-12}\,M^4_{\rm Pl}$ and with the minimum shifted to zero. The black line denotes the inflection point. The parameter choices are: $W_0=A_i=\xi=a_2=1$, $g_s=0.2$, $\gamma=2$ and $n=10.$}
\label{fig:potential_kmi}
\end{figure}

\begin{figure}
\begin{center}
\includegraphics[width=0.45\textwidth]{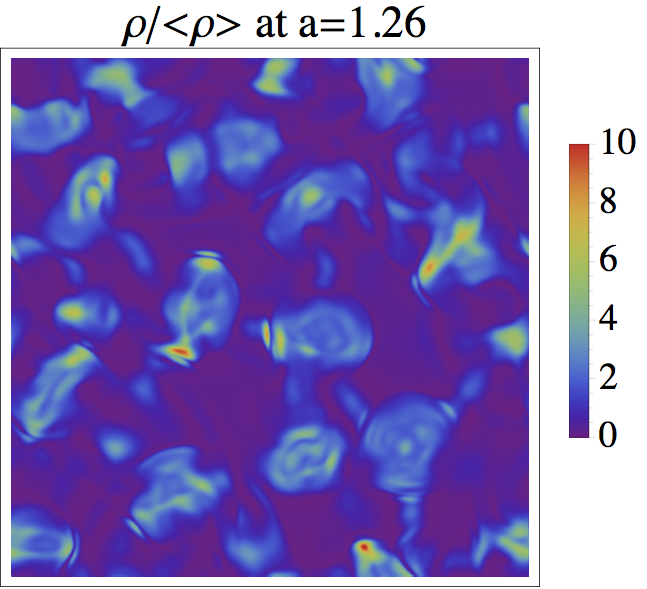}
\includegraphics[width=0.45\textwidth]{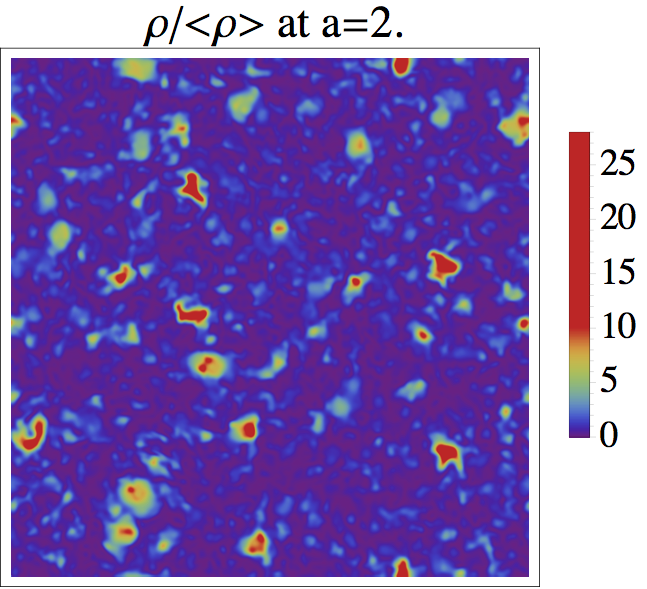}
\includegraphics[width=0.45\textwidth]{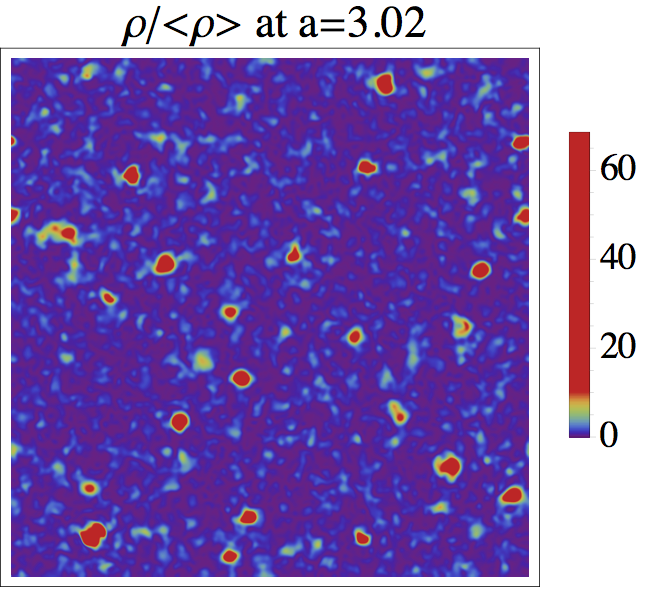}
\includegraphics[width=0.45\textwidth]{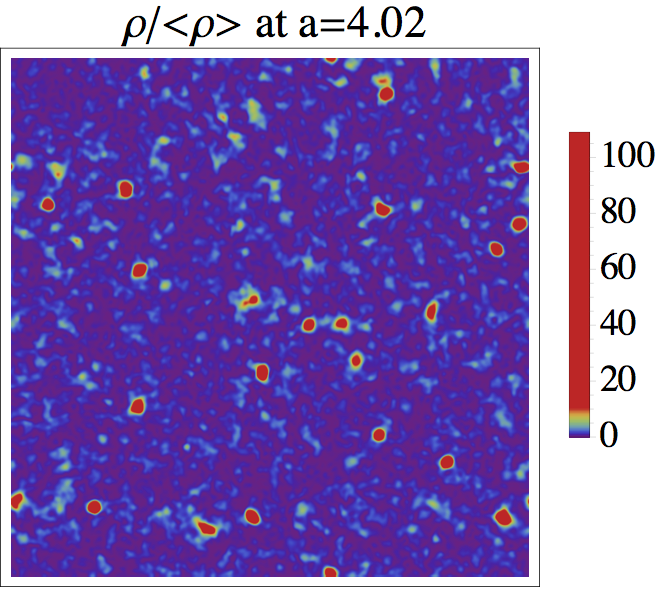}
\end{center}
\caption{Four snapshots of the energy density in a 2d simulation for our LVS blow-up modulus example at $a=1.26$, $a=2$, $a=3.02$ and $a=4.02$. Clearly, asymmetric oscillons are formed at $a\sim3$. Videos of the simulations can be found here~\cite{movieslink}.}
\label{fig:energydensity}
\end{figure}

\begin{figure}
\begin{center}
\includegraphics[width=0.45\textwidth]{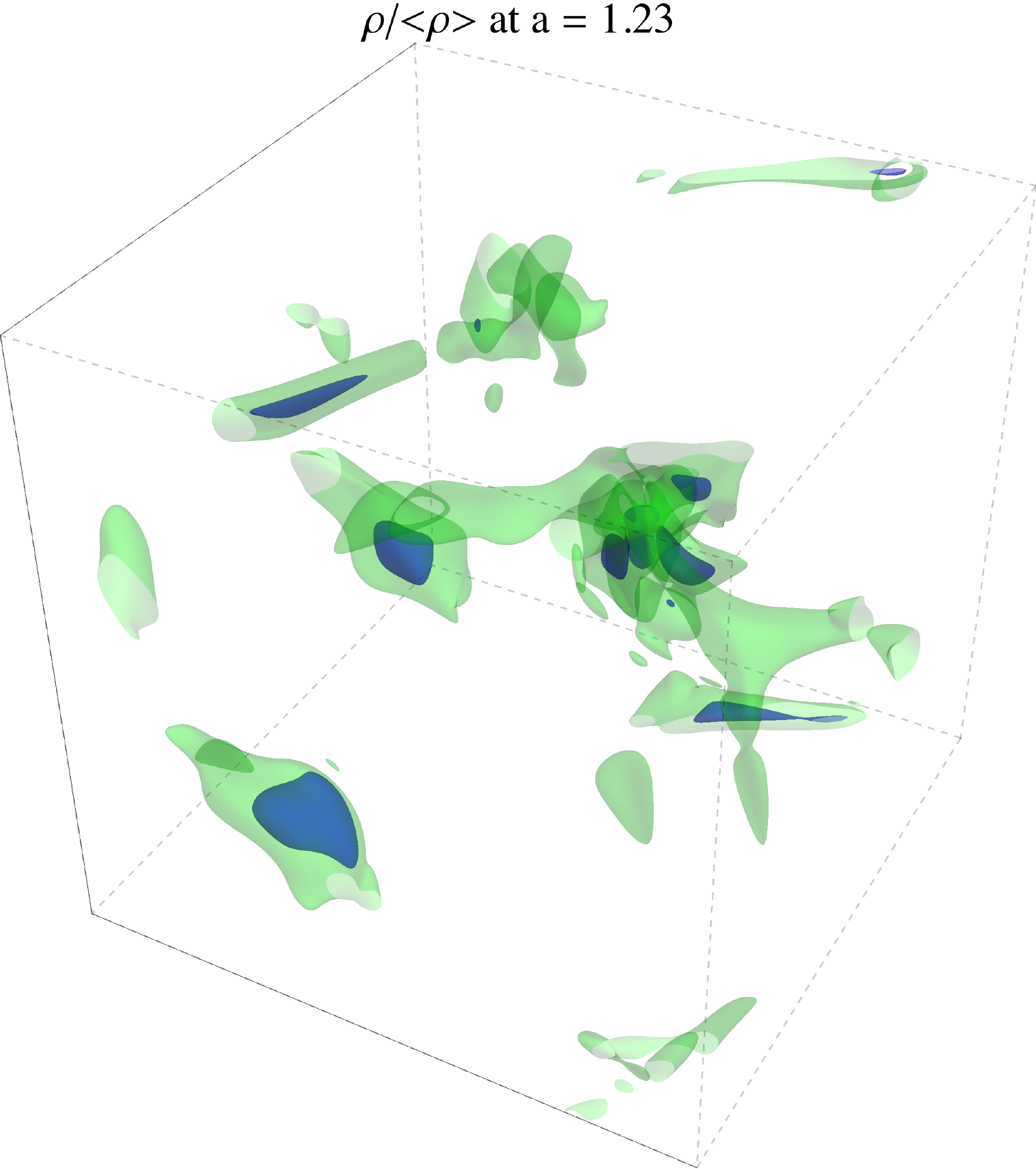}
\includegraphics[width=0.45\textwidth]{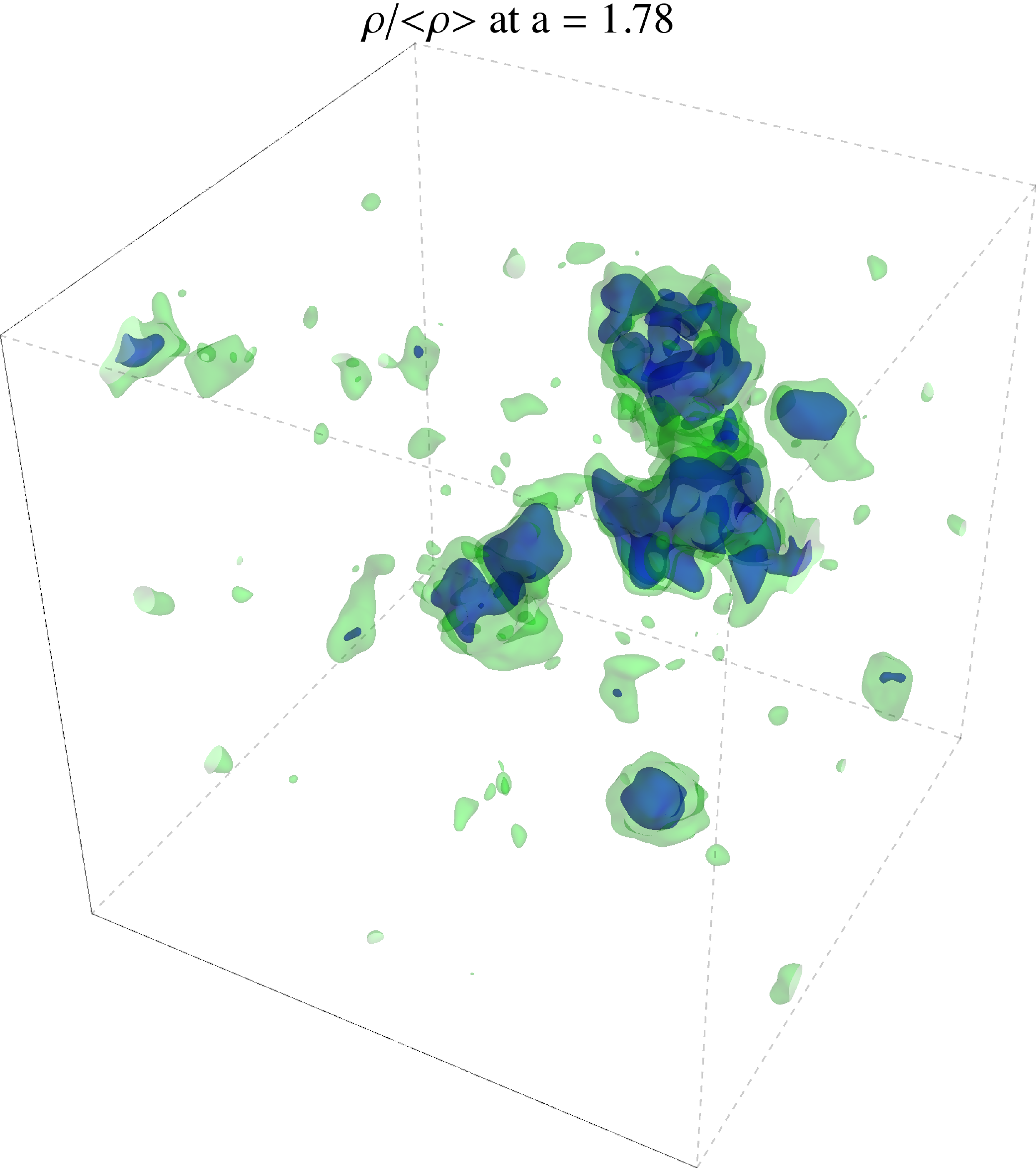}
\includegraphics[width=0.45\textwidth]{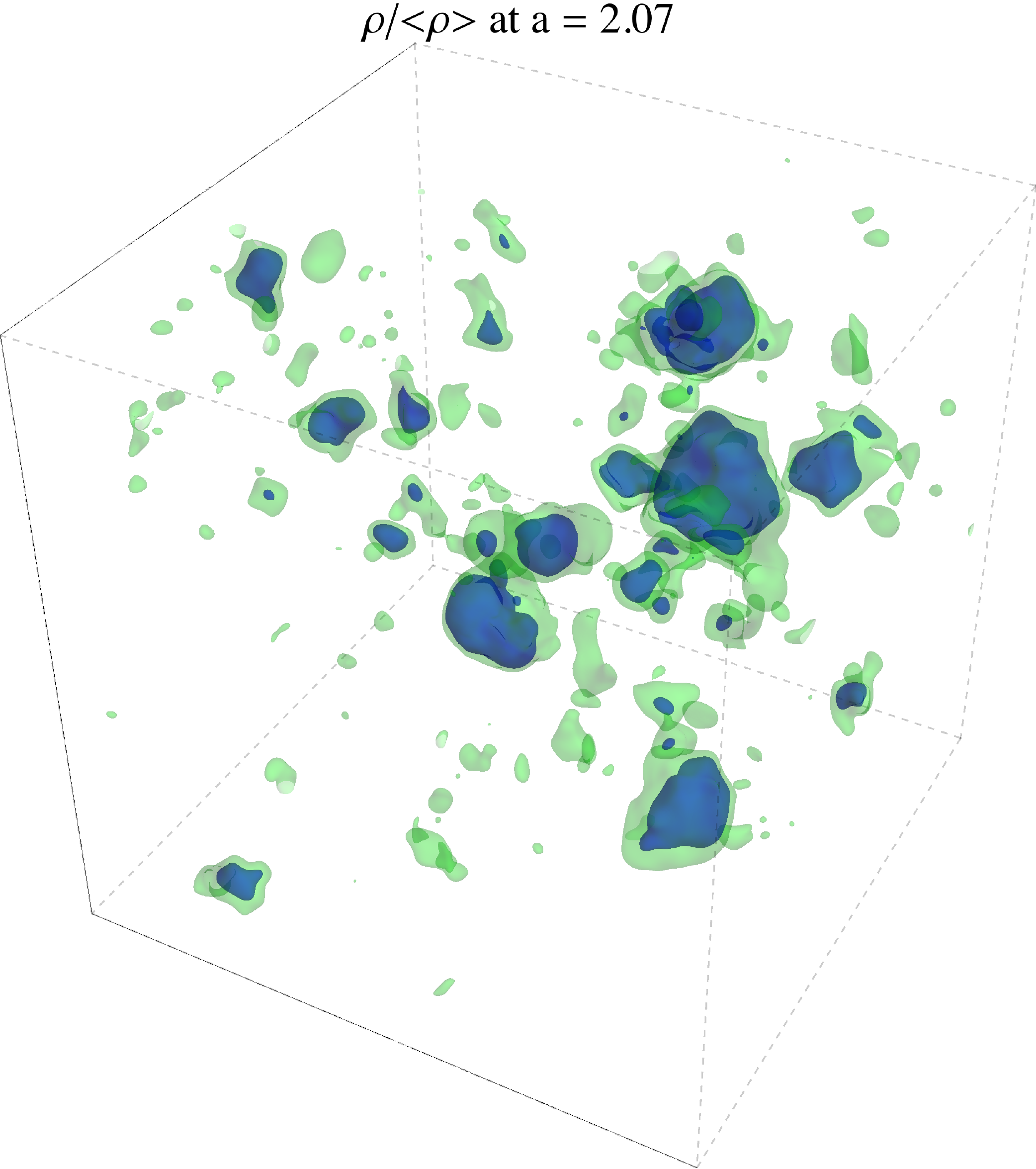}
\includegraphics[width=0.45\textwidth]{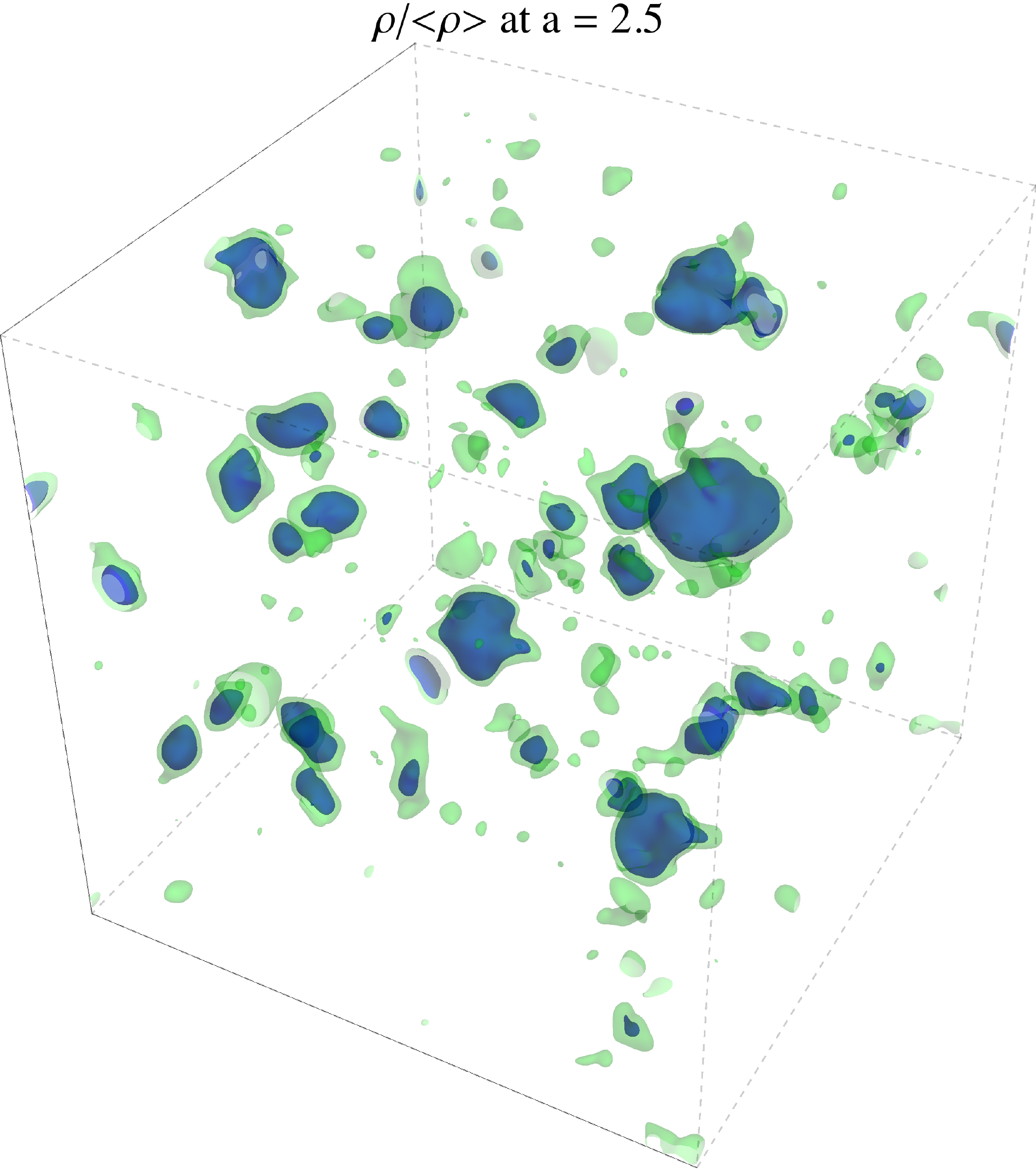}
\end{center}
\caption{Three dimensional energy density distribution in units of the average energy density. The green surfaces correspond to overdensities six times the average energy density, while the blue ones correspond to twelve times the average energy density.}
\label{fig:energydensity_3D}
\end{figure}

Another, probably the most interesting, result is the three dimensional distribution of the energy density. It is shown in Figure~\ref{fig:energydensity_3D} in units of the average energy density $\langle\rho\rangle$, for different moments in time. The green areas correspond to regions with $\rho = 6\,\langle\rho\rangle$ and the blue ones to regions with $\rho = 12\,\langle\rho\rangle$. One can see that the energy density starts fragmenting once non-linear interactions become important ($a\sim1.2$) leading to highly energetic regions. As the Universe expands the dynamics between field fluctuations eventually become less violent, leading to stable, highly energetic, bubbly regions. These bubbly regions represent localised, large amplitude oscillations in field space, i.e.\ oscillons. Although the snapshots shown in Figure~\ref{fig:energydensity_3D} are most likely still representing a phase of oscillon formation, one can see that for $a\ge1.78$ there are already bubbles which persist at least until the end of our simulation. 
Compared to the oscillons we found in KKLT (see Figure~\ref{fig:energydensity_KKLT}) the oscillons for blow-up moduli look more deformed and also many smaller inhomogeneities are present. Similar to the model studied in \cite{Antusch:2016con}, the tachyonic oscillations lead to violent field dynamics and the homogeneous mode decays quickly into inhomogeneous fluctuations within the first four oscillations of the condensate. 
The fluctuations in KKLT, however, are gradually amplified over many oscillations of the background and over a larger period of expansion. By looking at Figure~\ref{fig:energydensity_KKLT} and Figure~\ref{fig:energydensity_3D}, it seems as if inhomogeneities in KKLT are directly generated in the form of oscillons, while the formation of stable oscillons for blow-up moduli happens indirectly through the fragmentation of previously generated, unstable inhomogeneities.
Another impression of the formation and dynamics of oscillons can be seen in two-dimensional simulations where we show snapshots of the evolution of the energy density in Figure~\ref{fig:energydensity}.

\begin{figure}
\begin{center}\includegraphics[width=\textwidth]{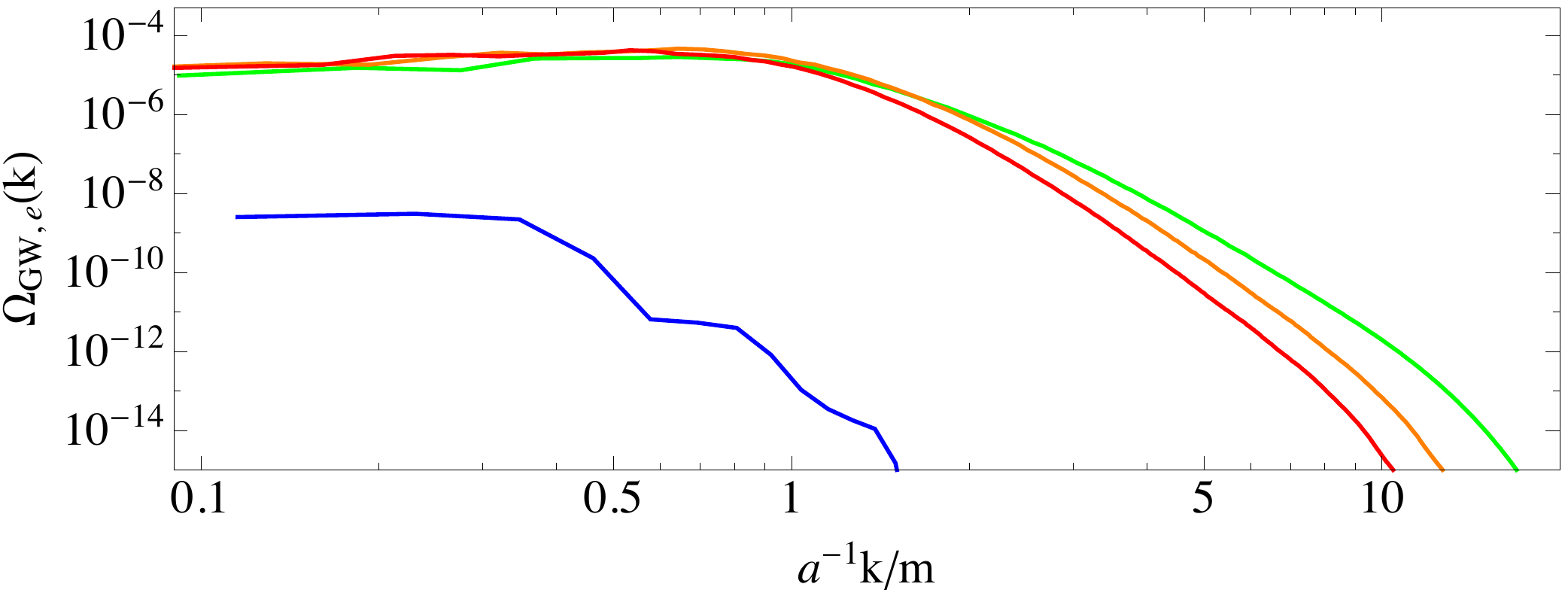}
\end{center}
\caption{Spectrum of Gravitational waves $\Omega_{\rm GW, \rm e}(k)$ as a function of the \textit{physical} momentum $a^{-1}\,k$. The spectrum is shown at different moments in time which correspond to: the end of linear preheating at $a\simeq1.16$ (blue), shortly after the beginning of the non-linear regime at $a\simeq1.45$ (green), at $a\simeq2.1$ (orange), and at the end of the simulation $a\simeq2.5$ (red).}
\label{fig:KMI_GW_spec}
\end{figure}

The rich dynamics within this blow-up moduli setup lead also to the production of GWs. The stochastic background of GWs produced during the early stage of preheating is shown in Figure~\ref{fig:KMI_GW_spec} at different moments in time. After the end of linear preheating $a\gtrsim1.16$ the spectrum forms a flat plateau for $k/a\lesssim m$ which falls off for larger values of $k/a$. The peaky structure in the GW spectrum, which is typically formed in the presence of oscillons, is not visible by the end of our simulation. This, however, does not mean that oscillons do not produce GW. One possible reason for the (yet) absent peaky structure could be that the latter is simply hidden by the stochastic background produced during and shortly after the tachyonic oscillations. This background is produced once during the early stage of preheating and is subsequently redshifted due to the expansion of the Universe. Oscillons, however, are an active source of GW production until they decay. If they live for a sufficiently long period and efficiently produce GWs, the peaky structure in the spectrum of GWs will eventually become visible at some later stage of the evolution. The final spectrum shown in Figure~\ref{fig:KMI_GW_spec} (red curve), is not expected to be the final result since oscillons continue to be produced. If the universe would instantly reheat at that time the frequencies of the plateau (corresponding to $a^{-1}k/m \sim 0.1-1$ in Figure~\ref{fig:KMI_GW_spec}) would lie today at
\be
f_0\sim10^8\,\rm{Hz} - 10^9\,\rm{Hz}\,,\qquad \textrm{with}\qquad\Omega_{\rm{GW},0} \sim 10^{-10} - 5\times10^{-10}\,.
\ee
Similar as in KKLT, an overall rescaling of the potential from complex structure moduli which is smaller than unity would also lead to lower frequencies. Altering, other model parameters could in principle also alter the frequencies of the stochastic GW background. Furthermore, the volume modulus being the lightest modulus in this scenario, will at some point start to dominate the energy density of the Universe. This, in turn, leads to an additional period of matter domination and thus pushing not only the frequencies but also $\Omega_{\rm{GW},0}$ to lower values.

\section{Conclusions and open questions}
\label{sec:conclusions}

Moduli fields may be the only stringy remnants that survive at low energies and particularly after a period of inflation. It is usually stated that the dilution effect of inflation makes it difficult to test any fundamental theory at scales larger than the inflation scale. Even though this is the case for string theory, moduli fields survive after inflation and can play a crucial role in the relatively late post-inflationary cosmology, improving the probability to allow us to identify cosmological signatures of string theory. 

In this article we studied the potential for moduli fields to produce oscillons during ``moduli preheating'', i.e.\ independent of the particular mechanism that drives inflation. We may summarise our results stating that:

\begin{itemize}
\item{}
Oscillons can be produced from moduli in string theory. For two very well motivated examples of moduli potentials, the overall volume in KKLT and the blow-up moduli in the LVS we find the production of oscillons. 

\item{} The mechanisms producing oscillons in our examples are distinct (parametric resonance and tachyonic oscillations) and lead to subsequently different oscillon dynamics. The shape of the gravitational wave spectra shown in Figure~\ref{fig:KKLT_GW_1e-5} and~\ref{fig:KMI_GW_spec} are different as well, which provides an -- in principle -- observable distinction of the moduli potentials. Compared to previous studies of blow-up moduli preheating, where oscillons were not discussed, we showed that oscillon production can take place via the tachyonic oscillation mechanism. On the contrary, oscillon production in KKLT happens via parametric resonance.  

\item{}
In both cases a distinct stochastic gravitational wave background is produced, which includes contributions from oscillon effects. Contrary to the case of GWs produced during preheating after inflation, the frequency of the GWs can in principle be lower due to the lower energies allowed in moduli potentials. For the examples studied in this paper the frequencies of the GWs are still high. It is an open question whether the relevant scales for other moduli potentials can be such that GWs are produced at frequencies within the reach of current and planned GW detectors such as LIGO \cite{TheLIGOScientific:2014jea} and the Einstein Telescope \cite{ET}.

\item{}
We also find that oscillon production is model dependent and it depends on the shape and the scale of the potential. In particular, for the considered parameter ranges, we find that no oscillons and sufficiently large inhomogeneities are produced for the volume modulus in the LVS and for fibration moduli in fibred Swiss cheese Calabi-Yau manifolds. Even though scalar  potentials for canonically normalised fibre moduli have the well studied exponential dependence with a flat plateau and an asymmetric structure around the minimum, due to the fact that the coefficient of the exponential is of order $\mathcal{O}(1)$, long-lived oscillons are not produced. Similarly, for the volume modulus for which the scalar potential is a sum of exponentials, with the coefficient of the exponential of order $\mathcal{O}(1)$, we do not have a production of long-lived oscillons. 

\item{}
The absence of large non-perturbative effects, in particular of oscillons, also implies the absence of ``overshooting effects'' of the modulus field into the decompactification region. This provides additional support for the stability of the metastable KKLT and LVS minima. It will be interesting to study whether overshooting happens in other moduli potentials and what the possible consequences are. 

\end{itemize}

These results are opening several future directions of research; let us mention a few:
\begin{itemize}

\item{} The general case of multi-field potentials needs further study. This will allow us to determine the phenomenological parameter capturing the phase of matter domination from the UV perspective. In particular the displacement of the volume modulus after the blow-up mode relaxes to its own minimum provides one example of such a dilution. Aiming further, one might ask what is the phenomenology of multi-field displacements. Which displacements are compatible with explicit models of string inflation. Concretely, is it possible to keep lighter moduli at their minimum?

\item{} Another aspect a study of multi-field potentials can answer is the coupling to other fields, i.e.~moduli and Standard Model degrees of freedom. This will allow to explicitly study the decay of oscillons. Global string models of chiral matter with inflation such as those recently studied in \cite{Cicoli:2017shd} could be a good laboratory for these investigations.

\item{} Note that in a situation where several moduli are displaced and undergo a significant growth of fluctuations, the possible gravitational radiation produced adds up. It might be worth investigating further on how generic it is that $\mathcal{O}(100)$ moduli are displaced from their minimum and what the generic size of such a gravitational wave signal would be.

\item{} In this article we have focused on potentials for K\"ahler moduli in type~IIB string theory. It will be very interesting to see whether other moduli potentials also support oscillons and which dynamics can be obtained. 
For example, they could lead to oscillons which might even trigger the formation of primordial black holes.

\end{itemize}
Clearly, further studies are needed to fully explore the  implications of oscillons in post-inflationary string cosmology. We hope to return to these exciting phenomenological and theoretical directions in the near future.

\section*{Acknowledgments}
We would like to thank Michele Cicoli, Joseph Conlon, Jin U Kang, David M.C. Marsh and Pramod Shukla for valuable discussions. SA and FQ thank the organisers of the III Saha Workshop on ``Aspects of Early Universe Cosmology''  for hospitality during the early stages of this project. SK and FM are funded in part by the European Research Council as Starting Grant 307605-SUSYBREAKING. This work has been supported by the Swiss National Science Foundation.

\appendix
\section{On inhomogeneities in other models}
\label{sec:discussion and other models}

Moduli displacement beyond the inflection point around a minimum seems to be fairly generic. Such a displacement makes a phase of tachyonic preheating unavoidable. As explained in Section~\ref{sec:oscillonproduction}, tachyonic preheating leads to the growth of perturbations. Independently of the formation of oscillons, the enhanced spectrum of perturbations leads to the production of a plateau-like gravitational wave spectrum. In this section we discuss this growth of perturbations for other moduli potentials.

\subsection{Volume modulus in LVS}
\label{sec:LVS}

Here we analyse the displacement of the volume modulus of the LVS beyond the inflection point in the scalar potential from Equation~\eqref{eq:LVSscalarpotential}. The structure of the potential for this field is different compared to the potential for the volume modulus in KKLT. In particular we observe that the height of the barrier and the mass of the field at the minimum are always roughly of the same order, with their ratio $\lesssim 20$. As a consequence, after the first tachyonic preheating phase, the field does not cross the inflection point anymore. We also find that parametric resonance is an inefficient mechanism for the growth of fluctuations in this scenario. 

The potential of the canonically normalised volume modulus $\phi$ in the LVS can be expressed as~\cite{Cicoli:2016olq}
\be
\frac{V(\phi)}{M^4_{\rm Pl}}=\frac{-3\,P\,W_0^2}{4}e^{-\frac{3\sqrt{3/2}\phi}{M_{\rm Pl}}}\left[2\,\left(\tfrac{3}{2}\right)^{3/4}\,\frac{\phi^{3/2}}{M^{3/2}_{\rm Pl}} -\frac{\hat{\xi}}{P} - 3\,\left(\tfrac{3}{2}\right)^{1/4}\,\frac{\phi_*^{1/2}}{M^{1/2}_{\rm Pl}}\,e^{\frac{\sqrt{3/2}(\phi -\phi_*)}{M_{\rm Pl}}}  \right]\,,
\label{eq:potential_LVS_1}
\ee
where the canonically normalised modulus $\phi$ is related to the volume of the CY manifold by
\be
\phi/M_{\rm Pl} = \sqrt{2/3}\log{\mathcal{V}}\,\quad\textrm{and}\quad P=\alpha\sum_{i}\alpha_i a_i^{-3/2}\,,
\ee
and $\phi_*$ is the minimum of the potential satisfying 
\be
\frac{\hat{\xi}}{P} = 2\,\left[\left(\frac{3}{2}\right)^{3/4}\frac{\phi_*^{3/2}}{M^{3/2}_{\rm Pl}} - \frac{3}{2}\left(\frac{3}{2}\right)^{1/4}\frac{\phi_*^{1/2}}{M^{1/2}_{\rm Pl}} \right]\,.
\label{eq:phi_star_xiP_relation}
\ee
By replacing $\hat{\xi}/P$ in Equation~\eqref{eq:potential_LVS_1} with the relation~\eqref{eq:phi_star_xiP_relation} we can express the volume modulus potential in a way that the only free parameters are the value of the minimum $\phi_*$ and the overall normalization of the potential $3\,P\,W_0^2/4$: 
\begin{equation}
\frac{V(\phi)}{M^4_{\rm Pl}}=3\left(\tfrac{3}{2}\right)^{1/4}V_0\,e^{-\frac{3\sqrt{3/2}\phi}{M_{\rm Pl}}}\,\left[\frac{\phi_*^{1/2}}{M^{1/2}_{\rm Pl}}\,e^{\frac{\sqrt{3/2}(\phi -\phi_*)}{M_{\rm Pl}}} - \,\left(\tfrac{2}{3}\right)^{1/2}\,\frac{(\phi^{3/2}-\phi_*^{3/2} )}{M^{3/2}_{\rm Pl}}- \frac{\phi_*^{1/2}}{M^{1/2}_{\rm Pl}}\right]\,,
\label{eq:potential_LVS_2}
\end{equation}
where we have defined
\be
V_0 \equiv M^4_{\rm Pl}\frac{3\,P\,W_0^2}{4}\,.
\ee
Since $V_0$ is simply an overall rescaling of the potential, it does not have an effect on the evolution of the homogeneous volume modulus. Generic values for the underlying UV parameters lead to a value for $V_0\lesssim  M^4_{\rm Pl}$. For the purpose of this paper we set $V_0=M^4_{\rm Pl}$. By fixing $V_0$, the only free parameter that remains is the value of the vacuum expectation value of the \Kahler modulus $\phi_*$.

\begin{figure}[tbp]
\centering
\includegraphics[width=\textwidth]{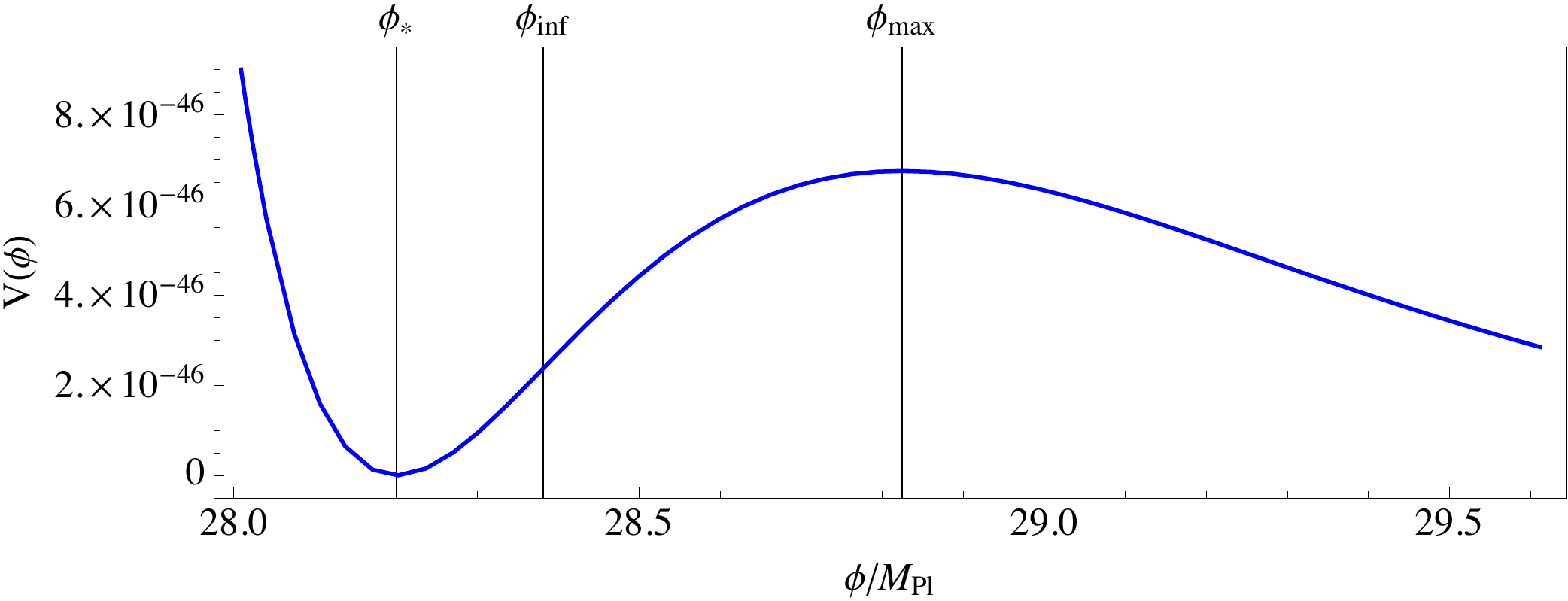}
\caption{The scalar potential in the LVS as a function of the canonically normalised volume modulus $\phi$ for $\phi_*\simeq28.2 M_{\rm Pl}$. In string units, the value of the volume in the minimum is $\mathcal{V}_{\rm min}=10^{15}$. The solid black lines denote the value of the canonically normalised volume modulus at the minimum of the potential $\phi_*$, at the inflection point of the potential $\phi_{\rm inf}$, and at the local maximum of the potential $\phi_{\rm max}$. }
\label{fig:LVS_potential}
\end{figure}

An example plot of the potential with $V_0=M_{\rm Pl}^4$ and $\phi_*\simeq28.2 M_{\rm Pl}$ is shown in Figure~\ref{fig:LVS_potential}. The solid black lines correspond to the field value at the local maximum of the potential $\phi_{\rm max}$, the field value at the inflection point of the potential $\phi_{\rm inf}$, and the field value at the minimum of the potential $\phi_*$.

\subsubsection{Floquet analysis}
\label{sec:floquet_LVS}
\begin{figure}
\centering
\subfigure{\includegraphics[width=7.5cm]{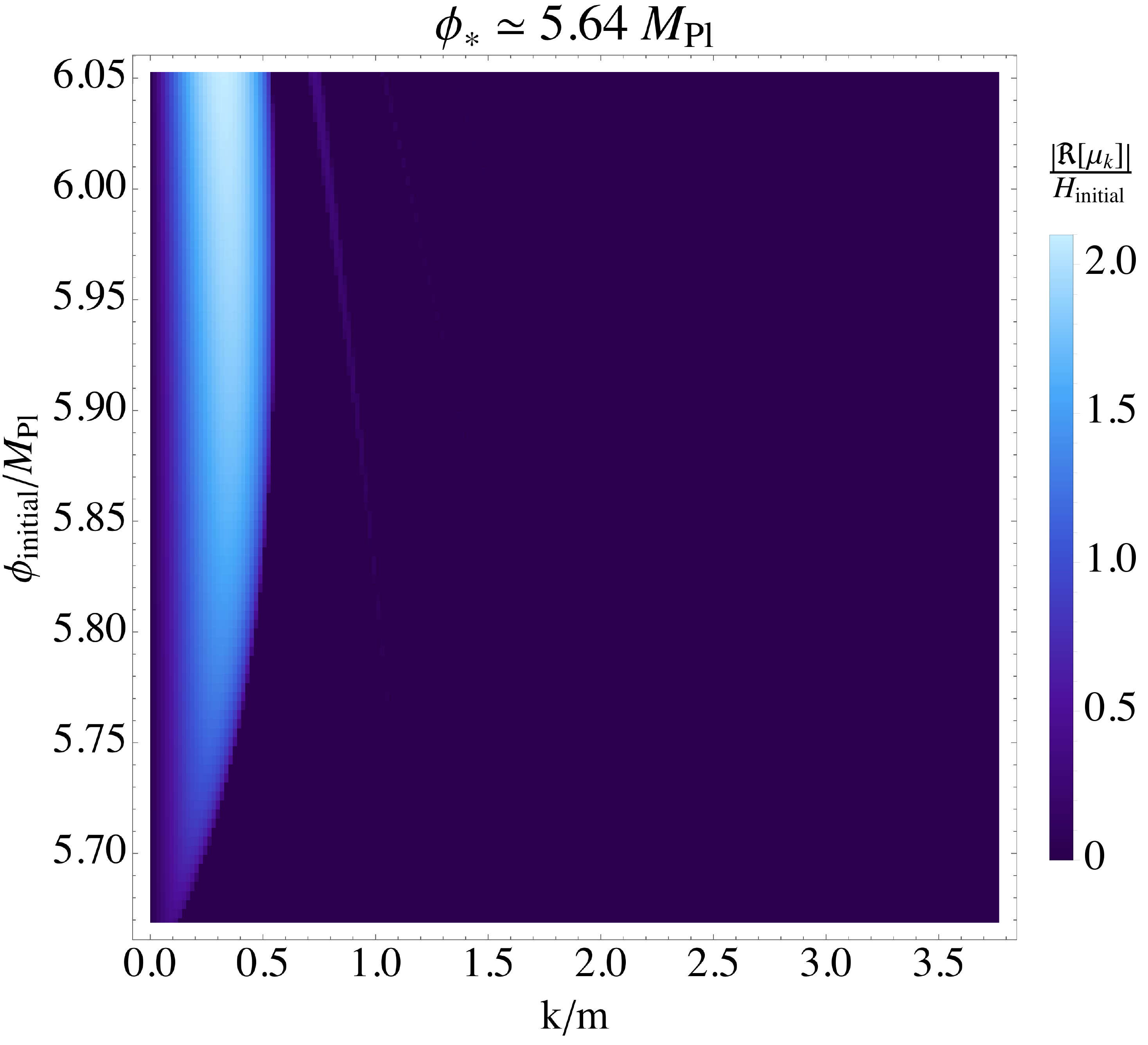}}
\hfill
\subfigure{\includegraphics[width=7.5cm]{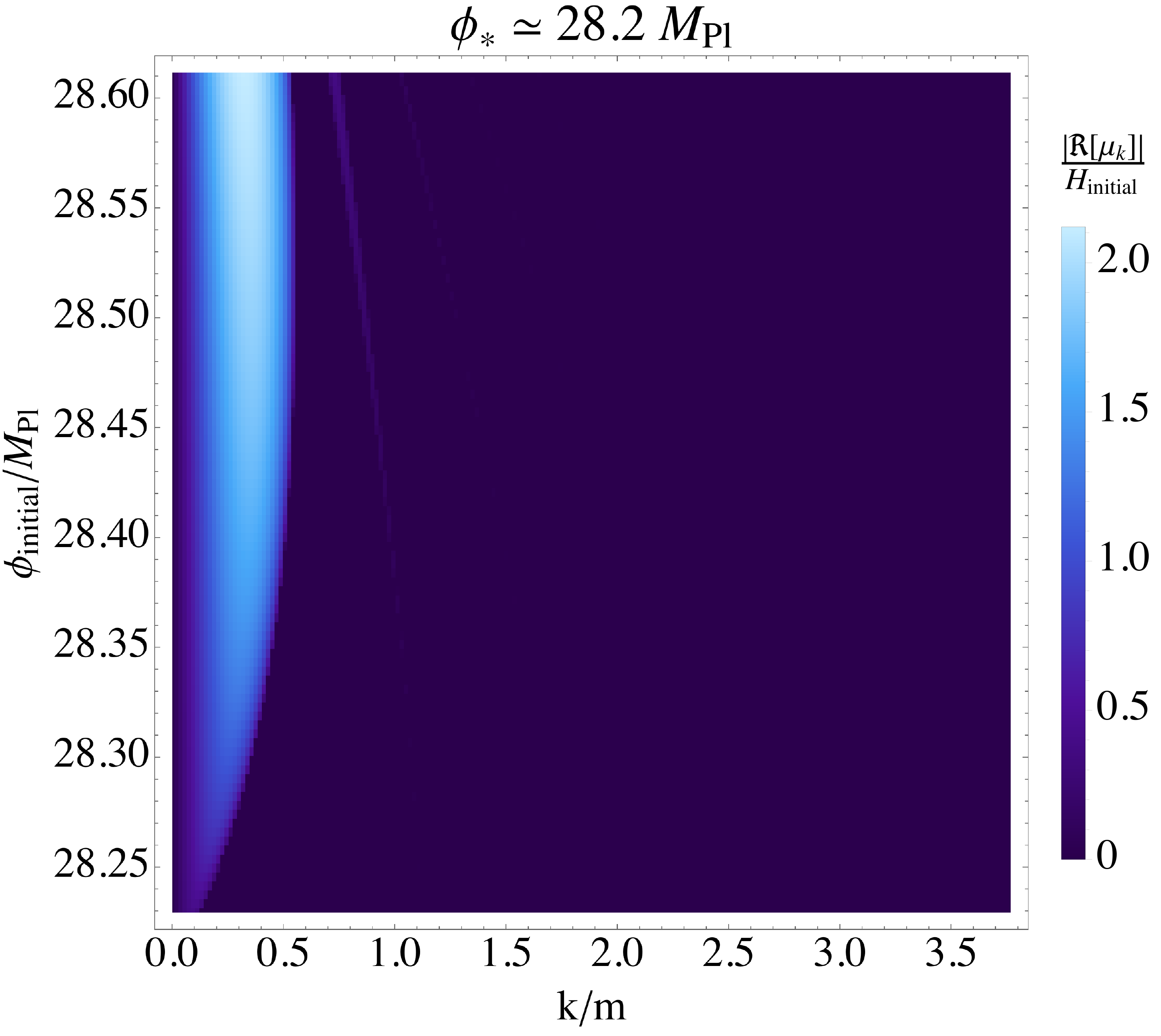}}
\caption{Instability bands for the volume modulus potential~\eqref{eq:potential_LVS_2} in the LVS. The instability bands are represented by the Floquet exponent $|\Re[\mu_k]|/H$ as a function of the initial field value $\phi_{\rm initial}$. The analysis was performed for two different values of $\phi_*$: $\phi_* \simeq 5.64 \,M_{\rm Pl}$ (corresponding to $\mathcal{V}_{\rm min}=10^{3}$) (left) and for $\phi_* \simeq 28.2 \,M_{\rm Pl}$ (corresponding to $\mathcal{V}_{\rm min}=10^{15}$) (right).}
\label{fig:floquet_LVS}
\end{figure}

For two different realizations of the potential~\eqref{eq:potential_LVS_2}, corresponding to two different choices for $\phi_*$, we performed a Floquet analysis in Minkowski space. We computed the real part of Floquet exponents $\Re[\mu_k]$ as a function of the initial field value of the homogeneous background field $\phi_{\rm initial}$. The results of our analyses are presented in Figure~\ref{fig:floquet_LVS}. The figure shows the real part of the Floquet exponent $|\Re[\mu_k]|$ in units of the Hubble parameter $H_{\rm initial}=1/M_{\rm Pl}\sqrt{V(\phi_{\rm initial})/3}$, for $\phi_* \simeq 5.64 \,M_{\rm Pl}$ (left) and for $\phi_* \simeq 28.2 \,M_{\rm Pl}$  (right).

The Floquet diagram exhibits three distinct instability bands, the strongest and broadest of which at values of $k/m\lesssim0.5$. On the basis of these results, we do not expect a strong growth of fluctuations, since the largest values of $|\Re[\mu_k]|/H_{\rm initial}$ are of order unity, which in turn means that the growth rate of the fluctuations is comparable to the expansion rate of the Universe. To confirm our expectations, we solved the linearised equations for the fluctuations in an expanding Universe.

\subsubsection{Homogeneous field evolution and linear perturbations}
\label{sec:linear_LVS}
We computed the evolution of the fluctuations for the potential~\ref{fig:LVS_potential} with $\phi_* \simeq 28.2\,M_{\rm Pl}$ by numerically solving Equation~\eqref{eq:perturbationEOM} with initial conditions for the fluctuations given by Equation~\eqref{eq:IC_fluctuations}. At the same time we solved Equation~\eqref{eq:homogeneousEOM} for the homogeneous component $\phi(t)$. We used the following initial conditions for the homogenous background field $\phi$, its velocity $\dot{\phi}$ and the Hubble parameter $H$:
\be
\phi_{\rm initial} \simeq 28.61\,M_{\rm Pl}\,, \qquad\dot{\phi}_{\rm initial} = 0\,\,,\qquad H_{\rm initial}=\sqrt{\frac{V(\phi_{\rm initial})}{3\,M^2_{\rm Pl}}} \simeq 1.39\times10^{-23}\,M_{\rm Pl}\,.
\ee

\begin{figure}
\centering
\includegraphics[width=\textwidth]{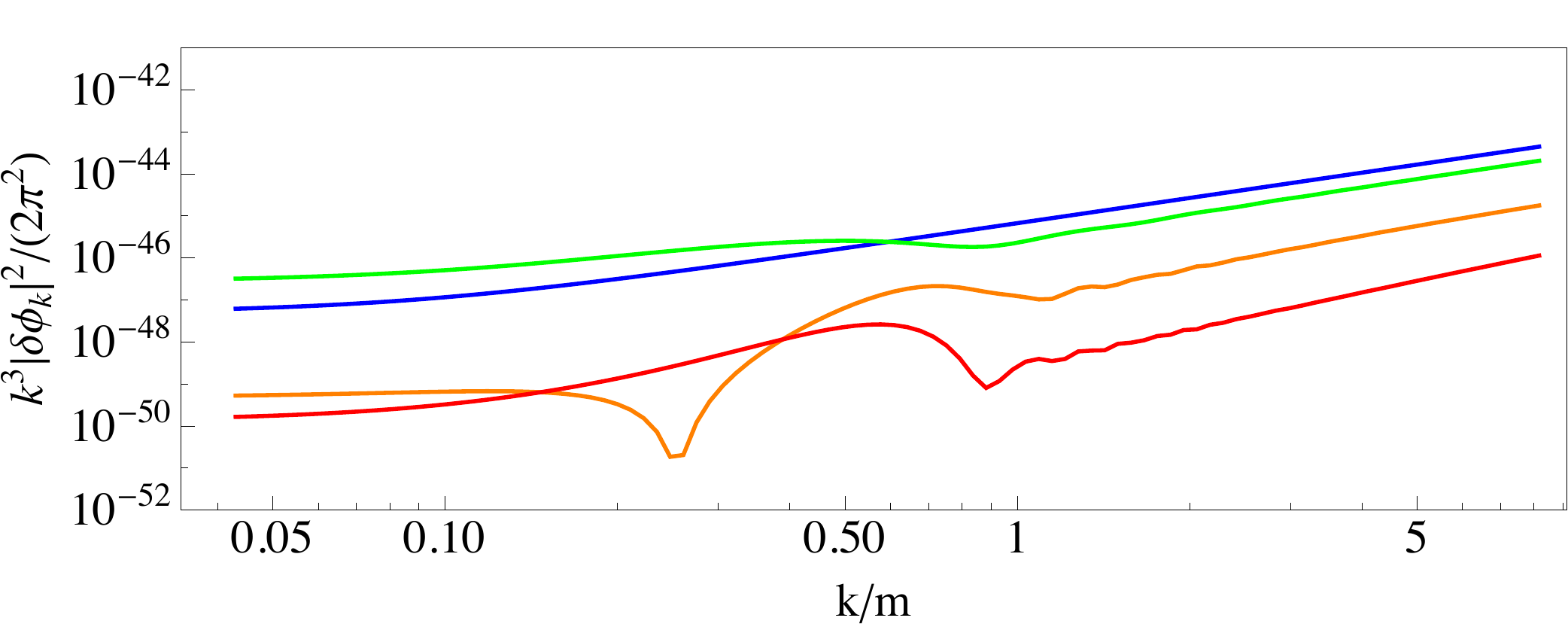}
\caption{The spectrum of the fluctuations $\delta\phi_k$ of the canonically normalised volume modulus in the LVS. The spectrum is shown as a function of the \textit{comoving} momentum $k$ at different moments in time: the initial spectrum (blue), before the first oscillation of $\phi$ at the end of tachyonic preheating (green), after eight oscillations (orange), and after 47 oscillations (red). }
\label{fig:LVS_spec}
\end{figure}

\begin{figure}
\centering
\subfigure{\includegraphics[width=7.4cm]{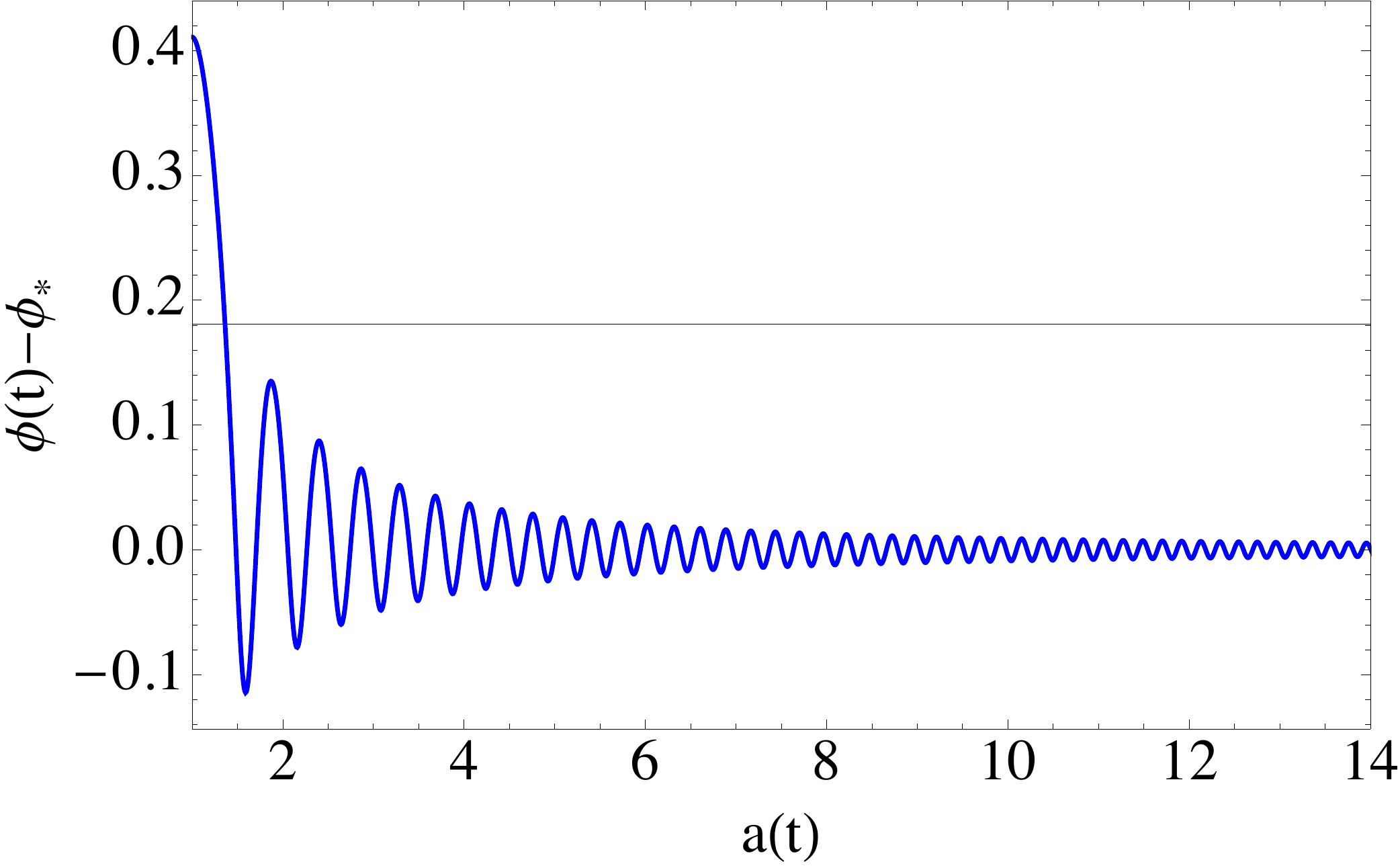}}
\hfill
\subfigure{\includegraphics[width=7.5cm]{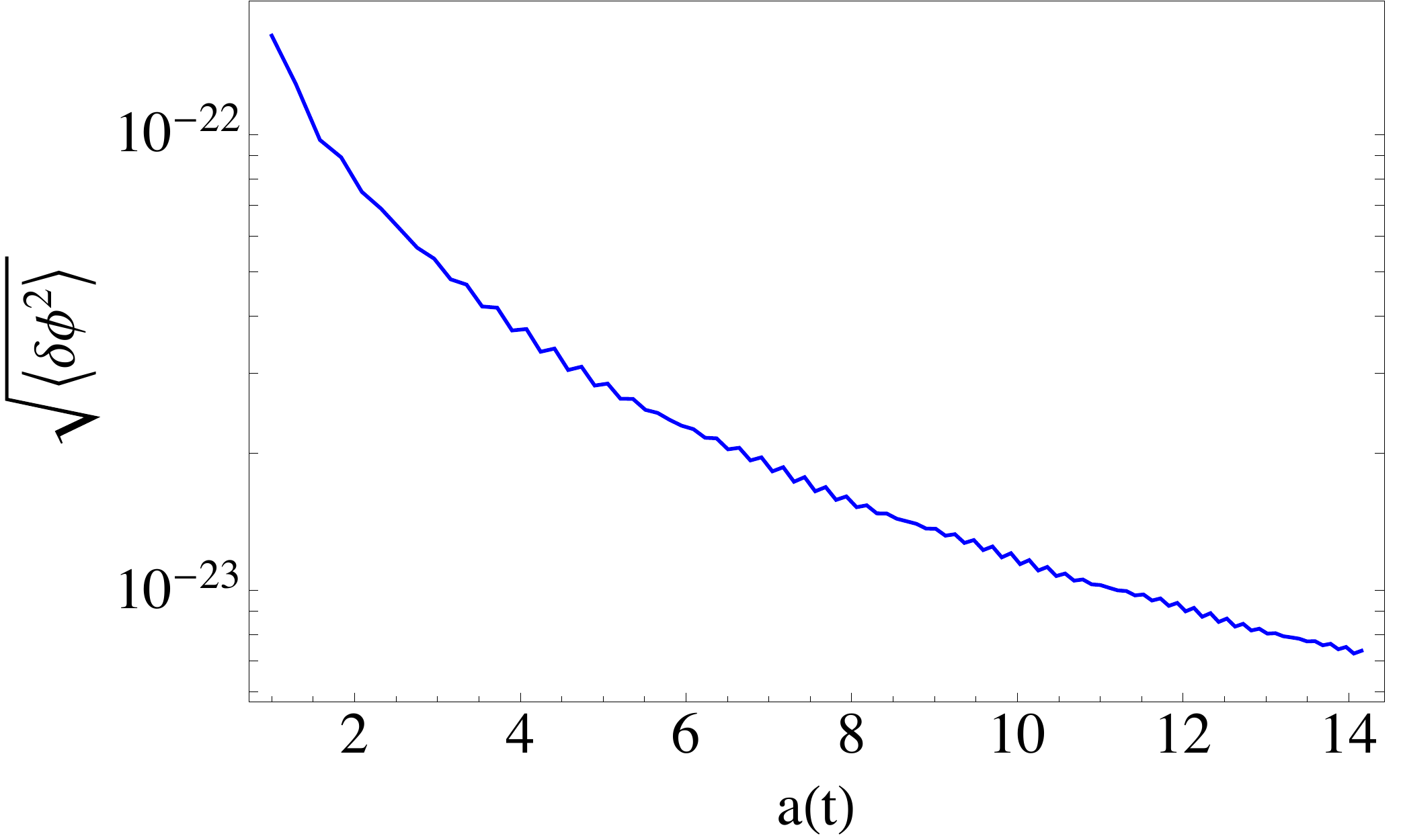}}
\caption{Numerical results from the evolution of the volume modulus in the LVS. \textit{Left:} The evolution of $\phi(t)$ as a function of the scale factor $a(t)$. The solid black line denotes the field value at the inflection point of the potential. One can see that no tachyonic oscillations occur. \text{Right}: Evolution of the variance $\sqrt{\langle\delta\phi^2\rangle}$ as a function of $a(t)$.}
\label{fig:LVS_hom_var}
\end{figure}

The numerical results are shown in Figure~\ref{fig:LVS_spec} and~\ref{fig:LVS_hom_var}. Figure~\ref{fig:LVS_spec} shows the spectrum of the fluctuations of the canonically normalised volume modulus as a function of the \textit{comoving} momentum $k$ at different moments in time: the initial spectrum (blue), before the first oscillation of $\phi$ at the end of tachyonic preheating (green), after eight oscillations (orange), and after 47 oscillations (red). 
Initially we have a phase of tachyonic preheating during which the infrared tail is amplified by roughly one order of magnitude. The phase of tachyonic preheating is subsequently followed by a weak self-resonance during which modes with $k\sim 0.5 m$ are amplified. The amplification of these modes leads to a tiny peak in the final spectrum shown in Figure~\ref{fig:LVS_spec}.

Figure~\ref{fig:LVS_hom_var} shows the evolution homogeneous field $\phi(t)$ (right) and the evolution of the variance $\langle\delta\phi^2\rangle$ (left) as a function of the scale factor $a(t)$. The variance was calculated from the spectrum of the linear perturbations according to
\be
\langle \delta\phi^2\rangle = \int d\log k\,\frac{k^3}{2\pi^2}|\delta\phi_k|^2 = \int dk\,\frac{k^2}{2\pi^2}|\delta\phi_k|^2\,.
\ee
One can see that the variance decreases continuously since no significant amplification of the perturbations takes place. However, since the homogeneous field $\phi(t)$ is damped and its amplitude of oscillation decreases as well, a net growth of fluctuations with respect to the amplitude of $\phi(t)$ is in principle possible. In order to estimate the relative growth of fluctuations with respect to the amplitude of the field we can consider the following quantity
\be
\gamma(t) \equiv \frac{\sqrt{\langle\delta\phi^2\rangle(t)}}{\Phi(t)}\frac{\Phi(0)}{\sqrt{\langle\delta\phi^2\rangle_{\rm vac}}}\,,
\label{eq:def_gamma}
\ee
where $\Phi(t)$ is the envelope of $\sqrt{[\phi(t)-\phi_*]^2}$ and $\langle\delta\phi^2\rangle_{\rm vac}$ is the variance obtained from the initial vacuum spectrum. As long as the evolution of the fluctuations is linear, $\gamma(t)$ is a measure for the growth of the fluctuations relative to the amplitude of the field which is independent of the absolute value of the vacuum fluctuations (i.e.\ independent of $V_0$).

\begin{figure}
\centering
\subfigure{\includegraphics[width=7.5cm]{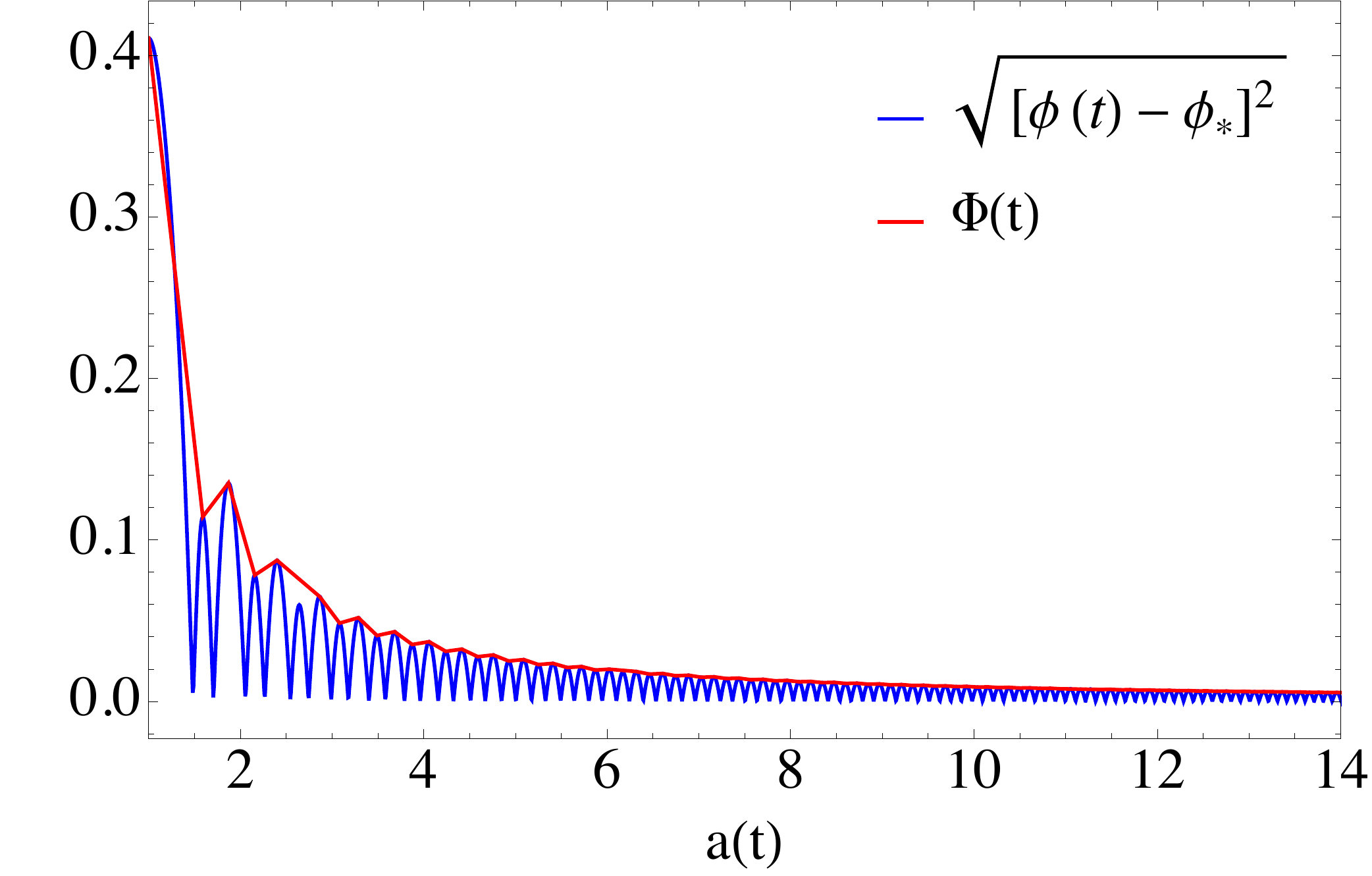}}
\hfill
\subfigure{\includegraphics[width=7.5cm]{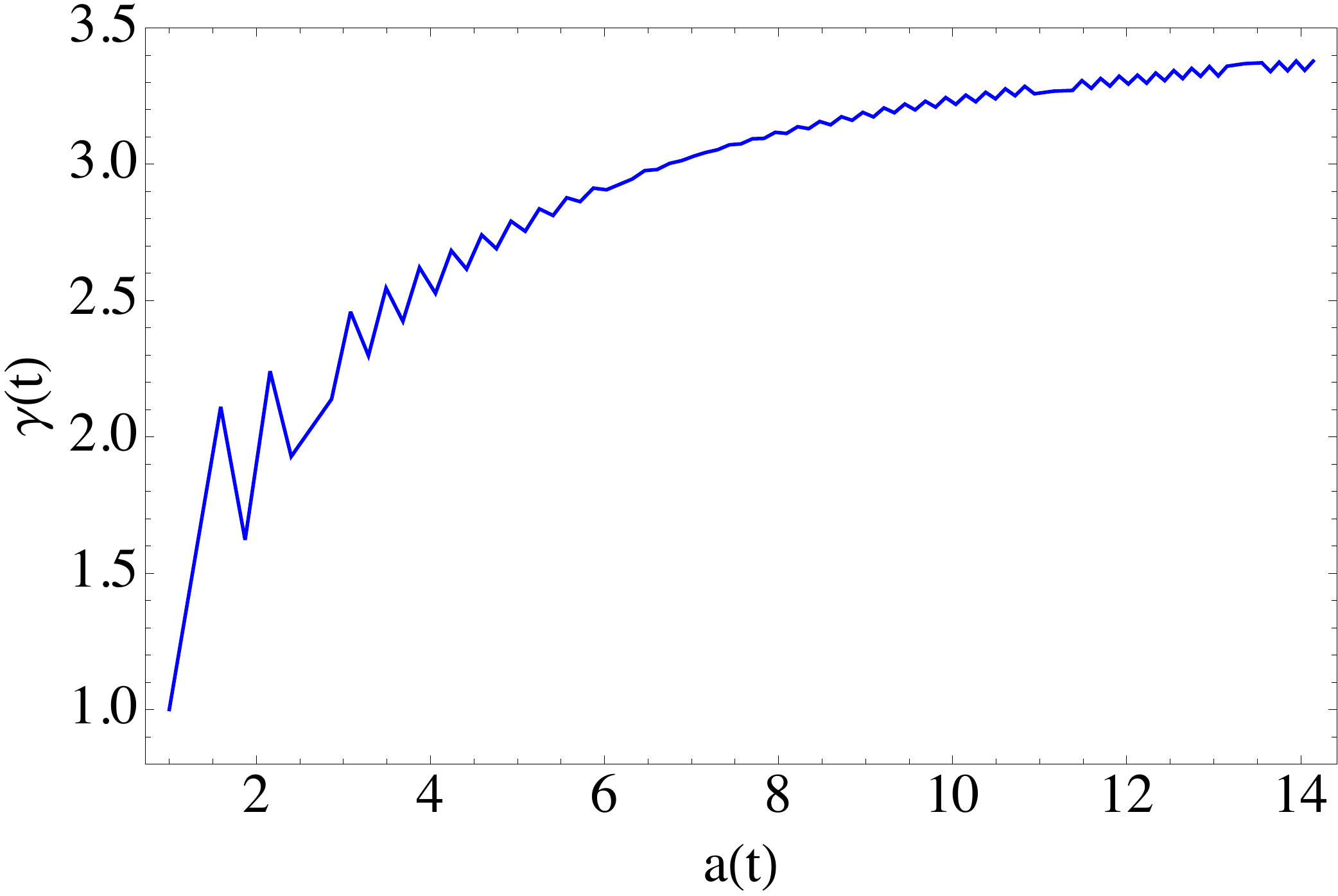}}
\caption{\textit{Left:} $\sqrt{[\phi(t)-\phi_*]^2}$ vs. $a(t)$ (blue) and its envelope $\Phi(t)$ computed at times at which $\dot{\phi}=0$. \text{Right}: Evolution $\gamma(t)$ (c.f.\ Equation~\eqref{eq:def_gamma}). $\gamma(t)$ reflects the growth of the fluctuations relative to the amplitude of the homogeneous background.}
\label{fig:LVS_gamma}
\end{figure}

The left part of Figure~\ref{fig:LVS_gamma} shows $\sqrt{[\phi(t)-\phi_*]^2}$ (blue) and $\Phi(t)$ (red) both as a function of $a(t)$, where $\Phi(t)$ was calculated at times at which $\dot{\phi}=0$. The right part of the Figure shows the evolution of $\gamma(t)$. One can see that in the case of the volume modulus the fluctuations grow only by a factor of $\sim3.5$ relative to the field amplitude and that the relative growth rate decreases with time. Provided that initially the fluctuations are small compared to the amplitude of homogeneous field $\phi(t)$, this means that always $\langle\delta\phi^2\rangle \ll\Phi(t)$. For the considered parameter ranges, we do not find significant inhomogeneities or non-perturbative effects. In turn, the absence of significant inhomogeneities implies that GWs are expected to be only marginally produced.

The mild and perturbative dynamics prevent the field from overshooting over the potential barrier into the decompactification minimum, thus supporting the stability of the LVS minimum.

\subsection{Fibre inflation}
\label{sec:fibre inflation}

The last example we want to consider is \textit{fibre inflation}~\cite{Cicoli:2008gp, Burgess:2016owb}. This is a model of large field inflation which gives rise to a tensor-to-scalar ratio in the experimentally accessible window $r \in (10^{-3}, 10^{-2})$~\cite{Cicoli:2016chb}.\footnote{A consistent string embedding of fibre inflation with an explicit orientifolded Calabi-Yau manifold and brane setup can be found in~\cite{Cicoli:2016xae}.} The inflaton corresponds to a K\"ahler modulus describing the `fibre' volume in a fibred Calabi-Yau manifold of Swiss-cheese type. In its simplest version the scalar potential of fibre inflation in terms of the canonically normalised field $\phi$ takes the form
\begin{equation}
V(\phi) = V_0 \left(c_0 + c_1 e^{-\frac{k \phi}{2}} + c_2 e^{- 2 k \phi} + \epsilon e^{k \phi}\right)\,,
\label{eq:fibre_potential}
\end{equation}
where $V_0$ and $\epsilon$ are coefficients that depend on the underlying parameters of the compactification, and can be written as
\begin{equation}
\label{eq:fibrecoefficients}
c_0 = 3 - \epsilon \,,\quad c_1 = -4 \left(1+ \frac{\epsilon}{6}\right)\,, \quad c_2 = \left(1 + \frac{2 \epsilon}{3}\right) \,.
\end{equation}
$\epsilon$ is naturally small (it is proportional to $g_s^4$), preserving the flatness of the inflationary plateau.
Similarly, $V_0$ is set by the overall stabilisation of the Calabi-Yau volume and hence hierarchically smaller than unity. The coefficients are adjusted such that $V(0) = 0$. $k = 2/\sqrt{3}$ is a coefficient that arises from the canonical normalization of the field $\phi$. 

\begin{figure}
\centering
\subfigure{\includegraphics[width=7.0cm]{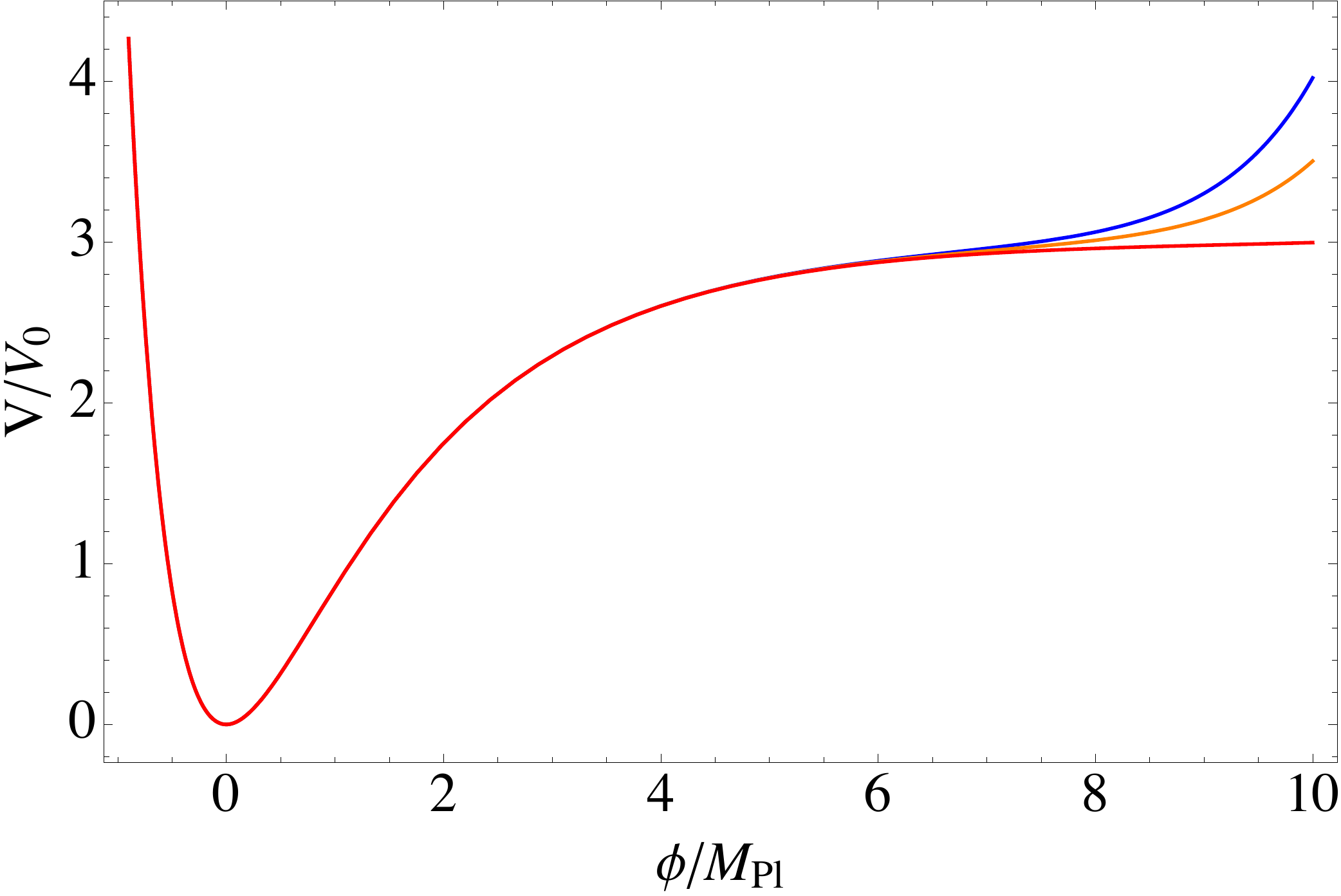}}
\hfill
\subfigure{\includegraphics[width=7.5cm]{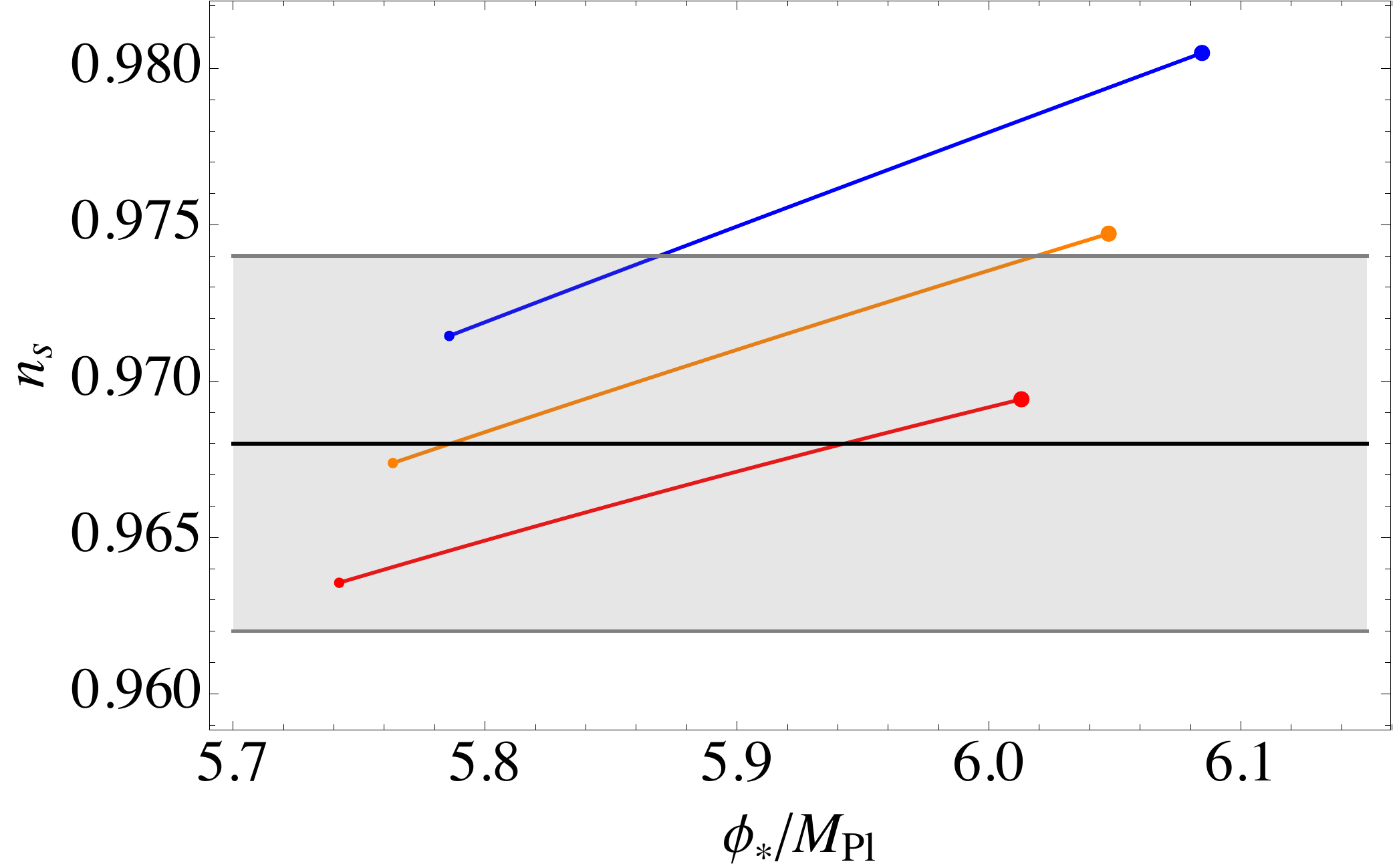}}
\caption{\textit{Left:} The fibre inflation potential~\eqref{eq:fibre_potential} in units of $V_0$ for $\epsilon = 10^{-7}$ (red), $\epsilon = 5\times10^{-6}$ (orange), and  $\epsilon = 10^{-5}$ (blue). \textit{Right:} Predictions for the spectral index $\ns$ for the three different values of $\epsilon$ (same colour coding as for the potential) as a function of the field value  at horizon exit $\phi_*$. Different values of $\phi_*$ correspond to different numbers of $e$-folds of slow-roll inflation $N_*$. The value of $N_*$ ranges from 50 (tiny dot) to 60 (large dot).}
\label{fig:potential}
\end{figure}

\subsubsection{Floquet analysis}
\label{sec:floquet_fibre}

For the fibre potential~\eqref{eq:fibre_potential} we performed two different Floquet analyses in Minkowski space. In the first analysis we kept the initial field value (i.e.\ the amplitude of oscillation) $\phi_{\rm initial}$ constant and computed the Floquet exponents  for different values of $\epsilon$ ranging from $10^{-7}$ to $10^{-1}$. A second analysis was  performed with a fixed value of $\epsilon$ where the Floquet exponents were computed as a function of $\phi_{\rm initial}$. 

\begin{figure}
\centering
\subfigure{\includegraphics[width=7.5cm]{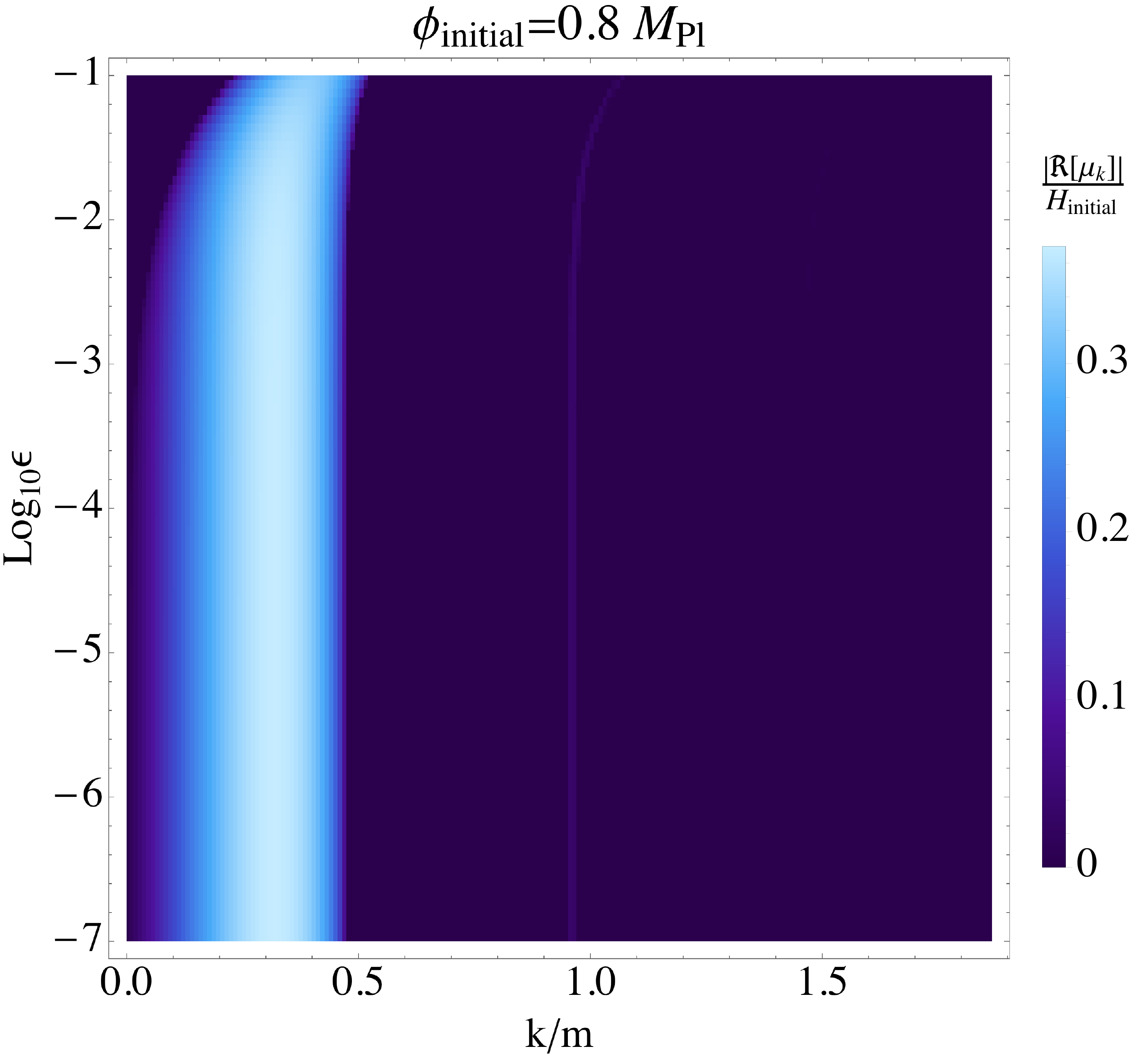}}
\hfill
\subfigure{\includegraphics[width=7.5cm]{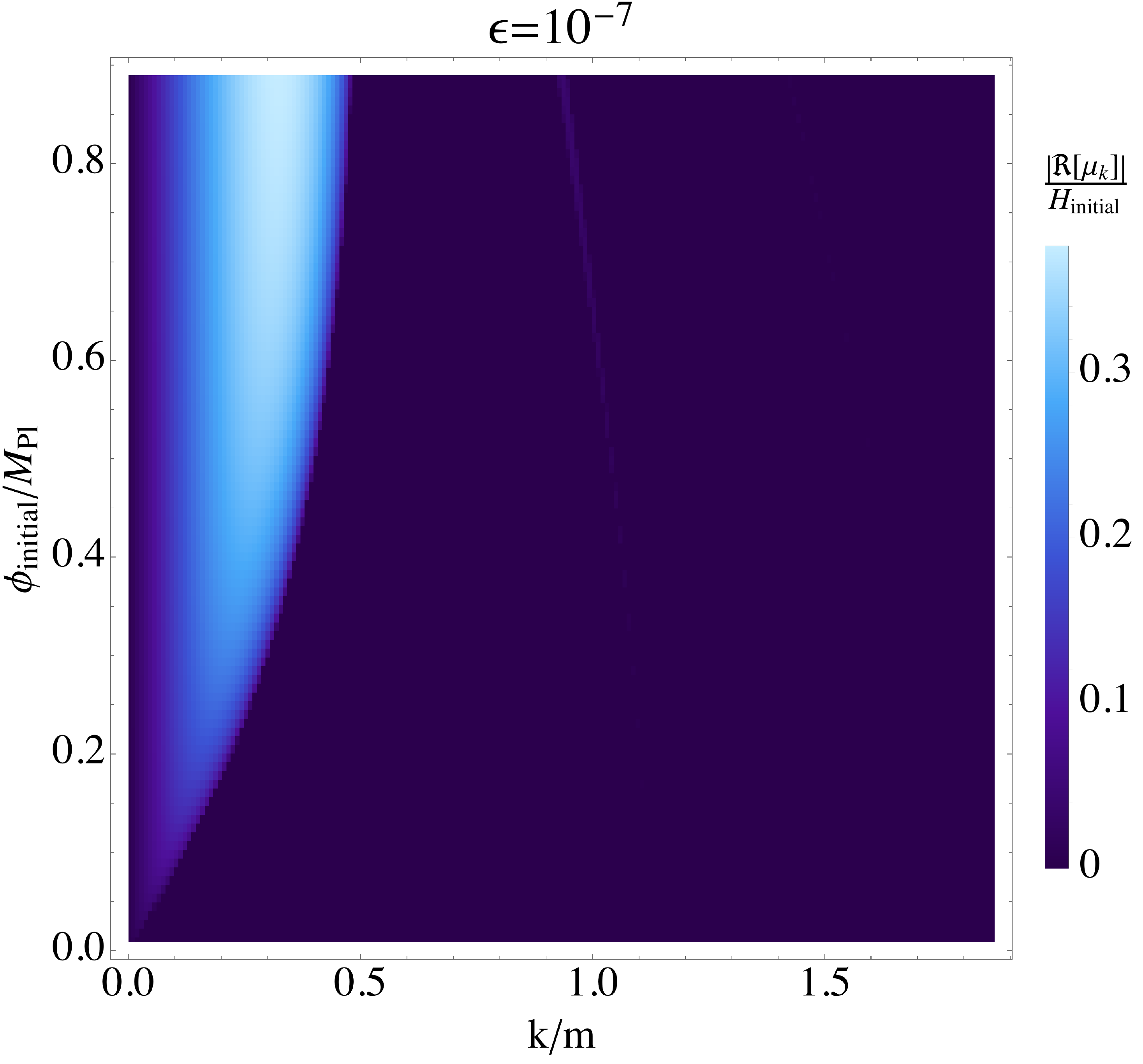}}
\caption{Results of a Floquet analysis of the Fibre inflation potential. \textit{Left:} The real part of the Floquet exponent $|\Re[\mu_k]|/H_{\rm initial}$ as a function of $k/m$ and $\epsilon$ \textit{Right:} $|\Re[\mu_k]|/H_{\rm initial}$ as a function of $k/m$ and the amplitude of oscillation of the field $\phi_{\rm initial}$. One can see that in none of the cases $|\Re[\mu_k]|/H_{\rm initial} > 0.4$ which is a strong indication that no significant growth of fluctuations occurs in an expanding Universe.}
\label{fig:floquet}
\end{figure}

The results of the analyses are shown in Figure~\ref{fig:floquet}. The left part of the figure shows the Floquet exponents $|\Re[\mu_k]|/H_{\rm initial}$ as a function of $k/m$ and $\epsilon$. On the right we show $|\Re[\mu_k]|/H_{\rm initial}$ as a function of the amplitude of oscillation $\phi_{\rm initial}$ for a fixed $\epsilon = 10^{-7}$. The Hubble parameter is calculated according to
\be
H_{\rm initial}=\sqrt{\frac{V(\phi_{\rm initial})}{3\,M^2_{\rm Pl}}}\,.
\ee
For the maximum $\phi_{\rm initial}$ we chose the value at the end of slow-roll inflation, when\footnote{$\varepsilon_\phi \equiv M^2_{\rm Pl}/2(V'(\phi)/V)^2\simeq1$ for $\phi \simeq 0.91$ holds for all values of $\epsilon$ between $10^{-7}$ and $10^{-5}$. For larger values of $\epsilon$ the potential becomes steeper. Hence, within our Floquet anlyses the field lies always outside the slow-roll regime.} $\varepsilon_\phi \equiv M^2_{\rm Pl}/2(V'(\phi)/V)^2\simeq1$, $\phi \simeq 0.91\ge\phi_{\rm initial}$.

One can see that the Floquet exponents are always small compared to the expansion rate of the Universe, since $|\Re[\mu_k]|/H_{\rm initial}\lesssim0.4$ for all values of $k$, $\phi_{\rm initial}$ and $\epsilon$. Therefore the fluctuations are not expected to grow significantly throughout the evolution of the field in an expanding Universe. In order to confirm this we compute the evolution of the perturbations $\delta\phi_k$ in an expanding Universe by solving the linearised equations of motion~\eqref{eq:perturbationEOM}.

\subsubsection{Homogeneous field evolution and linear perturbations}
\label{sec:linear_fibre}
We solved the equation of motion for the homogeneous field $\phi(t)$ Equation~\eqref{eq:homogeneousEOM} and at the same time the equations for the perturbations of the field Equation~\eqref{eq:perturbationEOM}. The initial conditions for the homogeneous field and the Hubble parameter are set to\footnote{We also checked explicitly that the results do not change significantly when replacing the initial field velocity with the slow-roll velocity $\dot{\phi}_{\rm initial} = -V'(\phi_{\rm initial})/(3\,H_{\rm initial})$.}
\be
\phi_{\rm initial} = 0.8\,M_{\rm Pl}\,,\qquad\dot{\phi}_{\rm initial} = 0\,\,,\qquad H_{\rm initial}=\sqrt{\frac{V(\phi_{\rm initial})}{3\,M^2_{\rm Pl}}} \simeq 0.46\,M_{\rm Pl}\,.
\ee
The fluctuations $\delta\phi_k$ are initialised as vacuum fluctuations according to Equation~\eqref{eq:IC_fluctuations}.

\begin{figure}[tbp]
\centering
\subfigure{\includegraphics[width=0.48\textwidth]{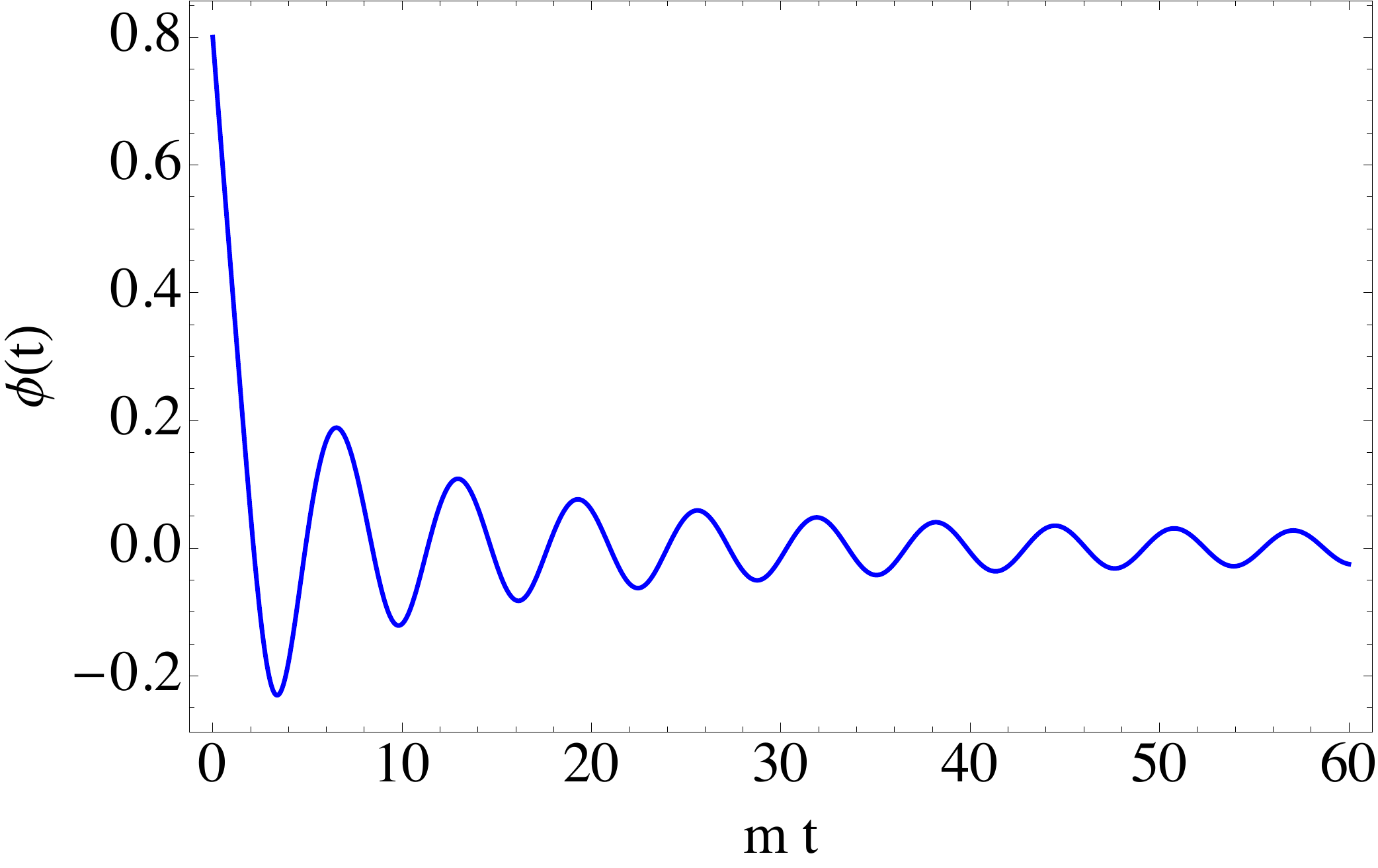}}
\hfill
\subfigure{\includegraphics[width=0.48\textwidth]{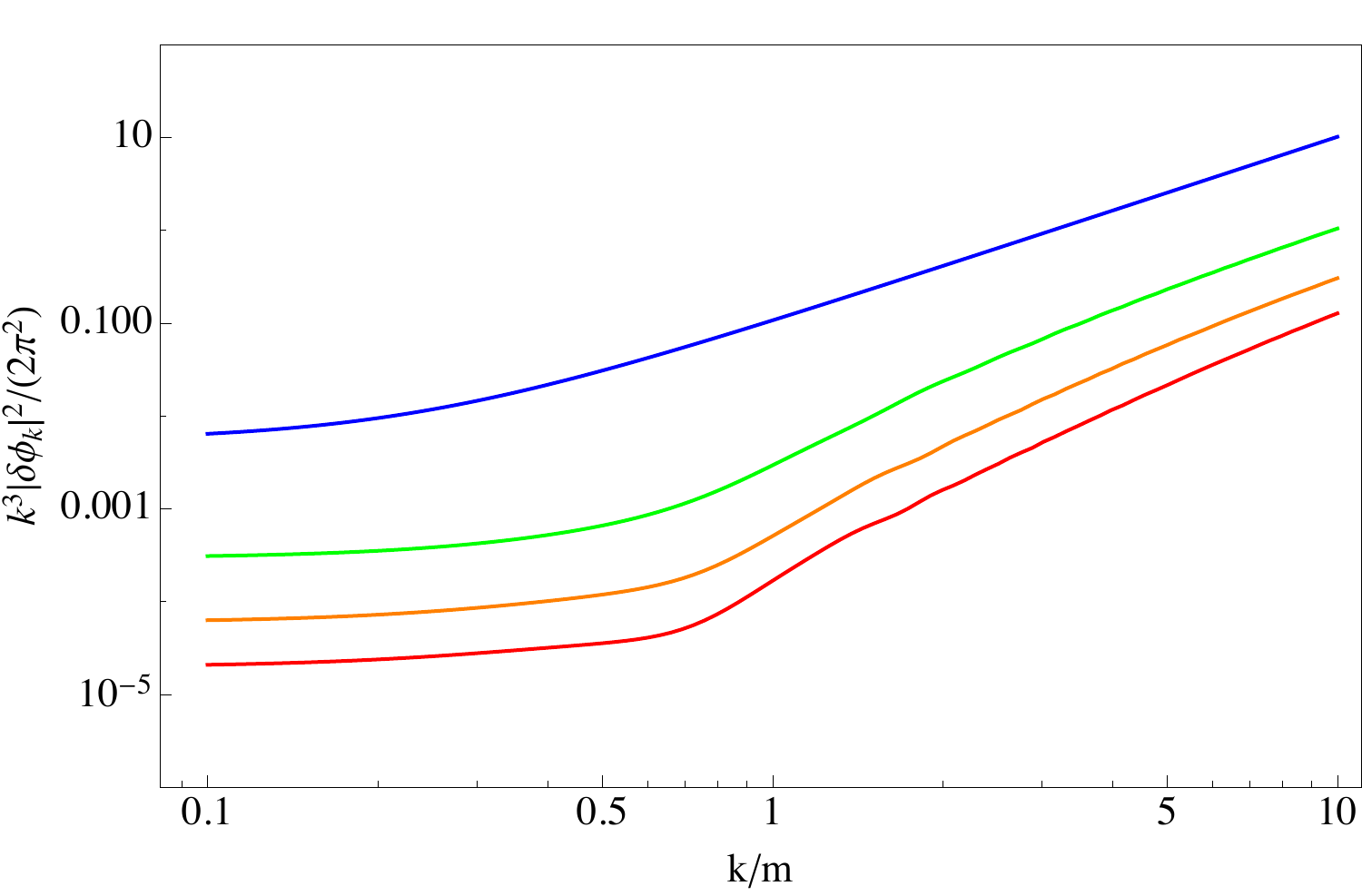}}
\caption{Numerical results from Fibre inflation. \textit{Left}: Homogeneous field evolution as a function of $m\,t$. \textit{Right:} Spectrum of the perturbations $\delta\phi_k$ at different moments in time: the initial spectrum (blue), after the first oscillation of $\phi$ (green), after four oscillations (orange), and after eight oscillations (red).} 
\label{fig:fld_and_specs}
\end{figure}

The results for the homogeneous field evolution and for the spectrum of the perturbations are shown in Figure~\ref{fig:fld_and_specs}. The left part of the figure shows the homogeneous field $\phi(t)$ as a function of the scale factor $a(t)$. One can clearly see that the amplitude of the field is strongly damped due to the expansion. On the right part of the figure the spectrum $\frac{k^3}{2\pi^2}|\phi_k|^2$ is shown in dependence of the \textit{comoving} wavenumber $k$ at different moments in time: the initial spectrum (blue), after the first oscillation of $\phi$ (green), after four oscillations (orange), and after eight oscillations (red). One can clearly see that none of the modes is significantly excited, as already expected from the Floquet analysis. The spectrum is redshifted due to the expansion and the field can be considered to stay practically homogeneous.

\bibliography{oscillon}{}

\providecommand{\href}[2]{#2}\begingroup\raggedright\begin{thebibliography}{10}

\bibitem{Coughlan:1983ci}
G.~D. Coughlan, W.~Fischler, E.~W. Kolb, S.~Raby, and G.~G. Ross, {\it
  {Cosmological Problems for the Polonyi Potential}},  {\em Phys. Lett.} {\bf
  B131} (1983) 59--64.

\bibitem{Banks:1993en}
T.~Banks, D.~B. Kaplan, and A.~E. Nelson, {\it {Cosmological implications of
  dynamical supersymmetry breaking}},  {\em Phys. Rev.} {\bf D49} (1994)
  779--787, [\href{http://xxx.lanl.gov/abs/hep-ph/9308292}{{\tt
  hep-ph/9308292}}].

\bibitem{deCarlos:1993wie}
B.~de~Carlos, J.~A. Casas, F.~Quevedo, and E.~Roulet, {\it {Model independent
  properties and cosmological implications of the dilaton and moduli sectors of
  4-d strings}},  {\em Phys. Lett.} {\bf B318} (1993) 447--456,
  [\href{http://xxx.lanl.gov/abs/hep-ph/9308325}{{\tt hep-ph/9308325}}].

\bibitem{Cicoli:2012aq}
M.~Cicoli, J.~P. Conlon, and F.~Quevedo, {\it {Dark radiation in LARGE volume
  models}},  {\em Phys. Rev.} {\bf D87} (2013), no.~4 043520,
  [\href{http://xxx.lanl.gov/abs/1208.3562}{{\tt 1208.3562}}].

\bibitem{Higaki:2012ar}
T.~Higaki and F.~Takahashi, {\it {Dark Radiation and Dark Matter in Large
  Volume Compactifications}},  {\em JHEP} {\bf 11} (2012) 125,
  [\href{http://xxx.lanl.gov/abs/1208.3563}{{\tt 1208.3563}}].

\bibitem{Allahverdi:2016yws}
R.~Allahverdi, M.~Cicoli, and F.~Muia, {\it {Affleck-Dine Baryogenesis in Type
  IIB String Models}},  {\em JHEP} {\bf 06} (2016) 153,
  [\href{http://xxx.lanl.gov/abs/1604.03120}{{\tt 1604.03120}}].

\bibitem{Allahverdi:2013noa}
R.~Allahverdi, M.~Cicoli, B.~Dutta, and K.~Sinha, {\it {Nonthermal dark matter
  in string compactifications}},  {\em Phys. Rev.} {\bf D88} (2013), no.~9
  095015, [\href{http://xxx.lanl.gov/abs/1307.5086}{{\tt 1307.5086}}].

\bibitem{Kane:2015jia}
G.~Kane, K.~Sinha, and S.~Watson, {\it {Cosmological Moduli and the
  Post-Inflationary Universe: A Critical Review}},  {\em Int. J. Mod. Phys.}
  {\bf D24} (2015), no.~08 1530022,
  [\href{http://xxx.lanl.gov/abs/1502.07746}{{\tt 1502.07746}}].

\bibitem{Kachru:2003aw}
S.~Kachru, R.~Kallosh, A.~D. Linde, and S.~P. Trivedi, {\it {De Sitter vacua in
  string theory}},  {\em Phys. Rev.} {\bf D68} (2003) 046005,
  [\href{http://xxx.lanl.gov/abs/hep-th/0301240}{{\tt hep-th/0301240}}].

\bibitem{Balasubramanian:2005zx}
V.~Balasubramanian, P.~Berglund, J.~P. Conlon, and F.~Quevedo, {\it
  {Systematics of moduli stabilisation in Calabi-Yau flux compactifications}},
  {\em JHEP} {\bf 03} (2005) 007,
  [\href{http://xxx.lanl.gov/abs/hep-th/0502058}{{\tt hep-th/0502058}}].

\bibitem{Conlon:2005ki}
J.~P. Conlon, F.~Quevedo, and K.~Suruliz, {\it {Large-volume flux
  compactifications: Moduli spectrum and D3/D7 soft supersymmetry breaking}},
  {\em JHEP} {\bf 08} (2005) 007,
  [\href{http://xxx.lanl.gov/abs/hep-th/0505076}{{\tt hep-th/0505076}}].

\bibitem{0804.3730}
M.~Gomez-Reino and C.~A. Scrucca, {\it {Constraints from F and D supersymmetry
  breaking in general supergravity theories}},  {\em Fortsch. Phys.} {\bf 56}
  (2008) 833--841, [\href{http://xxx.lanl.gov/abs/0804.3730}{{\tt 0804.3730}}].

\bibitem{0804.1073}
L.~Covi, M.~Gomez-Reino, C.~Gross, J.~Louis, G.~A. Palma, and C.~A. Scrucca,
  {\it {de Sitter vacua in no-scale supergravities and Calabi-Yau string
  models}},  {\em JHEP} {\bf 06} (2008) 057,
  [\href{http://xxx.lanl.gov/abs/0804.1073}{{\tt 0804.1073}}].

\bibitem{0706.2785}
M.~Gomez-Reino and C.~A. Scrucca, {\it {Metastable supergravity vacua with F
  and D supersymmetry breaking}},  {\em JHEP} {\bf 08} (2007) 091,
  [\href{http://xxx.lanl.gov/abs/0706.2785}{{\tt 0706.2785}}].

\bibitem{hep-th/0602246}
M.~Gomez-Reino and C.~A. Scrucca, {\it {Locally stable non-supersymmetric
  Minkowski vacua in supergravity}},  {\em JHEP} {\bf 05} (2006) 015,
  [\href{http://xxx.lanl.gov/abs/hep-th/0602246}{{\tt hep-th/0602246}}].

\bibitem{hep-th/0606273}
M.~Gomez-Reino and C.~A. Scrucca, {\it {Constraints for the existence of flat
  and stable non-supersymmetric vacua in supergravity}},  {\em JHEP} {\bf 09}
  (2006) 008, [\href{http://xxx.lanl.gov/abs/hep-th/0606273}{{\tt
  hep-th/0606273}}].

\bibitem{Gleiser:1993pt}
M.~Gleiser, {\it {Pseudostable bubbles}},  {\em Phys. Rev.} {\bf D49} (1994)
  2978--2981, [\href{http://xxx.lanl.gov/abs/hep-ph/9308279}{{\tt
  hep-ph/9308279}}].

\bibitem{Copeland:1995fq}
E.~J. Copeland, M.~Gleiser, and H.~R. Muller, {\it {Oscillons: Resonant
  configurations during bubble collapse}},  {\em Phys. Rev.} {\bf D52} (1995)
  1920--1933, [\href{http://xxx.lanl.gov/abs/hep-ph/9503217}{{\tt
  hep-ph/9503217}}].

\bibitem{Copeland:2002ku}
E.~J. Copeland, S.~Pascoli, and A.~Rajantie, {\it {Dynamics of tachyonic
  preheating after hybrid inflation}},  {\em Phys. Rev.} {\bf D65} (2002)
  103517, [\href{http://xxx.lanl.gov/abs/hep-ph/0202031}{{\tt
  hep-ph/0202031}}].

\bibitem{Broadhead:2005hn}
M.~Broadhead and J.~McDonald, {\it {Simulations of the end of supersymmetric
  hybrid inflation and non-topological soliton formation}},  {\em Phys. Rev.}
  {\bf D72} (2005) 043519, [\href{http://xxx.lanl.gov/abs/hep-ph/0503081}{{\tt
  hep-ph/0503081}}].

\bibitem{Farhi:2005rz}
E.~Farhi, N.~Graham, V.~Khemani, R.~Markov, and R.~Rosales, {\it {An Oscillon
  in the SU(2) gauged Higgs model}},  {\em Phys. Rev.} {\bf D72} (2005) 101701,
  [\href{http://xxx.lanl.gov/abs/hep-th/0505273}{{\tt hep-th/0505273}}].

\bibitem{Fodor:2006zs}
G.~Fodor, P.~Forgacs, P.~Grandclement, and I.~Racz, {\it {Oscillons and
  Quasi-breathers in the phi**4 Klein-Gordon model}},  {\em Phys. Rev.} {\bf
  D74} (2006) 124003, [\href{http://xxx.lanl.gov/abs/hep-th/0609023}{{\tt
  hep-th/0609023}}].

\bibitem{Graham:2006vy}
N.~Graham, {\it {An Electroweak oscillon}},  {\em Phys. Rev. Lett.} {\bf 98}
  (2007) 101801, [\href{http://xxx.lanl.gov/abs/hep-th/0610267}{{\tt
  hep-th/0610267}}]. [Erratum: Phys. Rev. Lett.98,189904(2007)].

\bibitem{Gleiser:2007te}
M.~Gleiser and J.~Thorarinson, {\it {A Phase transition in U(1) configuration
  space: Oscillons as remnants of vortex-antivortex annihilation}},  {\em Phys.
  Rev.} {\bf D76} (2007) 041701,
  [\href{http://xxx.lanl.gov/abs/hep-th/0701294}{{\tt hep-th/0701294}}].

\bibitem{Amin:2011hj}
M.~A. Amin, R.~Easther, H.~Finkel, R.~Flauger, and M.~P. Hertzberg, {\it
  {Oscillons After Inflation}},  {\em Phys. Rev. Lett.} {\bf 108} (2012)
  241302, [\href{http://xxx.lanl.gov/abs/1106.3335}{{\tt 1106.3335}}].

\bibitem{Achilleos:2013zpa}
V.~Achilleos, F.~K. Diakonos, D.~J. Frantzeskakis, G.~C. Katsimiga, X.~N.
  Maintas, E.~Manousakis, C.~E. Tsagkarakis, and A.~Tsapalis, {\it {Oscillons
  and oscillating kinks in the Abelian-Higgs model}},  {\em Phys. Rev.} {\bf
  D88} (2013) 045015, [\href{http://xxx.lanl.gov/abs/1306.3868}{{\tt
  1306.3868}}].

\bibitem{Gleiser:2014ipa}
M.~Gleiser and N.~Graham, {\it {Transition To Order After Hilltop Inflation}},
  {\em Phys. Rev.} {\bf D89} (2014), no.~8 083502,
  [\href{http://xxx.lanl.gov/abs/1401.6225}{{\tt 1401.6225}}].

\bibitem{Antusch:2015nla}
S.~Antusch, D.~Nolde, and S.~Orani, {\it {Hill crossing during preheating after
  hilltop inflation}},  {\em JCAP} {\bf 1506} (2015), no.~06 009,
  [\href{http://xxx.lanl.gov/abs/1503.06075}{{\tt 1503.06075}}].

\bibitem{Antusch:2015ziz}
S.~Antusch and S.~Orani, {\it {Impact of other scalar fields on oscillons after
  hilltop inflation}},  {\em JCAP} {\bf 1603} (2016), no.~03 026,
  [\href{http://xxx.lanl.gov/abs/1511.02336}{{\tt 1511.02336}}].

\bibitem{Bond:2015zfa}
J.~R. Bond, J.~Braden, and L.~Mersini-Houghton, {\it {Cosmic bubble and domain
  wall instabilities III: The role of oscillons in three-dimensional bubble
  collisions}},  {\em JCAP} {\bf 1509} (2015), no.~09 004,
  [\href{http://xxx.lanl.gov/abs/1505.02162}{{\tt 1505.02162}}].

\bibitem{Liu:2017hua}
J.~Liu, Z.-K. Guo, R.-G. Cai, and G.~Shiu, {\it {Gravitational Waves from
  Oscillons with Cuspy Potentials}},
  \href{http://xxx.lanl.gov/abs/1707.09841}{{\tt 1707.09841}}.

\bibitem{Antusch:2016con}
S.~Antusch, F.~Cefala, and S.~Orani, {\it {Gravitational waves from oscillons
  after inflation}},  {\em Phys. Rev. Lett.} {\bf 118} (2017), no.~1 011303,
  [\href{http://xxx.lanl.gov/abs/1607.01314}{{\tt 1607.01314}}].

\bibitem{1304.6094}
S.-Y. Zhou, E.~J. Copeland, R.~Easther, H.~Finkel, Z.-G. Mou, and P.~M. Saffin,
  {\it {Gravitational Waves from Oscillon Preheating}},  {\em JHEP} {\bf 10}
  (2013) 026, [\href{http://xxx.lanl.gov/abs/1304.6094}{{\tt 1304.6094}}].

\bibitem{Amin:2014eta}
M.~A. Amin, M.~P. Hertzberg, D.~I. Kaiser, and J.~Karouby, {\it
  {Nonperturbative Dynamics Of Reheating After Inflation: A Review}},  {\em
  Int. J. Mod. Phys.} {\bf D24} (2014) 1530003,
  [\href{http://xxx.lanl.gov/abs/1410.3808}{{\tt 1410.3808}}].

\bibitem{Barnaby:2009wr}
N.~Barnaby, J.~R. Bond, Z.~Huang, and L.~Kofman, {\it {Preheating After Modular
  Inflation}},  {\em JCAP} {\bf 0912} (2009) 021,
  [\href{http://xxx.lanl.gov/abs/0909.0503}{{\tt 0909.0503}}].

\bibitem{Amin:2010dc}
M.~A. Amin, R.~Easther, and H.~Finkel, {\it {Inflaton Fragmentation and
  Oscillon Formation in Three Dimensions}},  {\em JCAP} {\bf 1012} (2010) 001,
  [\href{http://xxx.lanl.gov/abs/1009.2505}{{\tt 1009.2505}}].

\bibitem{Amin:2013ika}
M.~A. Amin, {\it {K-oscillons: Oscillons with noncanonical kinetic terms}},
  {\em Phys. Rev.} {\bf D87} (2013), no.~12 123505,
  [\href{http://xxx.lanl.gov/abs/1303.1102}{{\tt 1303.1102}}].

\bibitem{Felder:2006cc}
G.~N. Felder and L.~Kofman, {\it {Nonlinear inflaton fragmentation after
  preheating}},  {\em Phys. Rev.} {\bf D75} (2007) 043518,
  [\href{http://xxx.lanl.gov/abs/hep-ph/0606256}{{\tt hep-ph/0606256}}].

\bibitem{Brax:2010ai}
P.~Brax, J.-F. Dufaux, and S.~Mariadassou, {\it {Preheating after Small-Field
  Inflation}},  {\em Phys. Rev.} {\bf D83} (2011) 103510,
  [\href{http://xxx.lanl.gov/abs/1012.4656}{{\tt 1012.4656}}].

\bibitem{Kofman:1994rk}
L.~Kofman, A.~D. Linde, and A.~A. Starobinsky, {\it {Reheating after
  inflation}},  {\em Phys. Rev. Lett.} {\bf 73} (1994) 3195--3198,
  [\href{http://xxx.lanl.gov/abs/hep-th/9405187}{{\tt hep-th/9405187}}].

\bibitem{Kofman:1997yn}
L.~Kofman, A.~D. Linde, and A.~A. Starobinsky, {\it {Towards the theory of
  reheating after inflation}},  {\em Phys. Rev.} {\bf D56} (1997) 3258--3295,
  [\href{http://xxx.lanl.gov/abs/hep-ph/9704452}{{\tt hep-ph/9704452}}].

\bibitem{Zhou:2013tsa}
S.-Y. Zhou, E.~J. Copeland, R.~Easther, H.~Finkel, Z.-G. Mou, and P.~M. Saffin,
  {\it {Gravitational Waves from Oscillon Preheating}},  {\em JHEP} {\bf 10}
  (2013) 026, [\href{http://xxx.lanl.gov/abs/1304.6094}{{\tt 1304.6094}}].

\bibitem{Baumann:2014nda}
D.~Baumann and L.~McAllister, {\em {Inflation and String Theory}}.
\newblock Cambridge University Press, 2015.

\bibitem{Cicoli:2016olq}
M.~Cicoli, K.~Dutta, A.~Maharana, and F.~Quevedo, {\it {Moduli Vacuum
  Misalignment and Precise Predictions in String Inflation}},  {\em JCAP} {\bf
  1608} (2016), no.~08 006, [\href{http://xxx.lanl.gov/abs/1604.08512}{{\tt
  1604.08512}}].

\bibitem{Landete:2017amp}
A.~Landete, F.~Marchesano, G.~Shiu, and G.~Zoccarato, {\it {Flux Flattening in
  Axion Monodromy Inflation}},  {\em JHEP} {\bf 06} (2017) 071,
  [\href{http://xxx.lanl.gov/abs/1703.09729}{{\tt 1703.09729}}].

\bibitem{Linde:1981mu}
A.~D. Linde, {\it {A New Inflationary Universe Scenario: A Possible Solution of
  the Horizon, Flatness, Homogeneity, Isotropy and Primordial Monopole
  Problems}},  {\em Phys. Lett.} {\bf 108B} (1982) 389--393.

\bibitem{Izawa:1996dv}
K.~I. Izawa and T.~Yanagida, {\it {Natural new inflation in broken
  supergravity}},  {\em Phys. Lett.} {\bf B393} (1997) 331--336,
  [\href{http://xxx.lanl.gov/abs/hep-ph/9608359}{{\tt hep-ph/9608359}}].

\bibitem{Izawa:1997df}
K.~I. Izawa, M.~Kawasaki, and T.~Yanagida, {\it {Dynamical tuning of the
  initial condition for new inflation in supergravity}},  {\em Phys. Lett.}
  {\bf B411} (1997) 249--255,
  [\href{http://xxx.lanl.gov/abs/hep-ph/9707201}{{\tt hep-ph/9707201}}].

\bibitem{Senoguz:2004ky}
V.~N. Senoguz and Q.~Shafi, {\it {New inflation, preinflation, and
  leptogenesis}},  {\em Phys. Lett.} {\bf B596} (2004) 8--15,
  [\href{http://xxx.lanl.gov/abs/hep-ph/0403294}{{\tt hep-ph/0403294}}].

\bibitem{Boubekeur:2005zm}
L.~Boubekeur and D.~Lyth, {\it {Hilltop inflation}},  {\em JCAP} {\bf 0507}
  (2005) 010, [\href{http://xxx.lanl.gov/abs/hep-ph/0502047}{{\tt
  hep-ph/0502047}}].

\bibitem{Antusch:2015vna}
S.~Antusch, F.~Cefala, D.~Nolde, and S.~Orani, {\it {Parametric resonance after
  hilltop inflation caused by an inhomogeneous inflaton field}},  {\em JCAP}
  {\bf 1602} (2016), no.~02 044,
  [\href{http://xxx.lanl.gov/abs/1510.04856}{{\tt 1510.04856}}].

\bibitem{Felder:2000hq}
G.~N. Felder and I.~Tkachev, {\it {LATTICEEASY: A Program for lattice
  simulations of scalar fields in an expanding universe}},  {\em Comput. Phys.
  Commun.} {\bf 178} (2008) 929--932,
  [\href{http://xxx.lanl.gov/abs/hep-ph/0011159}{{\tt hep-ph/0011159}}].

\bibitem{Polarski:1995jg}
D.~Polarski and A.~A. Starobinsky, {\it {Semiclassicality and decoherence of
  cosmological perturbations}},  {\em Class. Quant. Grav.} {\bf 13} (1996)
  377--392, [\href{http://xxx.lanl.gov/abs/gr-qc/9504030}{{\tt
  gr-qc/9504030}}].

\bibitem{Khlebnikov:1996mc}
S.~{\relax Yu}. Khlebnikov and I.~I. Tkachev, {\it {Classical decay of
  inflaton}},  {\em Phys. Rev. Lett.} {\bf 77} (1996) 219--222,
  [\href{http://xxx.lanl.gov/abs/hep-ph/9603378}{{\tt hep-ph/9603378}}].

\bibitem{GarciaBellido:2007af}
J.~Garcia-Bellido, D.~G. Figueroa, and A.~Sastre, {\it {A Gravitational Wave
  Background from Reheating after Hybrid Inflation}},  {\em Phys. Rev.} {\bf
  D77} (2008) 043517, [\href{http://xxx.lanl.gov/abs/0707.0839}{{\tt
  0707.0839}}].

\bibitem{Gukov:1999ya}
S.~Gukov, C.~Vafa, and E.~Witten, {\it {CFT's from Calabi-Yau four folds}},
  {\em Nucl. Phys.} {\bf B584} (2000) 69--108,
  [\href{http://xxx.lanl.gov/abs/hep-th/9906070}{{\tt hep-th/9906070}}].
  [Erratum: Nucl. Phys.B608,477(2001)].

\bibitem{Giddings:2001yu}
S.~B. Giddings, S.~Kachru, and J.~Polchinski, {\it {Hierarchies from fluxes in
  string compactifications}},  {\em Phys. Rev.} {\bf D66} (2002) 106006,
  [\href{http://xxx.lanl.gov/abs/hep-th/0105097}{{\tt hep-th/0105097}}].

\bibitem{LoaizaBrito:2005fa}
O.~Loaiza-Brito, J.~Martin, H.~P. Nilles, and M.~Ratz, {\it {Log(M(Pl) /
  m(3/2))}},  {\em AIP Conf. Proc.} {\bf 805} (2006) 198--204,
  [\href{http://xxx.lanl.gov/abs/hep-th/0509158}{{\tt hep-th/0509158}}].
  [,198(2005)].

\bibitem{movieslink}
{Link to simulations},
  ``\url{https://particlesandcosmology.unibas.ch/downloads/oscillons-from-string-moduli-movies.html}.''

\bibitem{TheLIGOScientific:2014jea}
{\bf LIGO Scientific} Collaboration, J.~Aasi {\em et.~al.}, {\it {Advanced
  LIGO}},  {\em Class. Quant. Grav.} {\bf 32} (2015) 074001,
  [\href{http://xxx.lanl.gov/abs/1411.4547}{{\tt 1411.4547}}].

\bibitem{ET}
{\it {\rm Einstein Telescope:} concept and design},  {\em
  {http://www.et-gw.eu/index.php/etdsdocument}}.

\bibitem{Cicoli:2017shd}
M.~Cicoli, I.~Garc"a-Etxebarria, C.~Mayrhofer, F.~Quevedo, P.~Shukla, and
  R.~Valandro, {\it {Global Orientifolded Quivers with Inflation}},
  \href{http://xxx.lanl.gov/abs/1706.06128}{{\tt 1706.06128}}.

\bibitem{Cicoli:2008gp}
M.~Cicoli, C.~P. Burgess, and F.~Quevedo, {\it {Fibre Inflation: Observable
  Gravity Waves from IIB String Compactifications}},  {\em JCAP} {\bf 0903}
  (2009) 013, [\href{http://xxx.lanl.gov/abs/0808.0691}{{\tt 0808.0691}}].

\bibitem{Burgess:2016owb}
C.~P. Burgess, M.~Cicoli, S.~de~Alwis, and F.~Quevedo, {\it {Robust Inflation
  from Fibrous Strings}},  {\em JCAP} {\bf 1605} (2016), no.~05 032,
  [\href{http://xxx.lanl.gov/abs/1603.06789}{{\tt 1603.06789}}].

\bibitem{Cicoli:2016chb}
M.~Cicoli, D.~Ciupke, S.~de~Alwis, and F.~Muia, {\it {$\alpha'$ Inflation:
  moduli stabilisation and observable tensors from higher derivatives}},  {\em
  JHEP} {\bf 09} (2016) 026, [\href{http://xxx.lanl.gov/abs/1607.01395}{{\tt
  1607.01395}}].

\bibitem{Cicoli:2016xae}
M.~Cicoli, F.~Muia, and P.~Shukla, {\it {Global Embedding of Fibre Inflation
  Models}},  {\em JHEP} {\bf 11} (2016) 182,
  [\href{http://xxx.lanl.gov/abs/1611.04612}{{\tt 1611.04612}}].

\end{thebibliography}\endgroup
\bibliographystyle{JHEP}
\end{document}